\documentstyle[12pt]{article}

\def\hybrid{\topmargin -20pt   \oddsidemargin 0pt
      \headheight 0pt   \headsep 0pt
      \textwidth 6.25in 
      \textheight 9.5in 
      \marginparwidth .875in
      \parskip 5pt plus 1pt   \jot = 1.5ex}

\hybrid

\def\beqa{\begin{eqnarray}}
\def\eeqa{\end{eqnarray}}
\def\x{\times}
\def\ox{\otimes}
\def\o+{\oplus}
\def\ra{\rightarrow}
\def\lra{\longrightarrow}
\def\Lra{\Longrightarrow}

\def\leria{\leftrightarrow}
\def\llra{\longleftrightarrow}

\def\hra{\hookrightarrow}
\def\exp{\mbox{exp}}
\def\length{\mbox{length}}
\def\tors{\mbox{tors}}
\def\back{\backslash}
\def\om{\omega}
\def\th{\theta}
\def\al{\alpha}
\def\be{\beta}
\def\ga{\gamma}
\def\de{\delta}

\def\la{\lambda}
\def\si{\sigma}
\def\Y{\Upsilon}
\def\pa{\partial}
\def\sign{\mbox{sign}}

\def\vol{\mbox{vol}}
\def\mod{\mbox{mod}}
\def\Im{\mbox{Im}}
\def\Re{\mbox{Re}}
\def\Sym{\mbox{Sym}}
\def\im{\mbox{im}}
\def\deg{\mbox{deg}}
\def\we{\wedge}

\def\u{\tilde{u}}

\def\C{{\cal C}}
\def\D{{\cal D}}

\def\F{{\cal F}}
\def\G{{\cal G}}
\def\H{{\cal H}}

\def\L{{\cal L}}
\def\M{{\cal M}}
\def\N{{\cal N}}

\def\S{{\cal S}}
\def\T{{\cal T}}

\def\De{\Delta}
\def\Ga{\Gamma}
\def\La{\Lambda}
\def\Om{\Omega}
\def\Si{\Sigma}

\newcommand{\resetcounter}{\setcounter{equation}{0}}

\begin{document}

\begin{titlepage}

\begin{flushright}
\hfill{HU-EP-02/58} \\
\hfill{hep-th/0212211}
\end{flushright}

\vspace{15pt}

\vspace{1cm}
\begin{center}

{\Large Superpotentials for $M$-theory on a $G_2$ holonomy manifold}

\vspace{.2cm}

{\Large  and Triality symmetry}

\vspace{30pt}

{Gottfried Curio}


{\it  Humboldt Universit\"at zu Berlin,
Institut f\"ur Physik,\\
Invalidenstrasse 110, 10115 Berlin,
Germany}
\vspace{20pt}


\end{center}

For $M$-theory on the $G_2$ holonomy manifold
given by the cone on ${\bf S^3}\x {\bf S^3}$
we consider the superpotential
generated by membrane instantons
and study its transformations properties,
especially under monodromy transformations
and triality symmetry. We find
that the latter symmetry is, essentially, even a symmetry
of the superpotential.
As in Seiberg/Witten theory, where a flat bundle
given by the periods of an universal elliptic curve over
the $u$-plane occurs,
here a flat bundle related to the Heisenberg group appears
and the relevant universal object over the moduli space
is related to hyperbolic geometry.

\end{titlepage}

\newpage

\section{\label{introduction}\large {\bf Introduction}}

The conifold transition among Calabi-Yau manifolds in type II
string theory has an asymmetrical character:
an ${\bf S^3}$ is exchanged with an ${\bf S^2}$.
When the situation is lifted to $M$-theory
the resulting geometries become completely symmetrical
[\ref{AMV}], [\ref{AW}]:
the two small resolutions given by the ${\bf S^2}$'s
(related by a IIA flop)
on the one side of the transition
become ${\bf S^3}$'s as well (Hopf fibred by the
$M$-theory circle ${\bf S^1_{11}}$; this concerns, in type IIA
language, the situation where
on this side one unit of RR flux on the ${\bf S^2}$ is turned on
and one has a $D6$ brane on the ${\bf S^3}$ of the
deformed conifold side; cf. also [\ref{Acha}]).
This symmetry of the three ${\bf S^3}$'s
is the triality symmetry $\Sigma_3$ of $M$-theory
on the corresponding non-compact
$G_2$ manifold which is a suitable deformation $X_7$ of a cone over
${\bf S^3}\x{\bf S^3}=SU(2)^3/SU(2)_D$
and comes naturally in three equivalent
versions $X_1, X_2, X_3$.

In [\ref{AW}] it was pointed out that for general reasons
the superpotential should have
an anti-invariant behaviour under the triality symmetry,
i.e. it should transform with the sign-character
of $\Sigma_3$ (cf. app. \ref{triality appendix}).
For this recall the action of the
order two element $\al$ coming from the interchange $g_2\leria g_3$
of $SU(2)$ elements in $(g_1,g_2,g_3)$, which
on $X_1={\bf R^4} \x {\bf S^3}$
(when gauging $g_3$ to $1$) is
given by $(g_1,g_2)\ra (g_1g_2^{-1}, g_2^{-1})$ and induces an orientation
reversing endomorphism on the tangent space (at a fixed point with $g_2=1$).
It acts as an R-symmetry under which the superpotential transforms odd.
Then a triality symmetric superpotential was considered [\ref{AW}]
which was suggested by global symmetry considerations on the moduli
space ${\cal N}={\bf P^1_t}$
(triality symmetric behaviour meaning here that
it transforms anti-invariantly, i.e. with the sign character).
The simplest possibility was
(with $\Sigma_3$
operating by $t\ra \om t$ where $\om =e^{2\pi i /3}$ and $t\ra 1/t$)
\beqa
W(t)=\frac{t^3-1}{t^3+1}
\eeqa

As we will show one can actually arrive at a closely related result
on a different route by considering
the actual non-perturbative superpotential
generated by membrane instantons. The analytic continuation of these
local (on ${\cal N}$) contributions gives an essentially
symmetric superpotential (cf. below).
Crucial will be a non-linear realization of the triality symmetry.
Under the following action of the triality group $\Sigma_3$
\beqa
\label{Sl2 realisation}
\begin{array}{ccccccc}
z & & & \be z = {\frac{1}{1-z}} & & & \be^2 z =\frac{z-1}{z} \\
\al z = \frac{1}{z} & & & \al \be z = 1-z & & & \al \be^2 z = \frac{z}{z-1}
\end{array}
\eeqa
the holomorphic observables given by the
variables $\eta_i$ form a ${\bf Z_3}$ orbit:
$\eta_{i-1}=\be \eta_i \;, \; \eta_{i+1}=\be^2 \eta_i$.
The variables $u_i$, given by the membrane
instanton amplitudes and constituting local coordinates at the
semiclassical ends of the global moduli space, are
(when globally analytically continued;
they are properly only first order variables)
holomorphically related to the $\eta_i$.
One has a relation
$\eta_3=\frac{1}{1-\eta_1}$ then also for the (global) $u_i$;
note the corresponding map in [\ref{AMV}] ($t$ and $V$
the sizes of ${\bf S^2}$ and ${\bf S^3}$ in type IIA; for $N=1$ there)
\beqa
\frac{1}{1-e^t}\sim e^V
\eeqa

We argue (in the framework of [\ref{AW}]) that
the full multi-cover membrane instanton superpotential
is given by the dilogarithm (cf. [\ref{AVII}])
\beqa
\label{dilog}
W(u)=Li(u)=\sum_{n=1}\frac{u^n}{n^2}
\eeqa

To accomplish this we reinterprete the treatment
of the one-instanton amplitude
in [\ref{AW}]. There the evaluation of the vev
$<\int_{D_i}C \cdot \int_{Q_i}*G>$ via an
auxiliary classical four-form field
$G$ was done in connection with the derivation of an
ordinary interaction from the superpotential.
The scalar potential computation for the superpotential
can be argued to describe not only the
one-instanton contribution
but the full instanton series.

Remarkably,
the actual superpotential given by the multi-cover membrane instantons
knows 'by itself', via its analytic continuation, that it entails a
triality symmetry.
For the function $W(u)$ satisfies the following symmetry
relations which will ensure that the dilogarithm superpotential
is {\em compatible with triality symmetry}
(i.e. that the local (on ${\cal N}$)
membrane instanton contributions fit together globally in this sense)
\beqa
\label{W trial trafo}
W(\frac{1}{u})&=&-W(u)-\zeta(2)-\frac{1}{2}\log^2(-u)\nonumber\\
W(1-u)&=&-W(u)+\zeta(2)-\log u \log (1-u)
\eeqa
The symmetry relations (given here for the transformation
under $\al$ and $\al\be$; from these all others are derived)
have the consequence that,
up to the elementary corrections provided by the products of two log's
and $\zeta(2)$, the $W(u)$ superpotential is invariant under
the transformations in the first line of (\ref{sym}) and transforms
with a minus sign under the mappings of the second line.
That is the 'local' superpotential transforms
(under the $Sl(2)$ action) up to the elementary corrections
with the sign character just as the global superpotential did (under
the linear action) and as it should a priori. In other words triality
symmetry is in this sense 'dynamical': it holds on the level of the
superpotential.

Similarly, and much more trivially, the geometric
world-sheet instanton series 'knows' about the ${\bf S^2}$-flop
transition. Note
the analogous behaviour of the instanton sums
$I_{ws}(\frac{1}{q})= -I_{ws}(q)-1$ and (\ref{W trial trafo})
describing the multi-coverings of the supersymmetric cycles
provided by the holomorphic ${\bf S^2}$ in the string world-sheet case
and the associative ${\bf S^3}$ in the membrane case, respectively.

Note that when (\ref{dilog}) is globally analytically continued over
the critical circle (the boundary of its convergence disk)
one gets monodromy contributions.
The monodromy representation of the fundamental group
$\pi_1 ({\bf P^1} \backslash  \{ 0,1,\infty \})$
will describe the multi-valuedness
$Li(z) \buildrel l_1 \over \lra  Li(z) -2\pi i \log z$.
The relevant local system is described by a bundle,
flat with respect to a suitable connection.
Like in the case of the logarithm where the
monodromy of $\log z$ around $z=0$ is captured by the monodromy matrix
$M(l_0)=\left(\tiny{\begin{array}{cc}1&2\pi i\\0&1\end{array}}\right)$
and the monodromy group is given by ${\cal U}_{\bf Z}\hra {\cal U}_{\bf C}$
where ${\cal U}$ denotes the upper triangular group
$\scriptsize{\left(\begin{array}{cc}1&*\\0&1\end{array}\right)}\subset
Sl(2)$
(the embedding of ${\cal U}_{\bf Z}$ in ${\cal U}_{\bf C}$
may include here the factor of $2\pi i$),
the corresponding generalisation in the case of the dilogarithm
involves upper triangular $3\x 3$ matrices,
i.e. one gets again admixtures
from 'lower' components when one considers constants,
ordinary logarithms and the dilogarithm all at the same time.
One then finds a function
\beqa
{\cal L}(z)=\Im  \; Li(z) - \Im \log \be z \; \Re \log z
\eeqa
which because of its $\pi_1$-invariance is {\em single-valued}.
Furthermore the quantity
${\cal L}$ now transforms {\em precisely} anti-invariantly under
$\Sigma_3$, i.e. without any correction terms
(just as the analogous single-valued cousin $\Re \log z $ of the logarithm
has anti-invariant transformation behaviour under the duality group
${\bf Z_2}$ with non-trivial element $\al:z\ra 1/z$). So both
deviations from
the expected transformation properties are cured at the same time.

We will describe a number of ways to understand this anti-invariant
transformation behaviour of $\L(z)$. Most importantly for
the interpretation via a string theory duality we
propose in the outlook
is the geometrical interpretation as the hyperbolic volume
\beqa
\label{intro vol}
\vol \, \Delta(z)={\cal L}(z)
\eeqa
of an ideal tetrahedron in hyperbolic three space ${\bf H_3}$ with vertices
$z_1,z_2,z_3, z_4$ lying on the boundary ${\bf P^1_C}$ of ${\bf H_3}$
which is manifestly independent of the numbering of the vertices
except that the orientation changes under odd renumberings,
showing the anti-invariant transformation behaviour (with $z$ the
cross ratio and $\Sigma_4 \ra \Sigma_3^{Sl(2)}$ after the gauging
$(z_1,z_2,z_3, z_4)\ra (0,1,\infty, z)$). Corresponding to this 3-volume
interpretation one has a 1-volume interpretation
(in the upper half-space model for ${\bf H_3}$)
\beqa
\label{intro len}
\length(\ga_z)=\Re \log z
\eeqa
Here $\ga_z$ is the path on the $j$-axis from $j$ to $|z|j$.
As described in the outlook it is natural to consider
(\ref{intro vol}) and (\ref{intro len}) together
(cf. (\ref{vol and len}) and (\ref{3 vectors})).

It may be worth mentioning that by
our description of the global relations on the quantum moduli space
we get two simple reinterpretations of the membrane anomaly
\beqa
\label{membrane anomaly}
\int_{D_1}C+\int_{D_2}C+\int_{D_3}C=\pi
\eeqa
First, the non-linear
$Sl(2,{\bf Z})$
realisation (\ref{Sl2 realisation})
of $\Si_3$, which connects the
three different quantities in question by a 'global' relation,
makes (\ref{membrane anomaly}) manifest (for $z$ being some $\eta_j$)
\beqa
\label{manifest betaprod}
z\cdot \be z \cdot \be^2 z = -1
\eeqa
Moreover (\ref{manifest betaprod}) is already a consequence of
having a global $\Si_3$ symmetry at all (cf. (\ref{betaprod})).
And secondly, in the dual hyperbolic model
a relation corresponding to (\ref{membrane anomaly})
(cf. (\ref{mapping of moduli spaces local version}))
becomes just the angle sum in an {\em euclidean} triangle
(the two explanations are related, cf. (\ref{euclid angle sum}))
\beqa
\label{euclid angle sum intro}
\al + \be + \ga = \pi
\eeqa

We also compare $W_{mem}$ to a superpotential
induced by $G$-flux
\beqa
W_G=\int_X (C+i\Y) \we G
\eeqa
A useful analogy is
provided by the mass breaking of $N=2$ $SU(2)$ gauge theory by the
tree-level term $W_{tree}=m \, tr \Phi^2$
whose quantum corrected version is
$W_{qu}=mu$ occuring in Seiberg/Witten theory:
it is given by a flux induced
superpotential $W_H=\int_{\tilde{Z}}\Om \we H_3$
on the Calabi-Yau $\tilde{Z}$ in type IIB (mirror dual
to the type IIA Calabi-Yau $Z$ which describes the string embedding
of the Seiberg/Witten theory)
in the double scaling limit [\ref{TV}], [\ref{CIV}] (crucial for
this reinterpretaion is that the field theory quantity $u$ occurs, in
the appropriate limit, among the Calabi-Yau periods).

Further we extend the theory to the case of singularities
of codimension four, describing
four-dimensional non-abelian gauge theories in different phases.
For some further relations to type IIA string theory and to
five-dimensional Seiberg/Witten theory see [\ref{5dim}].

For the speculative global interpretation developed in the outlook
think of $\Delta(u)$, or its generalisation to a
hyperbolic 3-manifold, as
playing the role of the Seiberg/Witten elliptic curve $E_u$
over the $u$-plane\footnote{think of a different copy of $H_3$
over each point $u$ in ${\bf P^1}$ as ambient space
for $\Delta(u)$ just as one has copies of the Weierstrass embedding plane
${\bf P^2}_{x,y,z}$ for $E_u$}.
The relevant non-perturbative quantity is
in both cases computed by a geometric period on the object varying
over the moduli space.
One might understand this as a computation of a $M$-theory
superpotential dual to the original membrane instanton sum
(analogous to the mentioned mass braking $W=mu$ in Seiberg/Witten, computed
in the stringy embedding from a flux superpotential
[\ref{TV}] where $u$ is a period of the type IIB mirror Calabi-Yau).
If the Calabi-Yau's are $K3$ fibered over a base ${\bf P^1}$ then the (single)
Seiberg/Witten curve can be understood as being fibered over the same
${\bf P^1}$ (with discrete fibre a spectral set related to $H^2(K3)$).
So in the general situation of $M$-theory on a $G_2$ holonomy manifold
$X_7$, $K3$ fibered over ${\bf S^3}$,
one may try to compare the quantum expression given by the membrane
instanton sum to a period coming from a ($G$-flux ?) superpotential
on a dual manifold $Y_7$
(a different $K3$ fibration over $B_3$, similar to some CY situations),
respectively to a period in the
'thinned out' (spectral)
version of $Y_7$ given by a hyperbolic 3-manifold $M_3$
(where the complexified Chern-Simons invariant is computed; this will
be described in more detail elsewhere [\ref{global paper}]).

The paper is organised as follows.
In {\em Section 2} we recall, following [\ref{AW}],
the $G_2$ holonomy manifold $X_7={\bf S^3} \x {\bf R^4}$, the quantum
moduli space ${\cal N}$ and the membrane anomaly.
In {\em Section 3} we describe the crucial non-linear realization
of the triality symmetry.
In {\em Section 4} we recall the treatment of the superpotential
$W(t)$ by global arguments on ${\cal N}$ and then
describe the local approach to the
superpotential by summing up the membrane instantons and
investigate its deviations from strict anti-invariance
with respect to the triality symmetry. Using the
study of the monodromy representation
(describing the Heisenberg bundle and the superpotential as its section)
we describe how one can, at the cost of introducing some
non-holomorphy, replace the notion of section by a function $\L$.
$\L$ is shown in four ways to transform anti-invariantly;
one of them uses hyperbolic geometry by giving $\L$ an interpretation as
a hyperbolic volume.
We also compare with a flux-induced superpotential.
In {\em Section 5} we extend to the case of singularities of codimension
four, describing
four-dimensional non-abelian gauge theories in different phases.
In the {\em Outlook} we compare the hyperbolic deformation moduli space
with the Seiberg/Witten set-up and interprete all findings as describing
a dual superpotential computation with the hyperbolic 3-simplex
playing the role of the Seiberg/Witten curve.
We indicate that the theory
should extend to cover general $K3$ fibered
{\em compact} $G_2$ manifolds (and global hyperbolic 3-manifolds).
In the {\em appendix} we study the representation of $\Si_3$
and give two  proofs of the anti-invariance of
$\L$. Furthermore we give some background concerning the monodromy
representation of $Li$, the hyperbolic geometry (including the volume
computation of an ideal tetrahedron),
and the cohomological interpretation.

\newpage
\section{\large {\bf The $G_2$ manifold over ${\bf S^3}\x {\bf S^3}$
and its moduli space }}
\resetcounter
The three manifolds $X_i$ (cf. [\ref{AW}]) are cones over
$Y= {\bf S^3}\x {\bf S^3}=SU(2)^3/SU(2)_D$
where $Y$ carries the (up to scaling) unique (Einstein)
metric
with $SU(2)^3$ (acting from the left)
and $\Sigma_3$ symmetry
($da^2$ stands for $-Tr (a^{-1}da)^2$ where $a=g_2g_3^{-1},
b=g_3g_1^{-1}, c=g_1g_2^{-1}$)
\beqa
d\Omega^2=\frac{1}{36}(da^2+db^2+dc^2)
\eeqa
The images $D_i$ in $Y$ of the three
$SU(2)$ factors fulfill the triality symmetric relation
\beqa
\label{sumrel}
D_1+D_2+D_3=0
\eeqa
(as three-cycles in homology)
indicating that there are actually only two ${\bf S^3}$'s.
The singular manifold $X^{sing}$ is a cone over
$Y$ ($r$ the radial coordinate)
\beqa
\label{cone}
ds^2=dr^2+r^2d\Omega^2
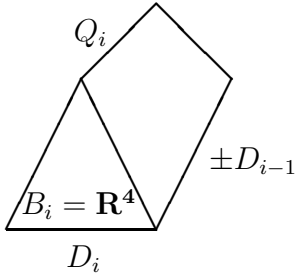
\eeqa
\unitlength1cm
\begin{picture}(15,3.5)
\thicklines
\put(0,0.5){\line(1,2){1}}
\put(0,0.5){\line(1,0){2}}
\put(2,0.5){\line(-1,2){1}}
\put(1,2.5){\line(1,1){1}}
\put(2,0.5){\line(1,2){1}}
\put(3,2.5){\line(-1,1){1}}
\put(0.9,3){$Q_i$}
\put(0.2,0.7){$B_i={\bf R^4}$}
\put(0.8,0){$D_i$}
\put(2.7,1.3){$\pm D_{i-1}$}
\put(5,3){\parbox[t]{10cm}{When embedded in one of the
$X_i\cong {\bf R^4} \x {\bf S^3}$
where the i'th $SU(2)$ is filled in to a $B_i={\bf R^4}$
one has $D_i\simeq 0$. The remaining three-sphere
which sits at the center of $X_i$, corresponding to the value
$0$ in the ${\bf R^4}$ or $r=r_0$ in (\ref{metric}), is
called $Q_i$. It is homologous to $\pm D_{i-1}\simeq \mp D_{i+1}$.}}
\end{picture}\\
The $G_2$ manifold $X$ has a covariant constant three-form $\Upsilon$
(resp. four-form $*\Upsilon$).
The modulus $\vol(Q_i)\sim r_0^3$ is not dynamical but more
like a coupling constant specified at infinity.
The deformed manifold $X={\bf R^4} \x {\bf S^3}$
(which near infinity is asymptotic to
(and for $r_0\ra 0$ reduces to) the cone (\ref{cone}))
has the $G_2$ holonomy metric ($r\in [r_0,\infty)$)
\beqa
\label{metric}
ds^2=\frac{dr^2}{1-(\frac{r_0}{r})^3}
+\frac{r^2}{36}\Bigl( da^2+db^2+dc^2
-(\frac{r_0}{r})^3 (da^2-\frac{1}{2}db^2+dc^2) \Bigr)
\eeqa
Let us examine the metric perturbations which preserve $G_2$ holonomy.
Using a new radial coordinate $y$ with
$\frac{dr^2}{1-(r_0/r)^3}=dy^2$,
provided at large $r$,
to the accuracy needed, by
\beqa
y=r\, \Bigl(1-\frac{1}{4}(\frac{r_0}{r})^3+{\cal O}( (\frac{r_0}{r})^6)\Bigr)
\;\;\; ,
\eeqa
one gets for the metric (with $(f_1,f_2,f_3)=(1,-2,1)$ and
up to terms $y^2{\cal O}( (\frac{r_0}{y})^6 )$)
\beqa
\label{y metric}
ds^2=dy^2+\frac{y^2}{36}\Bigl( da^2+db^2+dc^2
-\frac{1}{2} (\frac{r_0}{y})^3 (f_1\, da^2+f_2\, db^2+f_3\, dc^2)\Bigr)
\eeqa
At small $r_0$ or large $y$ one finds the conical metric with the full
$\Sigma_3$ symmetry; the first correction in the expansion of powers
of $r_0/y$ (at third order) is parametrized by the $f_i$.
So for $(f_i)$ a positive multiple of
$(1,-2,1)$ or its cyclic permutations
{\em or linear combinations}\footnote{\label{class meas}because regarding only
the lowest order term amounts to
linearization of the theory; this refers thus to the situation at
infinity; note that classical reasoning at infinity would
expect the $f_i$ nethertheless to be a positive multiple of
a cyclic permutation of $(1,-2,1)$
to fulfill the non-linear Einstein equations in the interior;
together with the other classical expectation $\int_{D_i}C=0$ (for
$D_i$ filled in) this would fix $\eta_i$ to its classical value $1$ at
$P_i$; but this behaviour is modified by quantum corrections [\ref{AW}]}
{\em of them} one gets $G_2$ holonomy, i.e. for
(the negative  $f_i$ indicates which
$D_i$ is filled in)
\beqa
\label{sum fi}
\sum f_i=0
\eeqa
One has for the volume of $Q_i$ and
the $y$-dependent volume of $D_i$
embedded in $X_i$
(at large $y$; with $\vol \, D_i$
given up to higher order terms in $(r_0/y)^3$)
\beqa
\label{vol Q}
\vol \, Q_i&= &2\pi^2 r_o^3\\
\label{vol D}
\vol \, D_i&=&\frac{2\pi^2}{27}y^3
\Bigl(1+\frac{3}{8}f_i (\frac{r_o}{y})^3
+{\cal O}( (\frac{r_0}{y})^6)\Bigr)\\
\label{vol D approx}
&\approx & \frac{2\pi^2}{27}y^3 + \frac{1}{72} \, f_i \, \vol Q_i
\eeqa
Here, the first correction to the divergent piece
is the finite volume defect
$\frac{1}{72}\; f_i \; \vol({\bf S^3_{r_0}})$.
Note that semiclassically the volume defects are
$\rho \, (1, -2, 1)$ or a permutation
(with $\rho \ra \infty$; one could choose $\rho =r_0^3$ or
absorb\footnote{In a measurement at infinity the parameter $r_0$
will not be known; one refers [\ref{AW}] also to $f_i$ as the volume
defect, when stating that the $f_i$ go to $\infty$ (in ratio
$(1, -2, 1)$ or a permutation) for $r_0\ra \infty$.}
$\rho$ giving $(f_i)\sim (1, -2, 1)$).
\\ \\
\noindent
{\em Quantum moduli space and observables}

In the quantum domain there is actually [\ref{AMV}], [\ref{AW}]
a smooth curve ${\cal N}={\bf P^1_C}$ (when compactified) of theories
interpolating between the three classical limits (large $r_0$)
given by the $X_i$ (given by three points $P_i$ of $t=\om^{i+1}$
with $t$ the global coordinate, $\om = e^{2\pi i /3}$).

A holomorphic observable on ${\cal N}$ must combine
as SUSY partners
the $C$-field period\footnote{which at large radial coordinate $r$
is independent of $r$ for a
$C$-field flat near infinity (to keep the energy finite),
entailing that the components of $C$
are of order $1/r^3$}
\beqa
\alpha_i=\int_{D_i}C
\eeqa
with an order $1/r^3$ metric perturbation
(w.r.t. the conical metric), as in
\beqa
y_i=\exp\Bigl( kf_i+i(\alpha_{i+1}-\alpha_{i-1})\Bigr )
\eeqa
(with $\prod_i y_i=1$ by (\ref{sum fi})).
Actually one works with the quantity
\beqa
\label{etadef}
\eta_i=\exp\Bigl( \frac{k}{3}(f_{i-1}-f_{i+1})+i\alpha_i\Bigr )
\eeqa
(so $\eta_i=(y_{i-1}^2y_i)^{1/3}\; , \;
y_i=\eta_{i+1} / \eta_{i-1}$,
cf. (\ref{minus transform})).
On the other hand one identifies $\eta_i$ by global considerations on
the genus zero moduli space ${\cal N}$ as
the following rational function
\beqa
\label{eta}
\eta_i=-\om \frac{t-\om^i}{t-\om^{i-1}}
\;\;\;\;\;\;\;\;\;\;\;\;\;\;\;\;\;\;
\;\;\;\;\;\;\;\;\;\;\;\;\;\;\;\;\;\;
\begin{array}{c|ccc}
   & P_1 & P_2 & P_3 \\
\hline
  \eta_1 & 1 & \infty & 0\\
  \eta_2 & 0 & 1 & \infty \\
  \eta_3 & \infty & 0 & 1 \\
\end{array}
\eeqa
We have also the
local coordinate $u_i$
at $P_i$ given by the membrane instanton amplitudes
\beqa
\label{u eta variables}
u_i&=&\exp \Bigl( -T \mbox{vol}(Q_i)+i\int_{Q_i}C \Bigr ) \nonumber\\
\eta_{i}&=&
\exp \Bigl( k \frac{f_{i}+2f_{i-1}}{3}+i\int_{D_{i}}C \Bigr )
\eeqa
The local parameter $u_i$ vanishes at $P_i$ due to the large volume of the
manifold $X_i$. We denote by $\Phi_i$ the physical modulus
to which it is related via $u_j=e^{w_j}=e^{i\Phi_j}$,
i.e. (where we have set $T=1$; $Q_i$ is an (isolated)
supersymmetric cycle so $\Upsilon|_{Q_i}$ is the volume form)
\beqa
\label{phi modulus}
\Phi_j=\int_{Q_j}C+i\Upsilon=\phi_j+i\; \mbox{vol}(Q_j)
\eeqa
\noindent
{\em \label{membrane anomaly section} The membrane anomaly}

To be well-defined the phase of the $\eta_i$ variable
must be modified [\ref{AW}] to
\beqa
e^{i\al_i}=\sign \, \mbox{Pf}(\D) \; e^{i\int_{D_i}C}
\eeqa
where $\D$ is the Dirac operator on ${\bf S^3}$ with values in
the positive spinor bundle of the normal bundle and $\mbox{Pf}$
denotes its Pfaffian (square root of the determinant) which occurs
in the fermion path integral and must be combined with the classical
phase factor $e^{i\int_{D_i}C}$ in the worldvolume path integral for a
membrane wrapping ${\bf S^3}$. Now
for a three-manifold $X_3$ which is the boundary of a (spin)
four-manifold $B$ one has [\ref{AW}]
(with $\D_B$ the $S(N_B)$ valued Dirac operator on $B$
with Atiyah-Patodi-Singer boundary conditions along $X_3$)
\beqa
\sign \, \mbox{ Pf}(\D)=:e^{i\pi \mu({\bf S^3})}
=e^{i\pi \mbox{ind} (\D_B)/2}
=e^{i\pi w_4(N_B)}=e^{i\pi \chi(B)}
\eeqa
If $B$ could be chosen to be smooth (as for a single $D_i$)
the correction would be ineffective, but for the union (relevant for
$\sum_i \al_i=\pi$) of the {\em intersecting} $D_i$ this cannot be the case.
Now one gets the result $\sum_i\al_i=\pi$ either from a union of $B_i$'s
respectively bounded by the $D_i$ or more directly from slightly
perturbing the $D_i$ as follows. For this let us recall that
${\bf P^2_H}=({\bf H^3}\back \{ 0 \}) /{\bf H^x}={\bf S^{11}}/{\bf S^3}$
has [\ref{AW}] $Y={\bf S^3}\x {\bf S^3}=SU(2)^3/SU(2)_D$
fibres over a triangle
$\De = \{ [\la,\mu,\nu] | \la, \mu, \nu \geq 0\}\subset {\bf P^2_R}$
from the quaternionic norm ${\bf P^2_H} \; \buildrel p \over \lra \De
=  {\bf P^2_R}/ {\bf Z_2^2}$ (where ${\bf Z_2^2}\cong
\Bigl( ( {\bf Z_2^{\la}}\x{\bf Z_2^{\mu}}\x{\bf Z_2^{\nu}})/
{\bf Z_2^{diag}} \Bigr)$) .
For a line $\bar{B}={\bf S^4}={\bf P^1_H}\subset  {\bf P^2_H}$
and $B=\bar{B}-\cup_{i=1}^3 D^4_{open}(p_i)$ for $p_i=\bar{B}\cap L_i$
($L_i\subset {\bf P^2_H}$ the coordinate lines,
$D^4_{closed}(p_i)\subset \bar{B}$ small 4-discs around $p_i$
of resp. boundary ${\bf S^3}\sim D_i$) one has
\beqa
\label{mem anom proof}
\pa B \simeq D_1 + D_2 + D_3
\eeqa
and finds $1$ for the self-intersection number (the Euler class
of the normal bundle, the mod $2$ relevant number).
We will refer then to the membrane anomaly as
\beqa
\label{the membrane anomaly}
\sum_i\al_i=\pi \;\; \;\;\;\;\;\; (mod \, 2\pi)\;\;\;\;\;\; \mbox{or}
\;\;\;\;\;\;\;\; \prod_i \eta_i = -1
\eeqa
\noindent
{\em The conifold transition in type IIA}

In a type IIA reinterpretation (cf. [\ref{AW}])
one divides by the circle ${\bf S^1_{11}}=U(1)\subset SU(2)_1$
giving for $X_1={\bf R^4}\x {\bf S^3}
=(SU(2)_1 \x {\bf R}^{\geq 0})\x {\bf S^3}$ the type IIA manifold
$({\bf S^2} \x {\bf R}^{\geq 0})\x {\bf S^3}={\bf R^3}\x {\bf S^3}$
with fixed point at the origin, i.e. the deformed conifold
$T^* {\bf S^3}$ with a D6-brane wrapping the zero-section.
For $X_2$ or $X_3$ one gets
${\bf R^4}\x {\bf S^3}/U(1)={\bf R^4}\x {\bf S^2}$, the two
small resolutions of the conifold together with a unit of RR
two-form flux on ${\bf S^2}$ (as ${\bf S^3}$ is Hopf fibered
by ${\bf S^1_{11}}$ over ${\bf S^2}$).
One may compare with the special Lagrangian
deformations [\ref{Joyce}] of the cone over $T^2$ with
different ${\bf S^1}$'s killed in homology; the fixed point set under
the $U(1)_D$ is $L={\bf S^1}\x {\bf R^2}\subset X={\bf S^3}\x {\bf R^4}$
(from ${\bf C}\subset {\bf H}$),
the ${\bf S^1}$ being the boundary (where the fibre shrinks)
of the disc $D^2={\bf S^3}/U(1)$.

The deviation of the metric from the conical form
being of order $(r_0/r)^3$ for large $r$
(so not square-integrable in seven dimensions),
$r_0$ is not free to fluctuate (the kinetic energy of the fluctuation
would be divergent). So in the four-dimensional low energy theory it is
rather a coupling constant than a modulus.
To have a normalizable (or at least log normalizable) mode,
one of the circles at infinity should approach a constant size (which
can happen in many ways related to the Chern-Simons
framing ambiguity [\ref{AKV}])\footnote{On $H_3(Y)={\bf Z}\oplus {\bf Z}$
one has a full $Sl(2, {\bf Z})$ operating but for the
'filled in' versions only the three spaces $X_i$
are allowed as the
closed and co-closed three-form $\Upsilon$ of class
$(p,q)\in {\bf Z}\oplus {\bf Z}$
corresponds to a regular metric just for the three cases
$(p,q)=(0,1) \, , (-1,0)$ or $(1,-1)$ (where the unbroken
${\bf Z_2}\subset\Sigma_3$
exchanges the ${\bf S^3}$ factors) [\ref{Brand}].
But for the ${\bf S^1}\x {\bf S^1}=T^2$
(relating to $L$ as $Y$ to $X$) not only the $\Sigma_3$ but
the full $Sl(2, {\bf Z})$ is allowed which expresses the framing
ambiguity [\ref{AKV}]. Cf. also the case ${\bf S^5 \x S^5}$
in [\ref{Wi 5brane}].}
which is not the case
for the ${\bf Z_3}$ symmetric point discussed in [\ref{AW}] and here.
In this sense the expression 'superpotential' has to be qualified;
one gets the actual superpotential for an ordinary modulus
if the local geometry $X_7$ is embedded in a compact $G_2$ manifold
(cf. sect. \ref{global}).
The metric (\ref{metric}) describes an $M$-theory lift of a type IIA
model with the string coupling infinite far form the $D6$ brane;
to have at infinity an $M$-theory circle of finite radius
one of the three $SU(2)$ symmetries of
(\ref{metric}) must be broken to $U(1)$ [\ref{BGGG}].

\newpage
\section{\label{nonlin action}The non-linear symmetry action}
\resetcounter

We will consider two actions of $\Si_3$ on ${\bf P^1_C}$.
In the first ('linear') action the ${\bf Z_3}$ sector
acted by multiplication with an element of ${\bf C^*}$;
in the second case this sector will act non-linearly
(the operation of $\al$ will be given in both cases by $z\ra 1/z$).
\\ \\
\noindent
{\em The linear action of $\Si_3$}

Here the 'rotation' subgroup ${\bf Z_3}$ is generated by the action
$t\ra \omega t$ on ${\cal N}={\bf P^1_t}$ and $\al$ acts by
$t\ra 1/t$, giving as images of $t$ under $\Sigma_3$
$\scriptsize{\left(\begin{array}{ccccc}
t & & \om \; t  & & \om^2 \; t \\
t^{-1} &  &  \om^2 \; t^{-1} & &\om \; t^{-1}
\end{array}\right)}$ (cf. (\ref{group elements})).
The involution $\iota: t\ra -t$ is an automorphism of
$({\bf P^1_t}, \Sigma_3)$, i.e.$\Sigma_3$ compatible: $\iota\ga=\ga \iota$.

{\em Degenerate orbits in the t-plane}

We treat the question of fixpoints or degenerate orbits. The structure
of $\Sigma_3$ leads one to look for two-element and three-element orbits.
The two-element orbit, whose elements are then fixed respectively by
the ${\bf Z_3}$ cosets, is
$\scriptsize{\left(\begin{array}{ccccc}
0 & & 0  & & 0\\
\infty &  &\infty & &\infty
\end{array}\right)}$.
In the other case the full $\Si_3$ orbit is already
covered by the ${\bf Z_3}$ orbit; the transforms under the
remaining (order two) elements from the non-trivial ${\bf Z_3}$ coset
will then just repeat the ${\bf Z_3}$ orbit in some order.
This gives the two possibilities
$\scriptsize{\left(\begin{array}{ccccc}
1 & & \om  & & \om^2  \\
1 &  &  \om^2  & &\om
\end{array}\right)}$
and
$\scriptsize{\left(\begin{array}{ccccc}
-1 & & -\om  & & -\om^2  \\
-1 &  &  -\om^2  & &-\om
\end{array}\right)}$. $\iota$ exchanges these two orbits
and fixes the elements of the two-element orbit.
\\ \\
\noindent
{\em The non-linear action of $\Si_3$}

Note that the action induced on the $\eta_i$ is as follows.
The cyclic permutation of the points $P_i$ ($i=1, \; 2, \; 3$),
which is described by the rotation transformation
$t\ra \om t$, produces the
corresponding cyclic permutation
$\eta_1 \ra \eta_3 \ra \eta_2 \ra \eta_1$ on the $\eta_i$,
as seen from (\ref{eta}).
Furthermore the inversion induces
$\eta_1\ra \eta_3^{-1}, \; \eta_2\ra \eta_2^{-1}$.
The $\eta_i$ which fulfill
the relation $\eta_1\eta_2\eta_3=-1$ (reflecting the membrane anomaly)
are actually related by
\beqa
\label{etarel}
\eta_3=\frac{1}{1-\eta_1} \;\; , \;\; \eta_1=\frac{1}{1-\eta_2} \;\; ,
\;\; \eta_2=\frac{1}{1-\eta_3}
\eeqa
So consider now instead of the linear action
of ${\bf Z_3}$  the non-linear action of it resp. of
the full symmetry group $\Sigma_3$ as $Sl(2,{\bf Z})/\Gamma(2)$.
As is well known from the theory of the Legendre $\lambda$ function,
the elements are given then as the
fractional linear transformations displayed below.
$\Si_3$ occurs not only as a quotient but also as a subgroup.
For this recall that
the holomorphic automorphism group of ${\bf P^1}$ is given by
$Aut({\bf P^1})=PGl(2,{\bf C})=Gl(2,{\bf C})/{\bf C^*}=
PSl(2,{\bf C})=Sl(2,{\bf C})/\{ \pm {\bf 1}_2 \} $
and that for two triples of points of ${\bf P^1}$ there
exists an automorphism mapping these two sets of elements onto each other.
In particular we will consider
the elements permuting the set $\{ 0,1,\infty\}$ which are then
given by transformations
$z, \be z, \be^2 z$ and $\al z , \al \be z , \al \be^2 z$
understood as mappings ${\bf P^1}\ra {\bf P^1}$, i.e.
$Aut({\bf P^1}, \{ 0,1,\infty \})=\Si_3$.
This leads {\em just formally} to the relation (which
restates (\ref{the membrane anomaly}))\footnote{This product of
the meromorphic functions $\be^i z$ has nothing to do
with the group multiplication given by composing mappings
which leads to $\Si_3^{Sl_2}$. The point of the argument is to
fix the sign-prefactors relevant for (\ref{betaprod});
to write down just transformations with the prescribed divisors
is of course easy.}
\beqa
\label{betaprod}
\prod_{i \in {\bf Z_3}}\be^i z =-1
\eeqa
That this product is a constant
follows already from the divisor relations\footnote{so for example
$(e \, z )=\underline{0}-\underline{\infty}$ indicates that
the function $z\ra z$ has a simple zero/pole at $0/\infty$.}
(note that the $\be^i$ are permutations
on the set $\{ 0,1,\infty \}$):
$(e \, z )=\underline{0}-\underline{\infty} \;\; , \;\;
(\be \, z)=\underline{\infty} - \underline{1} \;\; , \;\;
(\be^2 \, z)=\underline{1}-\underline{0}$
imply that their product is a nowhere vanishing
globally holomorphic function, so a constant $x\neq 0$.
Now, in the ${\bf Z_2}$ sector given by $\{e, \al\}$,
the transformation $\al$, mapping $0 \leria \infty$ and $1$ to itself,
will operate as multiplicative inversion
(i.e. $z \cdot \al z =1$).
So $x^2=\prod_{\ga \in \Si_3}\ga z = 1$
as $\prod_{i \in {\bf Z_3}}\be^i z = x = \prod_{i \in {\bf Z_3}}\be^i
\al z = \prod_{i \in {\bf Z_3}}\al \be^i z$. Then
$x=(\al z) \cdot \be (\al z ) \cdot \be^2 (\al z) =
z\cdot \al \be^2 z \cdot \be^2 z $ for an $\al$-fixpoint
(so $z=+1$ or $-1$) shows that $x=-1$.

One finds for the concrete functional form of the
transformations
\beqa
\label{sym}
\begin{array}{ccccccc}
z & & & \be z = {\frac{1}{1-z}} & & & \be^2 z =\frac{z-1}{z} \\
\al z = \frac{1}{z} & & & \al \be z = 1-z & & & \al \be^2 z = \frac{z}{z-1}
\end{array}
\eeqa

{\em Degenerate orbits in the $\eta$-plane}

Let us consider again the question of
degenerate orbits (now under the non-linear $Sl(2)$ action;
the cases will correspond to the descriptions in the $t$-plane under
(\ref{eta})).
Concerning first the {\em two-element orbits} note that their elements are
${\bf Z_3}$ fixpoints. Therefore the condition
(\ref{the membrane anomaly}) reads
in this case $\eta^3=-1$, so $\eta = -\om$ or $-\om^2$
(the solution $-1$ leads to another case, cf. below).
So up to permutation in the still running
factor ${\bf Z_2}=\{ e, \al\}$ one finds the case
$\scriptsize{\left(\begin{array}{ccccc}
-\om & & -\om  & & -\om  \\
-\om^2 &  &  -\om^2  & & -\om^2
\end{array}\right)}$.
To develop the {\em three-element orbits} note
that the three possibilities to repeat
a value $\eta=e(\eta)$)
from the first line in the second line
(which will then, as a set, repeat the first line)
lead to the cases
$\eta=1/\eta$ so $\eta=\pm 1$, or
$\eta=1-\eta$ so $\eta=1/2$ or $\infty$, or finally
$\eta=\eta/(\eta-1)$ so $\eta=0$ or $\eta=2$.
If one looks for the corresponding orbits one finds that up to cyclic
permutations (in the $P_i$ with which one starts) one has just the two
cases
$\scriptsize{\left(\begin{array}{ccccc}
0 & & 1  & & \infty  \\
\infty &  &  1  & & 0
\end{array}\right)}$
and
$\scriptsize{\left(\begin{array}{ccccc}
-1 & & 1/2  & & 2  \\
-1 &  &  2  & & 1/2
\end{array}\right)}$.\footnote{The $\pm 1$ occurring here
are $\al$-fixpoints just as the
elements of the two-element-orbit were ${\bf Z_3}$-fixpoints.}
$\iota$ exchanges these two orbits and leaves fixed the
two elements of the two-element-orbit.\footnote{${\bf P^1_{\eta}}$ has
(by transport from the ${\bf P^1_t}$) the involution
$\iota:\eta \ra \frac{\eta -2}{2\eta -1}$, which is
$\Sigma_3^{Sl(2)}$ compatible:
$\iota \ga = \ga \iota$.}
\\ \\
\noindent
{\em Relations of the variables}

Under the non-linear action (\ref{sym})
of the triality group $\Sigma_3$
the $\eta_i$ form a ${\bf Z_3}$ orbit
\beqa
\label{etai Z3 orbit}
\eta_{i-1}=\be \eta_i \;\;\; , \;\;\; \eta_{i+1}=\be^2 \eta_i
\eeqa
and the membrane anomaly (\ref{the membrane anomaly})
becomes manifest (for $z$ some $\eta_j$)
\beqa
\label{membrane anomaly manifest}
\prod_{i \in {\bf Z_3}}\be^i z =-1
\eeqa
As shown after (\ref{betaprod})
this is already a consequence of
having a global $\Si_3$ symmetry at all.

Recall that in a semiclassical regime with $D_i=0$ one has
$Q_i\simeq \pm D_{i-1}=\mp D_{i+1}$
\beqa
\label{semiclass QD relation}
Q_i=\mp D_{i+1}
\eeqa
>From the classical fact (\ref{semiclass QD relation}) one has
$\mp\int_{D_1}C =\int_{Q_3}C$, what is (with holomorphy) tantamount
to saying that $\eta_{i+1}\sim u_i$ to first order (as reflected
in the first order zero of $\eta_{i+1}$ at $P_i$
where $u_i$ vanishes to first order; the respective lower sign
in (\ref{semiclass QD relation}) is fixed at $P_i$).
We assume that such a relation persists\footnote{Thereby a relation
$\eta_{i-1}=\frac{1}{1-\eta_i}$ holds also for the $u_i$:
$u_3=\be u_1=\frac{1}{1-u_1}$; note the corresponding
map $e^t-1\sim \frac{1}{e^V}$ in [\ref{AMV}] (in the case $N=1$ there)}
(cf. footn.'s \ref{uglob footn} and \ref{interpret footn})
so that one has a relation  $\be u_i =\eta_i$:
the holomorphic relation
between the $u_i$ and the $\eta_i$ is fixed (up to a real factor)
by the relation
between their arguments (imaginary parts of logarithms), so
(\ref{u eta variables}), (\ref{semiclass QD relation}) imply
$u_i=\eta_{i+1}=\be^2 \eta_i$, i.e.
\beqa
\label{ui etai relations}
\be u_i=\eta_i
\eeqa

Note that the $u_i$ are, in contrast to the $\eta_i$, actually
only first order parameters; therefore the assertion made, where we
think of them as globally analytically continued,
has to be suitably interpreted\footnote{\label{uglob footn}In the
approach choosen the
relation could practically be taken as a definition of $u_i^{(glob)}$.
In principle it would be possible to rephrase the whole
discussion on superpotentials to follow w.r.t. the $\eta_i$ instead of
$u_i^{(glob)}$ but it would be much less intuitive.}. But note that
in [\ref{AVII}] indeed flat coordinates $-u_i$
are described (the $u,v$ there, being shifted by $\pi i$,
being the $w_i=\log u_i$ here).

(Concerning an ensuing relation (\ref{membrane anomaly manifest})
then also for the $u_i$ note
the deviation from the relation
$q\cdot q^{\om} \cdot q^{\om^2}=+1$
which one would have in classical variables for the linear action
on $q=e^s$ (cf. also footn. \ref{class lin act})
from $s\ra \om s$ (and which would be the analogue of $q\cdot q^{-1}=1$
in the ${\bf Z_2}$ case of the usual
IIA flop).)\footnote{This $s$ and the mentioned classical
linear action is not to be confused with the $t$ of the true global
quantum moduli space and its (quantum) linear action of the same
form (which corresponds under $t\leria \eta$ to the quantum
non-linear action).}

We recall below that the crucial argument in [\ref{AW}] for a
{\em one}-component {\em quantum} moduli space was the fact that
the classical statement $\al_i=\int_{D_i}C= \Im \log \eta_i =0$
is modified by membrane
instantons; such a membrane instanton contribution
will be present as soon as $u_i\neq 0$, i.e. away from the infinite
volume limit for $Q_i$ which would suppress the latter contribution.
Similarly the classical expectation of finding the $f_i$ in (a
permutation of) the ratio $(1, -2, 1)$ and therefore having
$\Re \log \eta_i=0$ for one of the $\eta_i$
gets modified (cf. footn. \ref{class meas}). So
one expects that actually
$u_i\neq 0 \Lra  \log \eta_i \neq 0$ as is the case in
(\ref{ui etai relations}).\\ \\
\noindent
{\em The critical circle and special points}

The 'critical' circle $|u_i|=1$ corresponds$^{\ref{interpret footn}}$
classically to the case
$\vol(Q_i)=0$ of a shrinking $Q_i$. Let us study the parameter $u_i$
on the critical circle.
One has ($\phi\in [0, 2\pi]$)
\beqa
\label{circle transform}
\log(\be \, e^{i\phi})=-\log(2\sin\frac{\phi}{2})+i(\pi - \phi)/2
\eeqa
So here (\ref{ui etai relations})
means for
$\phi_i=\int_{Q_i}C$ and\footnote{\label{interpret footn}The
geometrical interpretation
applies strictly only semiclassically, in general
the arguments apply just to the global variables. It is nevertheless
instructive to keep this interpretation in mind.}
$\al_i=\int_{D_i}C$
just $\al_i=(\pi-\phi_i)/2$ or (cf. (\ref{the membrane anomaly}))
\beqa
\label{angle relation}
\int_Q C + 2 \int_D C = \pi \;\;\;\;\;\; (mod \, 2\pi)
\eeqa

Note that by (\ref{circle transform})
the $\be$-transform of a parameter $z=re^{i\phi}=:e^{f+i\phi}$
on the critical circle $f=0$
stays there in the cases
($z$ being then a $\be$-fixpoint by (\ref{angle relation}))
\beqa
\label{circle orbit}
(f=0) \;\;\;\;\;\;\;\;\; \phi = \pm \pi / 3
\eeqa

There are some special points in moduli
space which are interesting to consider.
Let us consider first the phases where the ${\bf S^3}$ given by $Q_i$
has (seen classically) still
physical (non-negative) volume, i.e. the domain $|u_i|\leq 1$  or
$f_i\leq 0$ in parameter space$^{\ref{interpret footn}}$: here
$|u_i| < 1$ (i.e.$^{\ref{interpret footn}}$
$\vol (Q_i)>0$) or
$f_i<0$ is the region where $D_i$ is filled in
(i.e. $Q_i$ is not shrunken where $D_i$ is shrinkable).
So at the (classical) intersection of all three phases,
where all the three-spheres have to be shrinkable at the same time,
one finds $f_i=0$, i.e. the ${\bf Z_3}$ orbit given by
the $\eta_i$ triple lies on the unit
circle in the $\eta$-plane. This leads via (\ref{circle orbit}) to
the two possibilities\footnote{$\eta=-\om$ and $\eta=-\om^2$,
identified above:
the case where the $\eta_i$ constitute a completely degenerate
${\bf Z_3}$ orbit ($\eta_i=\be \eta_i =\be^2 \eta_i$), i.e. the
full $\Sigma_3$ orbit degenerates to a two-element orbit.}
\beqa
\label{intersection of three phases}
f_i = 0 \;\; , \; \; \al_i=\pm\pi/3
\eeqa
\newpage
\section{\label{local supo}\large {\bf The superpotential: the
local approach}}
\resetcounter

\noindent
{\em The global superpotential}

Let us recall first the global approach to the superpotential [\ref{AW}].
One expects the superpotential $W$ to vanish at the $P_i$.
If there are no further zeroes then $W$ will have exactly
three poles on the genus zero moduli space ${\cal N}$.
Both of these three element sets will have to be
complete $\Sigma_3$ orbits {\em by themselves}.
This leads to the two degenerate three-element orbits $\om^i$
and $-\om^i$. So the minimal solution is [\ref{AW}]
\beqa
\label{supoint}
W\sim \frac{t^3-1}{t^3+1}
\eeqa
Note that (under the linear action) it transforms with the sign character
\beqa
W(\ga t)=\sign (\ga) W(t)
\eeqa
i.e. $W(\om t)=W(t), \, W(t^{-1})=-W(t)$
(it transforms 'anti-invariantly')\footnote{In $W(t)= \frac{t^3-1}{t^3+1}$ an
underlying anti-invariant projection (\ref{proj anti inv})
is made manifest by using the relation
$W(t)\sim \prod_{i\in {\bf Z_3}}\frac{t-\om^i}{t+\om^i}
=\frac{1}{3}\sum_{i\in {\bf Z_3}}\frac{t-\om^i}{t+\om^i}$
giving
$W(t)\sim \sum_{\ga \in \Sigma_3}\sign(\ga)
\frac{1}{t+1}|_{\ga t}$.}.

Concerning the zeroes and poles of the global superpotential $W(t)$
for the ${\bf Z_3^{lin}}$ orbits of $1$ and $-1$ (in the $t$-plane),
respectively, note that one has corresponding ${\bf Z_3^{Sl_2}}$ orbits of
$\pm 1$ (now in the $\eta$-plane !) for the $\eta_i$, and so for the $u_i$:
$u_i=+1$ or $-1$, i.e.\footnote{but note here
the issue of $u_i^{(glob)}$ vs. $u_i^{(loc)}$, cf.
footn. \ref{interpret footn}
and remark after (\ref{ui etai relations})}
$\vol (Q_i)=0$ with $\int_{Q_i}C=0$ or
$\int_{Q_i}C=\pi$, so
the cases are\footnote{Note that
in case the independent
variables $z_i$ relevant for $W$ build a ${\bf Z_3}$ orbit
(like is the case for the $\eta_i$) the first case,
$z_i=0$ or $z_{i-1}=\be z_i=1$,
differs only in the $C$ field period,
from the second one.}:
first $u_i=0$, i.e. $\vol (Q_i)=\infty$  and $\int_{Q_i}C=0$
$\Lra$ zero for $W_t$; and, secondly,
$u_i=-1$, i.e. $\vol (Q_i)=0$ and $\int_{Q_i}C=\pi$
$\Lra$ pole for $W_t$
(this case may be compared with the case
of a vanishing ${\bf S^2}$ with
$\int_{{\bf S^2}}B=\pi$).\\ \\
\noindent
{\em The local superpotential}\\
The actual superpotential arises from the sum
of all the multi-cover membrane instantons
\beqa
\label{multicov mem inst}
W(u_i)=\sum_1^{\infty}a_nu_i^n
\eeqa

How do the non-perturbative
contributions from the local semiclassical informations near the $P_i$
fit together over the whole quantum moduli space ${\cal N}$ ?
Is $W(u_i^{(glob)}(t))=W(u_{i-1}^{(glob)}(t))$, i.e.
$W(u)=W(\be u)$: is $W$ (at least ${\bf Z_3}$)
triality symmetric ?

First note that the membrane instantons
make the deviation from the classical result $\alpha_i=0$ possible
([\ref{AW}] and recalled below),
just as sums of world-sheet instanton contributions in the case of the
type IIA string on a Calabi-Yau manifold give quantum corrections to a
classical (complexified) K\"ahler  volume. In the IIA case
of $N=2$ supersymmetry
one can answer the analogue of our question above by considering
the resummation of the geometric instanton series
$I_{ws}(q)=\sum_{n\geq 1} \frac{q^n}{n^0} = \frac{q}{1-q}$ for a flop
[\ref{W}]
where the K\"ahler parameter
$t=\int_{{\bf P^1}}B + i \; \mbox{area}({\bf P^1})$
in $q=e^{2\pi i t}$ is reflected as $t \ra -t$:
\beqa
\label{WS inst sum trafo}
I_{ws}(\frac{1}{q})= -I_{ws}(q)-1
\eeqa
The deviation from anti-invariance stems from change
in classical intersection numbers.

Now let us ask what the corresponding 'reflection' is on the
modulus $\Phi_j=\int_{Q_j}C+i\Upsilon=\phi_j+i\; \mbox{vol}(Q_j)$
in (\ref{phi modulus}) under which we should look for reasonable
transformation behaviour of the quantum corrections (reasonable
meaning a way of transformation so that the three contributions fit
together in a triality symmetric way over ${\cal N}$).

In the case of $M$-theory on our $G_2$ holonomy manifold
the question of flopping an ${\bf S^3}$,
instead of an ${\bf S^2}$ in type IIA on a Calabi-Yau manifold,
is more complicated as the K\"ahler moduli no lo longer fit
together naturally at the classical level [\ref{AW}].
Rather the metric moduli
$\mbox{vol}({\bf S^3_i})$ of the $X_i$, running classically over a half-line
$[0, \infty)$, are at angles $2\pi /3$ to one another
(in a copy of ${\bf R^2}$ containing the root lattice
$\Lambda$ of $SU(3)$); the $C$-field
periods measured at infinity on the different
$X_i$ take values in {\em different}
subgroups $E_i\cong H^3(X_i, U(1))\cong U(1)$ of $H^3(Y, U(1))\cong
U(1)\x U(1)$ (when restricted to $Y$).

So in view of the problems with the three
rays in ${\bf R^2}$ we will not consider the
transformation\footnote{\label{class lin act}generating
${\bf Z_3}$, or $\Sigma_3$ when combined with
complex conjugation on ${\bf R^2}\cong {\bf C}$
(or with inversion)}
given by rotation with $2\pi /3$ around the origin in this ${\bf R^2}$,
i.e. multiplication of the modulus with $\om$.
Rather one should now consider instead of this linear
action the non-linear action (cf. (\ref{sym})) of
$\Sigma_3$ represented as $Sl(2,{\bf Z})/\Gamma(2)$.
\\ \\
\noindent
{\em \label{one mem inst}The one-membrane instanton contribution}

Actually there is a {\em one-component}
moduli space comprising all the $P_i$ [\ref{AW}] as
quantum effects
given by membrane instantons cause a deviation from the classical
result $\alpha_i=0$.
For this recall that
to convert the interaction given by $u$, which is like a superpotential,
to an ordinary interaction one has to integrate over the fermionic collective
coordinates of the membrane instanton, i.e. to evaluate the chiral superspace
integral $\int d^2 \th \, u$
(there is also an integration $\int d^4 y$ over the
membrane position in ${\bf R^4}$ to be made).
As the fermion integral has the
properties of a derivation with respect to $w$
one gets\footnote{the contribution to $\int_{D}C$ of
a second term $2u\int d\th^1 w\int d\th^2 w$
occurring here is subleading for large $r$.}
($u=e^w$; $T=1$)
\beqa
\label{derivative nature}
\int d^2 \th \, u = u\int d^2 \th \, w = -2u \int d^2 \th \int_Q \Y
\eeqa
For this one gets the evaluation ($w=i\Phi$, cf. (\ref{phi modulus}))
\beqa
\label{oneinstcontrib}
\int d^2 \th \, w \sim \int_{Q_i}* G
\eeqa
In a second step one finds that the contribution
$\int_{{\bf R^4} \x Q_i}*G$
to the effective action induces a non-zero value of
$\alpha_i=\int_{D_i}C$ as one has\footnote{\label{aux G field} with a
classical $G$-field generated by a source $\int_{D_i}C$ as a means to
evaluate (\ref{deviation})}
\beqa
\label{deviation}
<\int_{{\bf R^4} \x Q_i}*_{11} G \cdot \int_{D_i}C > \neq 0
\eeqa
because the 'linking number' of the two three-spheres $Q_i$ and $D_i$ is
one (effectively given by the intersection number of $Q_i$ with
$B_i$).\\ \\
\noindent
{\em Derivation of the dilogarithm superpotential}

The evaluation (\ref{deviation}) occurred in the transition
from the superpotential $u$ to an ordinary interaction $\int
d^4y\, d^2\th \, u$: enhancing this argument
we want to argue that this
determines already the complete scalar potential
(also suggested by the form of a $G$ flux induced
superpotential, cf. subsect. \ref{flux supo subsect}), so
that one gets thereby
the {\em full} membrane instanton amplitude including all the higher
wrappings, i.e. the full quantum corrections.

Note first that more generally than in (\ref{derivative nature})
the derivative nature (w.r.t. $w=\log u$, cf.[\ref{AW}] p.60)
of the fermion integral gives
\beqa
\int d^2 \th \; W(u) \approx
\frac{dW}{d\log u} \int d^2 \th \; w
\eeqa
Now to determine the actual sum $W(u)$ of the
multi-cover membrane instantons we interpret (\ref{deviation})
as representing actually a relation (with $T=1$)
between the full
$\int d^2\th \; W(u)$ (cf. footn. \ref{scalar potential footn})
and
$\int_{D}C\cdot \int_{Q}*_7 G
=\int_{D}C\cdot  \int d^2\th \, w$
\beqa
\label{interpretation}
\int d^2\th \;W(u) \sim i \int_{D}C\cdot \int d^2\th \, w
\eeqa
For this note that
classically one has $\int_D C=0$ and of course also
$\int d^2 \th \, W = 0 $ as $W=0$. A shift $\De \, \int_D C \neq 0 $
away from the classical vanishing value was argued [\ref{AW}] to occur via a
contribution $\int d^2\th \, W \neq 0 $ from the membrane instantons.
Here we argue that actually the relation between
$\De \, \int_D C$ and $\De \, \int d^2\th \, W$ should be used to
show that $\int d^2\th \, W = \frac{dW}{d\log u} \int d^2 \th \; w
=\frac{dW}{d\log u} \int_Q *G = \frac{dW}{d\log u}$ {\em is}
(proportional to) $\int_D C= \Im \, \log \eta$ as in (\ref{interpretation}).
This leads with (\ref{ui etai relations})
to the differential equation\footnote{here we made
a holomorphic completion on the rhs
which the holomorphic lhs suggests}
for $W$
(cf. (\ref{flux supo derivation}))
\beqa
\label{first differential equation}
\frac{dW}{d\log u}=i\int_{D}C=i\,\Im \log \eta
\buildrel \mbox{h. c.} \over \lra = \log \eta = \log \be u
\eeqa
The two crucial inputs to get this were the two interpretations
(\ref{ui etai relations}) and (\ref{interpretation}).

With the differential equation
(\ref{first differential equation}) we get
for the full
superpotential (cf. [\ref{AVI}], [\ref{BGGG}])
\beqa
\label{Wrepres}
W(u_i)=-\int_0^{u_i}\frac{dt}{t}\log (1-t)
=\sum_{n=1}^{\infty}\frac{u_i^n}{n^2}=Li_{(2)}(u_i)
\eeqa
(cf. (\ref{Li relations}) for the polylogarithm).
Note that the integral
representation (\ref{Wrepres}) defines $W$ on the complex plane, cut
along the part $(1,\infty)$ of the positive real axis. In the series
representation for large volume $\vol(Q_i)\approx \infty$
the instanton contributions vanish, i.e. $W(0)=0$.
The function of the modulus $\Phi=-i w = -i \log u$ has, by
(\ref{first differential equation}),
a critical point exactly at $u=0$, the large volume
point $P_i$ (cf. (\ref{susy vacua})). So in
total\footnote{\label{just one end footn}because of the deviations
(cf. below) from strict
anti-invariance this captures just the $u=0$ end}
\beqa
\label{N=1 vacua}
\frac{\partial W}{\partial \Phi} =0 \Leftrightarrow u=0
\;\;\;\; , \;\;\;\;
W(\Phi)=0 \Leftarrow u=0
\eeqa
giving no proper supersymmetric vacuum
but the common decompactification runaway.

\subsection{The triality symmetry relations of the local superpotential}

\noindent
{\em Anti-invariance-with-correction-terms of the superpotential}

Now, remarkably, in the case of the actual membrane instanton superpotential,
the function $W(u)$ satisfies\footnote{Integrating the relation
$\frac{d}{du} W(\frac{1}{u})
=  - \frac{\log (1-\frac{1}{u})}{1/u} \cdot \frac{-1}{u^2}
= \frac{\log (1-u) - \log (-u) }{u}$
gives (\ref{symrel1})
(for the integration constant compare at $u=1$).
Partial integration gives (\ref{symrel2}):
$-\int_0^u \frac{dt}{t}\log (1-t)
= - \log u \log (1-u) -\int_0^u \frac{dt}{1-t}\log t$,
the last integral being $W(1-u)-W(1)$
(by the substitution $s=1-t$ in $W(1-u)$).}
the following symmetry relations which will ensure that
the local superpotential is compatible with triality symmetry
(almost)
\beqa
\label{symrel1}
W(\frac{1}{u})&=&-W(u)-\zeta(2)-\frac{1}{2}\log^2(-u)\\
\label{symrel2}
W(1-u)&=&-W(u)+\zeta(2)-\log u \log (1-u)
\eeqa
The symmetry relations entail that,
up to\footnote{The differential equation (\ref{first differential equation})
makes it technically clear that $W$ is not precisely anti-invariant
(cf. remark after (\ref{origterm})); what is remarkable is
that it is almost anti-invariant.}
the elementary corrections provided by the products of two log's
and $\zeta(2)$, the $W(u)$ superpotential
is invariant under the transformations in the first line of
(\ref{sym}) and transforms with a minus sign under the mappings of the
second line. That is the 'local' superpotential transforms
(under the $Sl(2)$ action)
up to the elementary corrections
with the sign character
just as the global superpotential did (under
the linear action) and as it should a priori.
The behaviour under ${\bf Z_3}$ shows how the local (on ${\cal N}$)
membrane instanton contributions fit together globally.

\noindent
{\em Relating invariance deviations by differentiation}

Let us compare the analogous behaviour of the instanton sums $Li_0$
and $Li_2$,
describing the multi-coverings of SUSY-cycles
provided by the holomorphic ${\bf S^2}$ in the string world-sheet case
and the associative ${\bf S^3}$ in the membrane case, respectively
(where the 'lower terms' $-\zeta(2)-\frac{1}{2}(-\pi^2\pm2\pi i w)$
are at most linear in $w=\log u$)
\beqa
\label{mem sum}
W_{mem}(\frac{1}{e^w})&=&
-W_{mem}(e^w)-\frac{1}{2}w^2 + \, lower \, terms\\
\label{sheet sum}
I_{ws}(\frac{1}{q})&=& -I_{ws}(q)-1
\eeqa
Note then that (\ref{mem sum}) corresponds after taking $\pa^2/\pa w^2$
via (\ref{Li relations}) to (\ref{sheet sum}).

Let us look on a related example concerning
the issue of corrections of polylogarithms.
By (\ref{WS inst sum trafo})
$Li_0(q)=\sum_{n\geq 1} q^n=\frac{q}{1-q}$ had the anti-invariant
transformation behaviour under ${\bf Z_2}$ up to mentioned correction.
Said differently, when one considers the full expression which
includes the classical contribution and the quantum corrections
one finds a smooth behaviour\footnote{i.e. start with the prepotential
as given by a cubic polynomial with the intersection numbers as
coefficients (possible lower polynomial terms are not relevant here)
plus the $Li_3$ term (including the instanton coefficients,
i.e. the number of rational curves in specific cohomology classes)
then take the third derivative (w.r.t. $t=\log \, q$)
and find the classical intersection number plus $Li_0(q)$
(again by (\ref{Li relations}))}.
The change in the classical intersection number will then be balanced
exactly by the change in the quantum contribution.

Now if a curve $C={\bf P^1}$ is flopped at a point $x_0$ along the
Horava/Witten intervall this is argued in [\ref{flop}] to cause a
$G=(\pm) \de_{C}$ contribution
(from $dG=(\pm) \de_{C} \de(x_{11}-x_0)dx_{11}$). This comes
as one has the usual anomaly balance
$dG=(tr F_{obs/hid}\we F_{obs/hid} - \frac{1}{2} tr R\we R)
\cdot \de(x_{11}-x_{obs/hid})dx_{11}$ at the boundaries,
but along the intervall, when crossing the flop point, the gravitational
contribution will have changed\footnote{The reason for the $1/2$ is
that $c_2(CY)\cdot S = c_2(S)-c_1^2(S)$ changes by $2$ as an blow-up
increases the Euler number of $S$ by one and the canonical class gets
a contribution from the exceptional divisor.}
[\ref{TianYau}], with $\de_{C}=\De_{flop} \frac{c_2}{2}$.
How can one have a jump between the endpoints of the flop transition
if these can also be smoothly related (when one does not go through
the singular point but encircles it by rotating the $B$-field) ?
As the latter process is not just classical geometry (as would be
comparing just $c_2$'s) one has to look at the quantum corrected
quantities where
a classical-quantum balance now takes
place at well. The relevant expression to look at is
\beqa
\label{cla qua bal}
12 \, F_1=(\frac{c_2}{2}\cdot J)t+Li_1(q)
\eeqa
and $\pa_t F_1\sim \frac{c_2}{2}\cdot J + Li_0(q)$ brings us
effectively back to the previous balancing argument.
\\ \\
\noindent
{\em Anti-invariance of the superpotential-with-correction-terms ?}

One might ask whether one could make the
superpotential anti-invariant by adding
suitable terms to $W$ so that
the troublesome remainder terms\footnote{From $R_{\al}$ and $R_{\al\be}$
in (\ref{symrel1}) and (\ref{symrel2}) one derives iteratively that
$R_{\be}=-2\zeta(2)-\log z \log \be z -\frac{1}{2}\log^2 (-\be z)
\; , \;
R_{\be^2}=-\zeta(2)-\log z \log \be z -\frac{1}{2}\log^2 z \; , \;
R_{\al\be^2}=-\frac{1}{2}\log^2 (1-z)$;
$R_{\al\be^2}$ follows also directly from integrating
$\frac{d}{dz}W(\frac{z}{z-1})=
-\frac{\log(1-\frac{z}{z-1})}{\frac{z}{z-1}}\frac{-1}{(z-1)^2}=
\log(1-z) \; \Bigl( \frac{1}{z}+\frac{1}{1-z} \Bigr)$.}
$R_{\ga}(z)$ (for ${\ga}\in \Sigma_3$) in
\beqa
\label{Wtrafo}
W(\ga \; z)=\sign(\ga) W(z) + R_{\ga}(z)
\eeqa
(the
log's and constants interfering with the precise anti-invariant
transformation behaviour) would be canceled.
For example the deviation in (\ref{sheet sum})
can be rectified that way,
the modified
$\tilde{I}_{ws}:=I_{ws}+1/2$ is
strictly anti-invariant: $\tilde{I}_{ws}(1/q)=-\tilde{I}_{ws}(q)$.
As for $W$ the deviations consist of quadratic
polynomials in the log's of $u$ and its ${\bf Z_3^{Sl_2}}$ transforms
one may try to  adjust $W$ by
adding a term of this type (as suggested by the fact that $I_{ws}$ and
$W_{mem}$ are related by taking a two-fold derivative
w.r.t. $\log\, u$, cf. above).

So one would make an ansatz to correct the
superpotential by additional terms\footnote{Reasonably
one has to demand that $C(z)$ is 'more elementary' than
$W(z)$ itself (like a polynomial in log terms) as one has always the
anti-invariant projection (\ref{proj anti inv}) with
$C=\sum_{\ga \in \Sigma_3, \ga \neq e}\sign(\ga)\ga W$. From
$C\sim \sum \sign(\ga)R_{\ga}$ as used above one finds that for this
it suffices that the $R_{\ga}$ are 'more elementary'.}
$C(z)$ which cancel the unwanted terms.
Including $C(z)$ one gets a modified superpotential
\beqa
\label{Wtilde}
\tilde{W}(z)=W(z)+C(z)
\eeqa
such that $\tilde{W}$ transforms just with the sign character.
One expects $\tilde{W}$ to have the structure
making manifest an underlying anti-invariant projection
(\ref{proj anti inv})  (cf. app., subsect. \ref{formal invar})
\beqa
\label{tildeWeta}
\frac{1}{6}\sum_{\ga \in \Sigma_3}\sign(\ga) W(\ga \cdot)
=\frac{1}{6}\sum_{\ga \in \Sigma_3}\sign(\ga)
[ \sign(\ga) W +R_{\ga} ]
=W+\frac{1}{6}\sum_{\ga \in \Sigma_3}\sign(\ga) R_{\ga}
\eeqa
>From this
one has for the sought-after
correction term\footnote{Coming with (\ref{minussigns}) from
$6C=-2\log z \log \be z +\log z \log\be^2 z +\log^2 (-z)
+\frac{1}{2}\log^2 \be z -\frac{1}{2}\log^2 (-\be z)$.}
$C=\frac{1}{6}\sum_{\ga \in \Sigma_3}\sign (\ga) R_{\ga}$
\beqa
\label{C eval2}
C=\frac{1}{6}\Bigl(-\frac{1}{2}\log^2 z-2 \log z \log \be z
+ \frac{3}{2}\log^2 \be z
+ 2 \log \be z \log \be^2 z  +\frac{1}{2}\log^2 \be^2 z \Bigr)
\eeqa
One finds also (with a subtle sign and remote similarity to the Rogers
modification (\ref{Rogers relations}))
\beqa
\label{Rogers footnote}
\tilde{W}(z)=Li(z)-\frac{1}{2}\log \be z \log z \;\;
- \frac{\pi^2}{6} \pm \frac{i \pi}{6} (\log z - \log \be z)
\eeqa
But a physical motivation for $C$ is unclear; one
wants to see the local individual
superpotentials in the three different phases
not added like in (\ref{tildeWeta}) but
rather naturally patching together
(cf. remark after (\ref{multicov mem inst})).

\noindent
{\em \label{critical circle}Behaviour of the local superpotential
on the critical circle}

Let us study $W$ on\footnote{which would
correspond classically$^{\ref{interpret footn}}$
to a shrinking $Q_i$ of $\vol(Q_i)=0$}
the critical circle $|u_i|=1$
(the boundary of the domain of convergence of the series (\ref{Wrepres})),
so just as a function of $\phi_i=\int_{Q_i}C$. Then
$W(e^{i\phi})-W(1)=-i\int_0^{\phi}d\chi\log(1-e^{i\chi})$
and (\ref{circle transform}) give the elementary evaluation
\beqa
\label{real part}
\Re \; W(e^{i\phi})=\sum_{n\geq 1}\frac{\cos n\phi}{n^2}
=\zeta(2)-\frac{1}{4}\phi(2\pi - \phi)
\eeqa
for the real part
and the non-elementary odd function $I(\phi)$
for the imaginary part
\beqa
\label{I}
I(\phi)=\sum_{n\geq 1}\frac{\sin n\phi}{n^2}=
-\int_0^{\phi}\log ( 2\sin \frac{\psi}{2} )d\psi
\eeqa
which is of period\footnote{If $\phi \notin [0, 2\pi ]$ one has to take
the absolute value inside the logarithm.}
$2\pi$.
As we will have opportunity to consider terms like
\beqa
\label{Lobachevsky}
\Pi (\phi):\, =\frac{1}{2}I(2\phi)=-\int_0^{\phi}\log ( 2\sin \psi )d\psi
\eeqa
let us note that\footnote{As $I(2\phi)=-2\int_0^{\phi}d\chi \log(2\sin \chi)
=-2\int_0^{\phi}d\chi \log(2\sin \frac{\chi}{2})
 -2\int_0^{\phi}d\chi \log(2\cos \frac{\chi}{2})
= 2I(\phi)
 +2\int_{\pi}^{\pi-\phi}d\xi \log(2\sin \frac{\xi}{2})
=2I(\phi)+2I(\pi)-2I(\pi-\phi)$.}
\beqa
\label{Irel2}
\frac{1}{2}I(2\phi)=I(\phi)+I(\pi + \phi)=I(\phi)-I(\pi - \phi)
\eeqa
In view of the membrane anomaly relation
$\al_1+\al_2+\al_3=\pi$ (for the $D_i$)
with symmetry for
$\al_i=\pi/3=\arg(-\om^2)$ or $\al_i=-\pi/3=\arg(-\om)$,
note\footnote{for the $Q_i$
but in view of the
identification$^{\ref{uglob footn}, \ref{interpret footn}}$
(\ref{ui etai relations})
of the $D_i$ and the $Q_i$ in a ${\bf Z_3}$ rotated phase}
that $I(\phi)$ becomes maximal at $\phi=\pi/3$
(as seen from solving $I'(\phi)=-\log(2\sin\frac{\phi}{2})=0$;
cf. (\ref{intersection of three phases}))
and that (by (\ref{Irel2}))
\beqa
\label{I relation}
I(\pi/3)=\frac{3}{2}I(2\pi/3)
\eeqa
So for the six
roots $e^{k\frac{2\pi i}{6}}$ ($k=1,\dots , 6$) one has
($I:=I(\pi/3)$, $\zeta:=\zeta(2)=\pi^2/6$)
\beqa
\label{Wvalues}
\begin{tabular}{c||c|c|c|c|c|c}
$e^{k\frac{2\pi i}{6}}
$ & $-\om^2$ & $\om$ & $-1$ & $\om^2$ & $-\om$ & $1$ \\ \hline
$W(e^{k\frac{2\pi i}{6}})$ &
$\frac{1}{6}\zeta + i I$ &
$- \frac{2}{6}\zeta +\frac{2}{3}i I$  &
$-\frac{3}{6}\zeta$ &
$-\frac{2}{6}\zeta -\frac{2}{3}i I$ &
$\frac{1}{6}\zeta - i I$ & $\zeta$
\end{tabular}
\eeqa
\\
{\em A ${\bf Z_N}$ symmetry property}

The result that $\sum_{{\bf Z_6}}Li(e^{k2\pi i/6})=\zeta/6=Li(1)/6$
leads to a more general observation
concerning the angular degree of freedom $\phi$ of $Li(z)$, more
precisely on the interrelation between entries equidistributed with
respect to $\phi$.
Generally one has\footnote{as $1-y^N=\prod_k(\om_N^k-y)=
\prod_k(1-\om_N^{-k}y)$ gives
$-\frac{1}{N}\int_0^{z^N}\log(1-t)d\log t
=-\int_0^{z}\log(1-y^N)d\log y
=-\sum_k\int_0^z\log(1-\om_N^{-k}y)d\log y
=-\sum_k \int_0^z \log(1-\om_N^kx)d\log x
=-\sum_k\int_0^{\om_N^kz}\log(1-t)d\log t$}
(with $\om_N=e^{2\pi i /N}$)
\beqa
\label{angular orbit}
\frac{1}{N}Li(z^N)=\sum_{k\in {\bf Z_N}}Li(\om_n^kz)
\eeqa
\newpage
\subsection{\label{monodromy subsection}
\large {\bf The monodromy representation}}

The multi-valuedness of $\log z$ and $Li(z)$
around $z=0,1$ and $\infty$ is described by the
monodromy representation of the fundamental group
$\pi_1 ({\bf P^1} \backslash  \{ 0,1,\infty \})$.
This describes for the generator loops
$l_i(t)$ ($i=0,1$, $t\in [0,1]$) which encircle (in the mathematically
positively oriented sense) $z=0$ and $z=1$, respectively
(then $l_{\infty} \circ l_1 \circ l_0=1$),
the increments (cf. also
app. \ref{monodromy appendix})
\beqa
\label{log multi valued}
\log z &\buildrel l_0 \over \lra &\log z + 2\pi i \\
\label{multi valued}
\log \be z &\buildrel l_1 \over \lra & \log \be z -2\pi i \;\; , \;\;
Li(z) \buildrel l_1 \over \lra  Li(z) -2\pi i \log z
\eeqa
The relevant local system is described by a bundle,
flat with respect to a suitable connection. In
the case of the logarithm the
monodromy (\ref{log multi valued}) is captured by the matrix
$M(l_0)=
{\scriptsize \left(\begin{array}{cc}1&2\pi i\\0&1\end{array}\right)}$
acting on the two-vector $(\log z, 1)^t$
and the monodromy group is given by ${\cal U}_{\bf Z}\hra {\cal U}_{\bf C}$
where ${\cal U}$ denotes the upper triangular group
$\scriptsize{\left(\begin{array}{cc}1&*\\0&1\end{array}\right)}\subset
Sl(2)$ (the embedding of ${\cal U}_{\bf Z}$ in ${\cal U}_{\bf C}$
may include the factor of $2\pi i$). The
generalisation in the case of the dilogarithm
involves the upper triangular $3\x
3$ matrices [\ref{Hain}], i.e. one gets again admixtures
from 'lower' components:
the hierarchical structure of the poly-logarithm $Li=Li_2$ with
respect to its predecessor $\log \be z = Li_1(z)$ entails that
its monodromy is not any longer given just by the addition of integers
(multiplied by $2\pi i$);
rather one has to consider constants,
ordinary logarithms and the dilogarithm all at the same time
and to consider the lower ones as monodromy contributions
of the next higher one.
One can organize this as follows. Analytic continuation about a loop $l$
in ${\bf P^1}\backslash \{ 0,1,\infty \}$ (based at $1/2$, say) leads
to the monodromy representation
\beqa
M: \pi_1 ({\bf P^1} \backslash  \{ 0,1,\infty \}) \ra Gl(3,{\bf C})
\eeqa

There are two equivalent ways to express this.
In a {\em vector picture} one assembles $Li$, the ordinary logarithm
and the constants to a three-vector $c_3$ and
finds for the images of the generator loops $l_i(t)$ ($i=0,1$)
the matrices $M(l_i)$ representing (\ref{log multi valued}),
(\ref{multi valued}) for $c_3$
\beqa
\label{monodromy matrices}
c_3=\left( \begin{array}{c}Li(z)\\ \log z \\ 1 \end{array} \right)
\; : \;\;\;\;
M(l_0)=  \left( \begin{array}{ccc}
                  1 &      0      &   0  \\
                  0 &      1      &   2\pi i   \\
                  0 &      0      &   1
                  \end{array} \right)
\;\; , \;\;
M(l_1)=  \left( \begin{array}{ccc}
                  1 &     -2\pi i       &   0  \\
                  0 &      1      &   0  \\
                  0 &      0      &   1
                  \end{array} \right)
\eeqa
Alternatively, in a {\em Heisenberg picture}, consider the complexified
Heisenberg group ${\bf {\cal H}_C}$
\beqa
\left( \begin{array}{ccc}
                  1 &      a      &   c   \\
                  0 &      1      &   b   \\
                  0 &      0      &   1
                  \end{array} \right)
\buildrel \cong \over \lra
(a,b \, | \, c)\in {\bf {\cal H}_C}
\eeqa
Instead of $c_3$ one considers here the expression
(a flat section of a suitable connection)
\beqa
\label{flat section}
\Lambda(z)=\Bigl( - \log \be z , \log z \, | \, - Li(z)\Bigr)
\eeqa
and left operation with ${\bf {\cal H}_Z}$ expresses
the multi-valuedness (\ref{multi valued}). More precisely
one finds for the
monodromy along the loops $l_i$ the representing left multipliers $h_i$
\beqa
\label{heis mono}
h_0=(0,1 \, | \, 0) \;\;\; , \;\;\;
h_1=(1 , 0 \, | \, 0) \;\;\; , \;\;\;
h_{\infty}=(-1 , -1 \, | \, 0)
\eeqa
So $h_0 \cdot (u,v \, | \, w) = (u,v+1 \, | \, w)$ and $h_1 \cdot
(u,v \, | \, w)=(u+1,v \, | \, w+v)$
give\footnote{Note that actually we consider $(a,b, c)\in {\bf Z^3}$
embedded in ${\bf {\cal H}_Z}$ via $(2\pi i a,2\pi i b \, | \,
(2\pi i)^2 c)$.} (\ref{multi valued}).\\
One has from
$(a,b \, | \, c)\ra (x,y)=(e^a, e^b)$ a bundle
projection with fibre
$(2\pi i)^2 {\bf Z}\backslash {\bf C_c} \buildrel \cong
\over \lra {\bf C^*}$ (via $c \ra S:=e^{c/2\pi i}$;
the entries of ${\bf {\cal H}_Z}$ are actually
from $((2\pi i) {\bf Z}, (2\pi i) {\bf Z} \, | \, (2\pi i)^2 {\bf Z}$))
\beqa
\label{heisenberg bundle}
&{\bf {\cal H}_Z}\backslash {\bf {\cal H}_C}& \nonumber\\
&\downarrow&\nonumber\\
&{\bf C^*_x}\x {\bf C^*_y}
=(2\pi i {\bf Z})^2 \backslash {\bf C^2_{a,b}}&
\eeqa
This carries a connection of curvature
$\frac{1}{2\pi i }  d\log x  \wedge \, d\log y$
coming from the connection
\beqa
\label{connection}
\nabla S
= \frac{1}{2\pi i} S \; (2\pi i \; d \, \log S - u \; dv)
= dS - S \, u \, dv /2\pi i
\eeqa
on the pullback\footnote{the bundle
$(2\pi i )^2{\bf Z}\backslash {\cal H} \ra {\bf C_a}\x {\bf C_b}$
of fibre $(2\pi i )^2{\bf Z}\backslash {\bf C_c}$;
so this is the complex analogue of (\ref{real Heisenberg sequence})}
of the bundle (\ref{heisenberg bundle}) to
${\bf C}\x {\bf C}$ along $(a,b)\ra (e^a, e^b)$. The latter trivialises the
bundle so that a section can be understood as a map
$S:{\bf C}\x {\bf C} \ra {\bf C^*}$.

Now consider the pullback (along the base map $z\ra (1-z,z)$)
of the bundle ${\bf {\cal H}_Z}\backslash {\bf {\cal H}_C}$ lying
over ${\bf C^*}\x {\bf C^*}$
to what we will call the {\it Heisenberg bundle}
$\underline{{\cal H}}$ over ${\bf P^1}\backslash \{ 0,1,\infty \}$
\beqa
\label{proper heisenberg bundle}
\begin{array}{ccc}
\underline{{\cal H}} & \lra & {\bf {\cal H}_Z}\backslash {\bf {\cal H}_C}  \\
\downarrow & & \downarrow \\
{\bf P^1}\backslash \{ 0,1,\infty \} & \buildrel (1-z,z) \over \lra
& {\bf C^*} \x {\bf C^*}
\end{array}
\eeqa
As the first two entries of a section $s$ of $\underline{{\cal H}}$ are fixed
by the construction (up to the indeterminacy caused by the coset)
$s$ has the form
$s(z)={\bf {\cal H}_Z}(- \log \be z , \log z \, | \, c)$.
Asking even for a {\em flat} section one finds
(undoing the fibre identification $c \ra e^{c/2\pi i}=S$) that
the flatness condition $dc=u\, dv$ (from (\ref{connection})) just expresses
(\ref{first differential equation}), the $Li$ integral, and that
the coset takes into account
the multi-valuedness (\ref{multi valued}).
So $\underline{{\cal H}}$ possesses
the {\it flat} section (\ref{flat section})
and the Heisenberg bundle just encodes the fact that
the 'function' $Li$ is a section.

To gain information about $Li$ itself by somehow projecting to it
is not straightforward
as the immediate extraction of $Li$ is obstructed
by the ${\bf {\cal H}_Z}$ coset. What actually can be extracted
is the (suitably adjusted) imaginary part of it as we
describe now.

\newpage
\subsection{\label{explic anti-inv of imag part}Anti-invariance
of the adjusted imaginary part ${\cal L}$ of $W$}

To motivate this note that the integral
representation (\ref{Wrepres}) defines $W$ on the complex plane, cut
along the part $(1,\infty)$ of the positive real axis where $W$ jumps
by $2\pi i \log z$ when crossing the cut;
so the expression $W(z)+i\arg (1-z) \log z$ is continuous; its
imaginary part coincides with ${\cal L}$ below.

Now note that the {\em complex-valued} 'function' $Li(z)$
of the {\em complex} variable $z$ is not well-defined as a
function according to the multi-valuedness expressed by the
increments $\Delta_i$ around the $l_i$ which follow from
(\ref{multi valued}), i.e. $\Delta_0=0\, , \, \Delta_1=-2\pi i \log z$.
Note that if we restrict the values by considering just
$\Im Li(z)$ then this {\em real-valued} 'function'
of the {\em complex} variable $z$
has still $\Delta_0=0\, , \, \Delta_1=-2\pi \Re \log z $.
Therefore, if we
go one step further and consider the {\em real-valued} 'function'
of the {\em real} degree of freedom $z=e^{i\phi}$ living
on the critical circle $|z|=1$, we get indeed a
well-defined function.

Now it is interesting to see that, with a
slight modification, we can actually do better.
Namely, to extrapolate this property beyond the critical
circle, consider the expression
$\psi=\log \be z \, \Re \log z$ (vanishing on $|z|=1$).
One finds that the real-valued combination
of a complex degree of freedom
${\cal L}(z)=\Im \, Li(z) - \Im \, \psi(z)$
is actually not only a well-defined function
i.e. $\pi_1({\bf P^1}\backslash \{0,1,\infty\})$-invariant,
but at the same time also $\Sigma_3$ anti-invariant, so it
transforms {\em precisely} with the
sign-character, i.e. without correction terms. Furthermore,
it is even expressible by a function depending just on a {\em real}
degree of freedom: the critical
circle.

This is the case
although ${\cal L}$ is not just depending only on the angular part
of the complex variable $z$; rather it depends on the value of
$I(\phi)=\Im \, Li|_{e^{i\phi}}={\cal L}|_{e^{i\phi}}$
on the angular parts of the different ${\bf Z_3}$ transforms
of $z$, which themselves
are not depending on the angular part of $z$ alone,
cf. (\ref{Kummer}).

So note first that the function
one finds
(which also satisfies
${\cal L}(\bar{z})=-{\cal L}(z)$)
\beqa
\label{adjust imag part}
{\cal L}(z)=\Im  \; Li(z) - \Im \log \be z \; \Re \log z
\eeqa
(cf. (\ref{adjustment combination}))
is $\pi_1$-invariant, i.e. {\em single-valued}
as is also easily checked from (\ref{multi valued}).

Now, just as the single-valued cousin $\Re \log z $ of the logarithm
has anti-invariant transformation behavior under the duality group
${\bf Z_2}$ (with $\al:z\ra 1/z$), we
will see that ${\cal L}$ transfroms anti-invariantly under
$\Sigma_3$. We give four arguments for this: the direct computational
check, an argument using representation theory,
a manifestly invariant rewriting and finally a geometric interpretation.

\noindent
{\em Anti-invariance of ${\cal L}$
(first argument): explicit evaluation (appendix \ref{explicit check})}

This is the brute force procedure given by the explicit check.

\noindent
{\em Anti-invariance of ${\cal L}$
(second argument): representation theory (appendix \ref{anti-invariant})}

More conceptually one has a representation theoretic argument
(cf. (\ref{dLi rep}), (\ref{psi rep})).

\noindent
{\em \label{Kummer section} Anti-invariance of ${\cal L}$
(third argument): rewriting to a manifestly invariant expression}

The third argument (going in essence back to Kummer) is by an explicit
rewriting (\ref{adjust-proj Z3}).
Consider the decomposition of $Li(z)$ in real and imaginary
parts as we did above
for its restriction on the critical circle $|z|=1$.
One finds with $z=re^{i\phi}$ that
\beqa
\label{decomp}
Li(z)&=&-\int_0^r\frac{\log(1-e^{i\phi}t)}{t}dt\nonumber\\
 &=&-\frac{1}{2}\int_0^r\frac{\log(1-2t\cos \phi + t^2)}{t}dt
    +i\int_0^r \arctan (\frac{t\sin \phi}{1-t\cos \phi}) \frac{dt}{t}
\eeqa
This gives\footnote{\label{arctan}Note that
$\Im \, \log z=\arctan \, \frac{\Im z}{\Re z}$ and
$\Im \, \log \be z=\arctan \, \frac{\Im z}{1-\Re z}$.}
with $\arctan \frac{t\sin \phi}{1-t\cos \phi}=:\chi$,
$\kappa:=\chi|_{t=r}=\Im \, \log \be z$
and the inversion relation
$t=\frac{\sin \chi}{\sin (\chi+\phi)}$
(considering $\phi$ as a parameter)
the evaluation
(using (\ref{Lobachevsky}), (\ref{P+ log}))
\beqa
\Im \, Li(z)&=&
\int_0^r \chi \, \frac{dt}{t}=\chi \log t|_0^r
-\int_0^{\kappa} \log t \, d\chi
=\kappa \log r
- \int_0^{\kappa}  \log \frac{\sin \chi}{\sin (\chi+\phi)}d\chi\nonumber\\
&=&\kappa \log r +\frac{1}{2}\Bigl( I(2\phi)+I(2\kappa)-I(2\phi+2\kappa)\Bigr)
\nonumber\\
\label{Kummer}
&=&\Im \, \log \be z\, \Re \log z +\frac{1}{2}
\Bigl( I(2\, \Im \, \log z)+I(2\, \Im \, \log \be z)
+I(2\, \Im \, \log \be^2 z)\Bigr)\nonumber\\
\eeqa
This shows that the 'function' $\Im \, Li(z)$, which a priori is a
non-elementary real 'function' of a {\em complex} variable, is
actually already determined (up to the elementary logarithmic product term)
by the real 'function' $I(\phi)$ of a {\em real} degree of freedom
(cf. remark above).

Using the notation
$z\lra e^{i\phi(z)}=z/|z|=\exp\{i\, \Im \, \log  z\}$
for the ($\al$-compatible) operation of taking the angular part,
one has by (\ref{Kummer}), (\ref{Lobachevsky})
\beqa
\label{adjust-proj Z3}
{\cal L}(z)=
\sum_{i \in {\bf Z_3}}\Pi( \phi( \be^i z) )
\eeqa
which shows that ${\cal L}(z)$ is $\Sigma_3$ anti-invariant
(as $\phi(\al z)=-\phi(z)$ and $I(\phi)$ is odd).
And
${\cal L}(z)=
\frac{1}{4i}\sum_{\ga \in \Sigma_3}\sign(\ga)\,Li(e^{2i \phi (\ga z)} )$
, making a
$\Sigma_3$ anti-invariant projection in ${\cal L}(z)$ manifest,
follows\footnote{Note the $\al$-anti-invariant projection in
$I(\phi)=\Im \, Li(e^{i\phi})=
\frac{1}{2i}\Bigl( Li(e^{i\phi})-Li(e^{-i\phi})\Bigr)$
making $I$ odd.}
with
$\Pi(\phi)=\frac{1}{2}I(2\phi)=\frac{1}{4i}\Bigl( Li(e^{2i\phi})-
Li(e^{-2i\phi})\Bigr)$ (cf. (\ref{angular orbit}) for $N=2$).\\

\noindent
{\em \label{hyperbolic volume}
Anti-invariance of ${\cal L}$ (fourth argument):
volume of an ideal hyperbolic tetrahedron}

This approach uses a geometric interpretation
(\ref{volume tetrahedron}).
The idea is to interpret the transformation behaviour
of ${\cal L}(z)$ (under the $\Sigma_3$ operation on $z$)
geometrically in the following sense: $z\in {\bf P^1}$ is interpreted
as being actually a cross ratio
(cf. (\ref{cross ratio}); the definition is normalized
so that $z=cr\{ \infty , 0, 1, z \}$)
\beqa
\label{z cross ratio}
z=cr\{ z_1,z_2,z_3, z_4 \}=\frac{z_1-z_3}{z_1-z_4}/\frac{z_2-z_3}{z_2-z_4}
\eeqa
of four points $z_1,z_2,z_3, z_4$ in ${\bf P^1}$  and the operation
of $\Sigma_3$ as the residual effect of the original $\Sigma_4$
on the $z_i$ (cf. (\ref{cross ratio table}), (\ref{cross ratio
isomorphism}));
then ${\cal L}(z)\in {\bf R}$ is understood as a geometrical
quantity which transforms under $\Sigma_4$ with the $\sign$ character
(of $\Sigma_4$ which induces the corresponding character on
$\Sigma_3$). For this geometrical quantity one takes the hyperbolic volume
of the ideal tetrahedron in hyperbolic three space ${\bf H_3}$
(cf. app. \ref{tetrahedron}) with vertices
$z_1,z_2,z_3, z_4$ lying on the boundary ${\bf P^1_C}$ of ${\bf H_3}$.
This is then manifestly independent of the numbering of the vertices
except that the orientation changes under odd renumberings
showing the anti-invariant transformation behaviour.

Here an ideal tetrahedron is a tetrahedron $\Delta$
(bounded by geodesic faces and geodesic
edges)\footnote{The geodescis
are vertical lines and semi-circles (in vertical planes) with
endpoints in the boundary ${\bf C}\cup \{ \infty \}$;
geodesic planes are vertical planes and hemispheres
(over ${\bf C}$ and bounded by geodesics).}
with vertices ${z_1, z_2,z_3, z_4}$ on  the boundary
${\bf C}\cup \{ \infty \}$. One has
$\vol \, \Delta = {\cal L}(z)$ with $z=cr \{z_1, z_2,z_3, z_4 \}$
(cf. app. \ref{tetrahedron})
or, equivalently\footnote{as the $z_i$ can be transformed to $(\infty, 0
, 1 , z)$ by an element of $Sl(2,{\bf C})$ on ${\bf P^1_{\bf C}}$,
an isometry of ${\bf H_3}$}
\beqa
\label{volume tetrahedron}
\vol \, \Delta(z)={\cal L}(z)
\eeqa
for an
ideal tetrahedron $\Delta(z)$ with vertices $(\infty, 0 , 1 , z)$.
As a check note that $\Delta(z)$ degenerates
if one of the faces degenerates, i.e. not only for $z=0, 1, \infty$
but also for $z$ being on a line with $0$ and $1$,
i.e. for $z$ real; in all these cases (\ref{adjust imag part})
vanishes as well.

Let us give an example. The symmetric hyperbolic three-simplex $\Delta_{sym}$
(with vertices on ${\bf P^1_C}$ and having all six dihedral angles
equal to $\pi /3$) is in the 'circle gauge' (cf. app.)
given by the vertices
$\infty, -1, -\om^2 , -\om$, so $z=-\om=e^{i\pi/3}$
and (cf. (\ref{intersection of three phases}))
\beqa
\label{symmetric simplex}
\ga_i=\pi/3
\eeqa
Now the volume (\ref{Lobachevsky volume}) of a hyperbolic three-simplex
becomes maximal\footnote{In ${\bf H_2}$
$area (\Delta_2)=\pi - \sum \al_i$ becomes maximal
($=\pi$) for $\al_i=0$ (like for the fundamental domain).}
for the symmetric case (\ref{symmetric simplex}) (cf.
the corresponding remark about $I(\phi)$ in subsection \ref{critical circle})
\beqa
\label{symmetric simplex volume}
\vol (\Delta_{sym})=3\Pi (\pi/3)
\eeqa
which, being equal to $\frac{3}{2}I(2\pi/3)$,
equals indeed ${\cal L}(z)=\Im Li(z)=I(\pi/3)$ by (\ref{I relation}).

\subsection{\label{lin modif subsect} Linear Modifications}

We consider now some possible slight modifications
which can occur from different prespectives but all have
a somewhat similar flavour.
Recall that we found (\ref{Rogers footnote})
for the formal triality
symmetric (anti-invariant) modification $\tilde{W}$ of $W$
\beqa
\tilde{W}(z)=R(z)\;
\pm \frac{i \pi}{6} (\log z - \log \be z) - \frac{\pi^2}{6}
\eeqa
So $\tilde{W}$ was a less than quadratic modification
of\footnote{In many respects (cf. sect. \ref{Outlook}) the Rogers modification
$R(z)=Li(z)-\frac{1}{2}\log \be z \log z$ (cf. (\ref{Rogers})),
which itself may be described
as a quadratic (in the log's of $z$ and its ${\bf Z_3}$ transforms)
modification of $Li$,
is a conceptually more natural object to consider.}
$R$, i.e. up to a constant just a {\em linear} modification.
We want to point here to other occurrences of such linear modifications.

For this let us recall the differential equation
(\ref{first differential equation}) for the
superpotential {\em before} doing the holomorphic completion
(which the aim to get a proper superpotential suggested)
\beqa
\label{antiholo diff equ}
\frac{dW}{d\log u} = \Im \log \be u
= \frac{1}{2i}(\log \be u - \log \be \bar{u})
\eeqa
(where we absorbed a factor $i$ into $W$).
Considering $z$ and $\bar{z}$ as two degrees of freedom like $\Re \, z$
and $\Im \, z$, the antiholomorphic part of the rhs of
(\ref{antiholo diff equ}) is independent of the differentiation
variable and one finds by giving up the holomorphicity demand on $W$
a superpotential $W_{anom}$ with a holomorphic anomaly
\beqa
W_{anom}(u)=\frac{1}{2i}Li(u)-\frac{1}{2i}\log \be \bar{u} \, \log u
\eeqa
which is a linear modification in the holomorphic coordinate
(of course quadratic when considered non-holomorphically). For
easier comparison we display some relations
\beqa
\Re \, W_{anom}(u)&=&\;\;\,\frac{1}{2}\Im \, Li(u) - \frac{1}{2}
(\Re \, \log \be u \, \Im \, \log u - \Im \, \log \be u \, \Re \, \log u)\\
\Im \, W_{anom}(u)&=&-\frac{1}{2}\Re \, Li(u) + \frac{1}{2}
(\Re \, \log \be u \, \Re \, \log u + \Im \, \log \be u \, \Im \, \log u)\\
\frac{1}{2}\L(u) &=&\;\;\, \frac{1}{2}\Im \, Li(u) - \frac{1}{2}
\Im \, \log \be u \, \Re \, \log u
\eeqa
Further, allowing [\ref{AVII}] in (the rhs of)
the differential equation (\ref{first differential equation}) of $W$
for an additional additive constant $-\log \beta u_*$,
one finds again a similar modification but with $u_*$ constant
\beqa
\label{Vafa}
W^{var}(u)=Li(u) - \log \be u_* \log u
\eeqa

Finally,
including codimension four singularities like in
${\bf R^4}\x {\bf S^3} \x {\bf R^4}/{\bf Z_N}$
leading to non-abelian gauge symmetry on
${\bf R^4}\x {\bf S^3}$ (cf. sect. \ref{co 4}) one has for the
full superpotential\footnote{cf. footn. \ref{full supo footn}; here $S$
is the superfield $tr\, W^{\al}W_{\al}$ of highest
component $\int d^2\th \, S = tr\, (F^2+iF\we F)$}
\beqa
W_{YM, mem}= c\,  W(u_{1,k}) + S \underline{\Phi}
= cN\, Li(e^{2\pi i k/ N}\underline{u}^{1/N}) - i S \log \underline{u}
\eeqa

\newpage

\subsection{\label{flux supo subsect}
\large{{\bf Comparison with a flux superpotential}}}

We now want to compare the membrane instanton superpotential
$W\sim Li$ with a flux induced
superpotential\footnote{\label{dual footn}strictly speaking this may
concern in general a 'dual' $G_2$ holonomy manifold; the difference
may concern in our case of the $M$-theory conifold just a
transition to a phase with the role of ${\bf S^3}$'s exchanged}
$W_G=\int_{W_7} G \we (C+i \Upsilon )$.
A non-trivial $G$-flux turned on (as classical background)
will break supersymmetry [\ref{AS}], [\ref{CK1}]. So the
mentioned comparison is possible only because of
the absence (\ref{N=1 vacua}) of proper susy vacua,
i.e. one can have a non-trivial $G$-flux
just$^{\ref{just one end footn}}$
for all $u\neq 0$. One tries to {\em choose}$^{\ref{comp footn}}$
$\int_B G :=\log \beta u $ (mimicking the quantum vev).
Recall that the notion of superpotential was somewhat improper
because of the non-compactness of $X_7$; similarly one does not have
a proper K\"ahler potential\footnote{Actually $W$
is a section of a line bundle
$L$ of $c_1(L)=\frac{1}{2\pi i } \pa \bar{\pa}K$
over the moduli space $\M$.}
in the infinite volume case, or a flux with support on a closed cycle
(here $B$ has effectively the boundary\footnote{compactification of $X$
gives the closed cone of boundary $Y={\bf S^3_Q}\x {\bf S^3_D}$;
for $M$-theory on manifolds with boundary [\ref{Hor/Wit}]
{\em open} membrane instanton contributions to the superpotential
become important [\ref{CK2}]} $D$). In the end
all of this should be embedded in
compact $G_2$ holonomy manifolds. But at least it is suggestive to see
how the form of the membrane instanton superpotential may reappear here.
Let us recall first the flux
superpotentials [\ref{TV}], [\ref{G}], [\ref{AS}].

\noindent
One has, schematically,
the flux-generated superpotential in type IIB on a
Calabi-Yau
\beqa
\label{superpot flux CY}
W_H=\int_{CY}H\wedge \Omega \;\;\;\; , \;\;\;\;
V_H=\int_{CY}H\wedge * H \;\;\;\;\;\; (\,\;\; + \,\;\;
n \,\;\; )
\eeqa
(with holomorphic three-form $\Omega$ and $H=H^R_3+\tau H^{NS}_3$,
cf. [\ref{TV}]) the associated scalar potential $V_H=\int d^2\th \, W_H$
can be obtained from a Kaluza-Klein
reduction of the kinetic term $H^2$
(including a topological integer
$n \sim \int_{CY}H^{NS}\we H^R$).
Similarly one has, schematically, on a $G_2$ holonomy manifold $X$
with covariant constant three-form $\Upsilon$
[\ref{AS}], [\ref{BW}]
\beqa
\label{scalpot flux G2}
W_G=\int_X G\wedge (C+i\Upsilon)\;\;\;\; , \;\;\;\;
V_G=\int_{X}G\wedge *G + (\int_X G \we C)^2
\eeqa
and gains\footnote{Precise normalizations [\ref{BW}] give
$\vol (X)=\frac{1}{7}\int \Upsilon \we * \Upsilon$,
$\th=\frac{1}{4\pi}\int G\we C  \; \in \; {\bf R}/2\pi {\bf Z}$,
$e^K=\frac{(2\pi)^3}{\vol(X)^3}$,
$W_G=\frac{1}{8\pi^2}\int G \we (\frac{1}{2}C+i\Upsilon)
\in {\bf C}/\frac{1}{2}{\bf Z}$,
$V_G\sim e^K \Bigl( \vol(X) \int |G|^2 + (2\pi \th)^2 \Bigr)$
(kinetic term $e^{K/3} |\pa_i W|^2\sim \int|G|^2$)}
in the scalar potential
the Kaluza-Klein reduction of the kinetic term $G^2$.
$V_G$ contains, now in our non-compact
case\footnote{\label{scalar potential footn}The two effective supersymmetry
transformations of the double fermionic integration in
$V=\int d^2\th \, W$
lead for $u$ [\ref{AW}] from the volume (metric) in
$\int_Q \Y \sim \int_Qd^3x\sqrt{g}$ ($Q$ is supersymmetric)
first to the gravitino
$\psi$ and then again to the bosonic field $C$, more precisely
to $\int_{Q_i}*G$; so
symbolically one gets from the $ \int G \we \Y $ in $W_G$
the $\int G \we *G  $ in $V_G$ via
$\int d^2\th \, \int_Q \Y \sim \int_Q *G$;
just as with $\Om$ and $*H$ for a Calabi-Yau.}
$X_7$, a term of the form
considered in (\ref{deviation})
\beqa
\label{scalpot flux G2 contrib}
\int_{X}G\wedge *G  =
\int_{B_i} G \; \int_{Q_i}*G \leria^{\ref{comp footn}}
\int_{D_i}C \; \int_{Q_i}*G
\eeqa
(at least in a schematic product ansatz where also
$\th \sim \int_{B_i}G \, \int_{Q_i} C\leria^{\ref{comp footn}}
\int_{D_i} C \, \int_{Q_i} C$
so one has still $V_G\sim \int_{D_i} C$).
Now we want to argue for the $P_i$ as
representing supersymmetric vacua.
For this note that the relations $ u_i=0$ and $\int_{D_i}C=0$
(which hold at the semiclassical end $P_i$ where $r_0 \approx \infty$)
give $W|_{P_i}=0$ and $\pa W|_{P_i}=0$: the first from
$\int_B G= \log \be u \ra 0$,
the latter by
$V\sim |\pa W|^2$ (at $P_i$ where $W|_{P_i}=0$)
together
with
$V_G\sim \int_{D_i}C$ by (\ref{scalpot flux G2 contrib})
(conversely
(\ref{deviation}), (\ref{scalpot flux G2 contrib})
would give $P\neq P_i\Lra \int_D C \neq 0 \Lra \pa W\neq 0$)
\beqa
\label{susy vacua}
P=P_i \Lra W|_P=0=\pa W|_P
\eeqa
Note further that (specialising to our
non-compact
$X_7$)
\beqa
\label{flux supo derivation}
\frac{\pa W_G}{\pa \Phi_i}=\int_X G\we \de_{B_i}
=\int_{B_i}G
\eeqa
{\em suggesting}\footnote{\label{comp footn}
In the compact case note that $G$-flux,
on a {\em compact} $K3$ fibre, say, of a $K3$ fibered $W_7$,
is quantised (in units of $2\pi$ over tension),
so constant over the moduli space
(and the duality [\ref{AchW}] with the heterotic string
might then be obstructed as for type IIA [\ref{CKKL}])).
Here $\int_B G = \int_D C$, being zero
classically, becomes in the quantum domain the varying expression
$\Im \, \log \eta$, which may now be {\em mimicked} (!)
by prescribing for each $u$ a corresponding classical
flux background (which entails the ensuing formal similarities).}
with (\ref{ui etai relations})
again the differential equation
(\ref{first differential equation}) (and (\ref{susy vacua})
by (\ref{N=1 vacua}))
\beqa
\label{flux derived diff equ}
\frac{dW_G}{d\, \log u_i} = \log \be u_i
\eeqa
\noindent
{\em Remark:} There is a formal similarity between
two invariance-adjustments of the superpotential $W_G=-i \, Li$.
The adjusted imaginary part $\L$, depending on $u=e^{i\Phi}$,
is invariant under monodromy from the
$\pi_1({\bf P^1_u}\backslash \{0, 1, \infty\})$ action (with
{\em additive} shifts $\De_{l_0}\Phi = 2\pi$,
$\De_{l_1}W_G=2\pi i \Phi$);
a term like $DW$ (now dependent on $\Phi$)
is invariant under
K\"ahler transformations (with {\em multiplicative}
shift $W\ra W e^{-F}$ and $K\ra K + F + \bar{F}$; so the section $W$
is adjusted to a well-defined function in  $e^G=e^K|W|^2$).
>From the general relation
$\pa_i K = \frac{i}{2}\frac{1}{\vol(X)}\int \chi_i \we * \Y$
(where $\Y=\Im \Phi^i \; \chi_i$ is the harmonic decomposition)
[\ref{BW}] one gets
in our local situation of $X_7$ with one $3$-cycle $Q$,
$\chi=\de_B$ and $\vol(X)=\Im \Phi \int_B * \Y$
the finite expression
$\pa_i K  =\frac{i}{2}\frac{1}{\vol(X)} \int_B * \Y
=\frac{i}{2}\, \frac{1}{\Im \Phi_i}$, which gives
for the covariant derivative
$DW =  \pa W + \frac{i}{2}\, \frac{1}{\Im \Phi_i} \, W $
(so that the difference to the ordinary derivative $\pa W$
vanishes when approaching via $\vol \, Q \ra \infty$ the end).
So one can compare (with $Li(u=e^{i\Phi})$)
\beqa
\Im \Phi \cdot \Im D W_G &=& \frac{1}{2} \Im \Big( iW_G \Big)
                          + \Im \Phi \, \Im \frac{\pa (iW_G)}{\pa (i \Phi)}\\
\L &=& \Im \, Li + \Im \, \Phi \, \Im \frac{\pa Li}{\pa (i\Phi)}
\eeqa
which leads, up to the factor $1/2$,
with $iW_G=Li$ to a certain formal parallelism.

\newpage
\section{\label{co 4}codimension $4$ singularities}
\resetcounter
When $X={\bf R^4}\x{\bf S^3}$ is divided through by a finite subgroup
$\Ga$ of $SU(2)$ one obtains
$X_{1, \Ga}\cong {\bf R^4}/ \Ga \x {\bf S^3}$
and $X_{2, \Ga}\cong X_{3, \Ga}\cong {\bf S^3}/ \Ga\x {\bf R^4}$
leading as effective four-dimensional field theories
to an  $ADE$ gauge theory and to a theory without massless
fields, respectively
(the latter explaining the conjectured mass gap of the former)
($\Ga$ operates always on
the first factor of $SU(2)^3/SU(2)_D$, the $i$ in $X_i$ denotes which
factor is filled in).\\ \\
{\em Relations between the observables given by the $\eta_i$-variables}

In more detail [\ref{AW}]
let $N=|\Ga|$ and $Y_{\Ga}=\Ga\back {\bf S^3} \x {\bf S^3}
=\Ga\back SU(2)^3/SU(2)$ where $\Ga$ acts on the {\em first} factor.
Clearly the triality symmetry $\Sigma_3$ is broken down to ${\bf Z_2}$.
The three-cycles $D_i$ project to the $D_i^{'}$ with
$D_1\simeq ND_1^{'}$ and $D_i\simeq D_i^{'}$ for $i>1$
and one has $ND_1^{'}+D_2^{'}+D_3^{'}=0$ and
$N\al_1^{'}+\al_2^{'}+\al_3^{'}=N\pi$ (with $\al_i^{'}=\int_{D_i^{'}}C$)
from the consideration of the membrane anomaly. The ensuing relation
$\eta_1^N\eta_2\eta_3=(-1)^N$ for the variables
$\eta_1=\exp \{ k \frac{2f_3+f_1}{3N}+i\al_1^{'} \}$ and
$\eta_i=\exp \{ k \frac{2f_{i-1}+f_i}{3}+i\al_i^{'} \}$ for $i>1$ entails
that now the $\eta_i$ are not a simple ${\bf Z_3}$ orbit
with $\eta_3=\be \eta_1, \eta_2=\be^2 \eta_1$,
as in (\ref{etai Z3 orbit}), but that rather
\beqa
\label{eta relations A series}
A-series\hspace{3.9cm}
\eta_2= (\be^2 \eta_1)^N \;\;\;\; , \;\;\;\;\; \eta_3= (\be \eta_1)^N
\hspace{3.5cm}
\eeqa
This is actually only the case if $\Ga$ is just a cyclic group,
corresponding to the $A$-series (in the type IIA
reinterpretation this means that one has wrapped $N$ $D6$-branes on the
${\bf S^3}$ respectively has $N$ units of Ramond flux on the ${\bf S^2}$).
For the two different types
of $D_n$ singularity in $M$-theory ($n\geq 4$), with gauge group
$SO(2n)$ and $Sp(n-4)$ (of dual Coxeter numbers $h=2n-2$ and $h'=n-3$,
fulfilling $N=h+2h'$), respectively, where the latter 'exotic' case
leads to the new semiclassical limit point $\eta_1=-1$ (beyond $0,1,
\infty$) of the quantum modulio space $\N_{\Ga}$, the relations are
\beqa
D-series\hspace{1.5cm}
\eta_2=\Bigl( \be^2 \eta_1\Bigr)^h \Bigl( \be^2 (-\eta_1)\Bigr)^{2h'}
\;\;\;\; , \;\;\;\;\;
\eta_3= \Bigl( \be \eta_1\Bigr)^h \Bigl( \be (-\eta_1)\Bigr)^{2h'}
\hspace{1.5cm}
\eeqa
Finally for the $E$-series one has (with $\om_t=e^{2\pi i /t}$)
new classical limits at $\eta_1=\om_t^{\mu}$: the different
$E$-singularities in $M$-theory are indexed by an integer $t$ dividing
some of the Dynkin indices $k_i$ of the $E$-group and an integer
$\mu$ with $1\leq \mu < t$ and $(\mu, t)=1$ (for $t\geq 2$; for $t=1$
is $\mu=0$). With $h_t=\frac{1}{t}\sum_{t|k_i}k_i$
the dual Coxeter number of the gauge group
$K_t$ in $M$-theory at a $G$-singularity of index $t$ one has (in general)
\beqa
E-series\hspace{2.6cm}
\eta_2=\prod_{t, \mu}\Bigl( \be^2 \om_t^{\mu}\eta_1\Bigr)^{t\, h_t}
\;\;\;\; , \;\;\;\;\;
\eta_3=\prod_{t, \mu}\Bigl( \be \om_t^{\mu}\eta_1\Bigr)^{t\, h_t}
\hspace{2.5cm}
\eeqa
\newpage
\noindent
{\em Relation to the instanton expansion parameters $u_i$}

After the mutual $\eta_i$ relations let us also give the analogues
of the relation (\ref{ui etai relations})
between the $\eta_i$ and the instanton parameters $u_i$.
Let us restrict
us to the $A$-series (in general a membrane wrapped on
${\bf S^3_Q}\subset X_{1, \Ga}$ corresponds to $t$ instantons in $K_t$.).
The different $X_{i,\Ga}$ are defined by the 'filling in' condition
$D_i^{'}\simeq 0$.
Consider the two cases $i=1$ resp. $i>1$ separately. At the
center of $X_{1,\Ga}={\bf R^4}/\Ga \x {\bf S^3}$ lies
${\bf S^3}=Q_1^{'}\simeq \pm D_{i>1}^{'}$
(as the membrane instanton corresponds in the four-dimensional
supersymmetric $SU(N)$ gauge theory to a point-like Yang-Mills instanton,
note that, because of chiral
symmetry breaking as detected by the gluino condensate,
the local parameter at $P_1$ is not
$u_1=\exp \{ i(\int_{Q_1^{'}}C+i\Upsilon) \}$ but rather $u_1^{1/N}$).
As earlier one gets $u_1=\eta_2$ or with
(\ref{eta relations A series})
\beqa
\label{u1 eta1 relation}
\be (u_1^{1/N})=\eta_1
\eeqa
(again footn.'s \ref{uglob footn}, \ref{interpret footn} apply).
For $i>1$ one has at the center of $X_{i, \Ga}={\bf R^4}\x {\bf S^3}/\Ga$
lying ${\bf S^3}/\Ga=Q_i^{'}=\pm D_1^{'}$ (this time $u_i$ is a good local
parameter at $P_i$). $u_3=\eta_1$ leads now to
\beqa
\label{u3 eta3 relation}
\be u_3=\eta_3^{1/N}
\eeqa
{\em Superpotential}

Let us consider the
ensuing superpotential evaluations (\ref{first differential equation})
(again for the $A$-series).
On $X_{i,\Ga}$, where $i=1$ or $3$, one gets for
$dW / d \log u_i$ now $\int_{{\bf S^3}/\Ga}C=\Im \log \eta_1$
and $\int_{{\bf S^3}}C=\Im \log \eta_3$, respectively,
and finds (after holomorphic completion) with (\ref{u1 eta1 relation}),
(\ref{u3 eta3 relation})($k\in {\bf Z_N}$)
\beqa
W(u_{1,k})&=&\int_0^{u_1} \log \be (t^{1/N})d\log t=
N Li(\om_N^k u_1^{1/N})\\
W(u_3)&=&\int_0^{u_3} \log (\be t)^N d\log t=N Li(u_3)
\eeqa
($u^{1/N}$ principal value, $\om_N=e^{\frac{2\pi i}{N}}$).
$u_1^{1/N}=\be^2 u_3$ by $u_1=\eta_2$, $u_3=\eta_1$ and
(\ref{eta relations A series}).
By (\ref{angular orbit})
\beqa
\label{supo sum relation}
\sum_{k\in {\bf Z_N}}W(u_{1,k})&=&Li(u_1)\\
W(u_3)&=&N Li(u_3)
\eeqa

Now consider the four-dimensional interaction
$\Im \, \int_{{\bf R^4}} d^4y \, \int d^2 \th \,
S \, \underline{\Phi}$
with $S=tr \, W^{\al}W_{\al}$
the 'glueball' chiral superfield of lowest
component the gaugino bilinear $tr \, \la^2$ and highest component
$\int d^2 \th \, tr \, W^2 = F^2 + i F \we F$
($W^{\al}$ the field-strength superfield
of highest component $F+i*_4 F$)
and $\underline{\Phi}$  the superfield
of lowest component $\Phi=\int_Q C+i\Y$
and highest component $\int_Q *_{(X_7)}G$ by (\ref{oneinstcontrib}).
Just as the coupling constant in front of
the kinetic term of the
seven-dimensional gauge fields on ${\bf R^4}\x {\bf S^3}$
gets rescaled by $\vol({\bf S^3})=\Im \, \Phi$ in four dimensions,
one has an interaction
$\int_{{\bf R^4}\x {\bf S^3}}tr F\we F \we C$
(gauge theory instantons carry membrane charge)
so $\Re \, \Phi $ leads to the four-dimensional theta-angle $\th$.
On the other hand one has the interaction
$\int_{{\bf R^4}\x {\bf S^3}}d^4y \, tr \la^2 \we *_{(X_7)}G$
and as $tr \la^2$ gets a vev $\sim \La^3 e^{\frac{i}{N}\th}\om_N^k$
($k\in {\bf Z_N}$, the $N$ different vacua from chiral symmetry breaking)
one finds [\ref{AW}] again (cf. (\ref{oneinstcontrib}))
the interaction proportional to $\int_{{\bf S^3}}*G$.

Taken together
(with relative weight factor\footnote{by a shift
$\Phi \ra \Phi +\Phi_0$ the $c$
can be identified with a shift in the bare coupling constant, so
$c\sim e^{\frac{i}{N}\Phi_0}$;
further there is an order $N^2$ factor
[\ref{V flux supo}], [\ref{W QCD str}]}
$c\sim e^{-1/g^2_{YM}}e^{\frac{i}{N}\th}$)
with the membrane instanton contribution $u$
one finds as superpotential
$W_{YM, mem^{(1)}}=S \underline{\Phi} + N^2 c\, u_1^{1/N}$
of critical point $S=-iNc\, u_1^{1/N}$.
This gives $\Phi=-i \log (\frac{i}{Nc}S)^N$ and
$W_{eff}(S)=-iS\log(\frac{i}{Nc}S)^N + iNS$
of critical point $(\frac{i}{Nc}S)^N=1$ or
$tr \, W^2=-iNc\, \om_N^k$.
We will consider elsewhere
the critical point and the effective
superpotential for the full
\beqa
\label{full supo footn}
W_{YM, mem}=S \underline{\Phi} + Nc\, W(u_{1,k})
\eeqa
\noindent
{\em Remark:}
There is another ${\bf Z_N}$ relation (\ref{Rogers CFT relation})
besides (\ref{supo sum relation}) (i.e. (\ref{angular orbit}))
which would be interesting to relate
with the ${\bf Z_N}$ of $SU(N)$ or to provide a gauge-theoretic meaning.

The expression
$R(z)=\frac{1}{2}\Bigl( Li(z)-Li(1-z) \Bigr) +\frac{\pi^2}{12}$,
cf. (\ref{Rogers}), is more suitable for expressing some $Li$ relations.
Actually one has a relation (cf. appendix, sect. \ref{Rogers CFT section})
\beqa
\label{Rogers CFT relation}
\sum_{l=1}^{N-1}R\Bigl( \frac{\sin^2\frac{\pi}{N}}
{\sin^2 l\frac{\pi}{N}}\Bigr)=
R(1)+\sum_{i=1}^{N-2} R \Bigl(
\frac{\sin^2 \frac{\pi}{N}}{\sin^2 (i+1)\frac{\pi}{N}} \Bigr)
=\frac{\pi^2}{6}\Bigl( 1+\frac{3(N-2)}{N}\Bigr)
\eeqa
Here the argument is $1/Q_{i0}^2$
(where ${\bf i+1}=Sym^i \, {\bf 2}$ with action
$\mbox{diag}(z^i, z^{i-2}, \dots , z^{-i})$)
\beqa
Q_{i0}=\frac{\sin (i+1) \frac{\pi}{N}}{\sin \frac{\pi}{N}}
=|\frac{1-\om_{N}^{i+1}}{1-\om_{N}}|=tr_{{\bf i+1}} \, \om_N
\eeqa
with ${\bf Z_{N}} \hra Sl(2,{\bf C})$
via $\mbox{diag}(z, z^{-1})$  for $z=\om_{N}$.

(\ref{Rogers CFT relation}) is interpretable
[\ref{Dupont Sah Conformal}] as the evaluation
(cf. (\ref{hatC2 evaluation}))
of a Cheeger Chern-Simons class on a generator of
$H_3({\bf Z_{N}}, {\bf Z})$:
consider the embedding ${\bf Z_{N}}\hra PSl(2, {\bf R})$
given by
\beqa
\label{Rogers ZN}
\left(\begin{array}{cc}\cos \frac{\pi}{{N}} & -\sin \frac{\pi}{{N}} \\
\sin \frac{\pi}{{N}} & \cos \frac{\pi}{{N}} \end{array}\right)
\eeqa
Furthermore one has a geometrical interpretation [\ref{CGT}] that
$\sum_{j=1}^k R(t_j)=\frac{\pi^2}{6}n$ for a certain integer $n$
when a 3-manifold $M$ is triangulated by $k$ oriented tetrahedra $T_j$.
Here for each vertex $i$ ($i=1, \dots , N$) a real number $x_i$ is
given and one has associated to the tetrahedron $T_j$ the corresponding cross
ratio $t_j=cr\{ x_a, x_b, x_c, x_d \}$. As the set of tetrahedra forms
a triangulation the boundary $\pa \sum_{j=1}^k T_j=0$
of the associated 3-chain is zero and this implies the relation
$\sum_{j=1}^k t_j \we (1-t_j)=0$
which then implies (\ref{Rogers CFT relation}).
(Cf. app., sect. \ref{cohomological interpretation})

\newpage

\section{\label{Outlook}Interpretation and Outlook}

The preceding investigations cause
three sorts of questions. First, one may dwell on
some of the points touched already:
the identification of the relevant coordinates ($u_i, \eta_i$)
(cf. discussion around (\ref{ui etai relations})),
especially the globalization question with possibly a
direct connection to the type IIA approach [\ref{AVII}] which
uses special flat coordinates
of $N=2$; the question of
holomorphic completion of the superpotential
in $\frac{\pa W}{ \pa \log u} = \Im \log \be u \ra \log \be u$,
respectively the holomorphy violation
(by the boundary; cf. the $E_2$ anomaly,
even in superpotential contexts);
the interpretation of
the transformation rules of $W$ (section of Heisenberg bundle,
or even a balancing argument as in (\ref{cla qua bal});
$W_G$ monodromy from the $C$ field shifts);
also to see directly, before evaluation, the connection
$X_7 \ra \De(z)$ (and that a $G$ flux (?) evaluates $W_G$ on $X_7$
to the (complexified, cf. below) invariant $\vol(\De)$).

The second type of questions concerns an interpretation
of the results obtained so far (sect. \ref{local interpret}).
Finally their posssible extension to more generic (compact) cases
and  placement in a greater conceptual context
(sketched in the more speculative sect. \ref{global}, described in
more dateil elsewhere [\ref{global paper}]).
(For relations to type IIA string theory cf. [\ref{5dim}].)

\subsection{\label{local interpret}Local interpretation}
{\em The universal object over the quantum moduli space}

We now want to compare the structures found
with corresponding constructions in the
description of pure $N=2$ $SU(2)$ gauge theory as given in the
Seiberg/Witten (S/W) set-up [\ref{SW1}]. There
one was interested in the section
$\left( \begin{array}{c} \pa_u a_D \\ \pa_u a \end{array} \right)
=\left( \begin{array}{c}
\int_{\be} \pa_u \la \\ \int_{\al} \pa_u \la \end{array} \right)
=\left( \begin{array}{c}
\int_{C^3_{\be}} \Omega \\ \int_{C^3_{\al}} \Omega \end{array} \right)$
of the flat bundle given by the first cohomology of the
universal elliptic curve ${\cal E}\ra {\bf P^1_u}$ over the quantum
moduli space ${\bf P^1_u}=\Gamma(2)\backslash {\bf H_2}$;
the fibre over $u$ is $H^1(E_u,{\bf C})$
with the elliptic curve $E_u$ given as two-fold covering of ${\bf P^1}$
branched at $\infty, 0, 1, u$ (degenerating
for $u$ being one of the three points $\infty, 0, 1$;
encircling the corresponding singularities gives monodromy elements
generating $\Gamma(2)$).

Recall the
relation\footnote{where $\C$ is built up as a covering over ${\bf P^1}$
'the same way' (replacing the intersection lattice $H^2(K3,{\bf Z})$
by a zero-dimensional spectral set)
that $W$ is built up as $K3$ fibration over ${\bf P^1}$}
of the (meromorphic) periods of the
S/W curve $\C$ (describing a gauge theory engineered
on a $K3$-fibered Calabi-Yau $X$ in type IIA)
and the (holomorphic) periods of the mirror Calabi-Yau $W$;
one has with cycles $C_3\subset W, \, C_1\subset \C$
\beqa
\label{CY period reduction}
\int_{C_3}\Om \sim \int_{C_1} \la
\eeqa

Now replace this universal family of elliptic curves
by a family of three-manifolds,
or rather (in our local case) three-simplices:
we associate to $z\in {\bf P^1_z}$ the hyperbolic geodesic three-simplex
given by the ideal tetrahedron $\Delta(z)$ in\footnote{think of a
different copy of ${\bf H_3}$ over each point $z\in {\bf P^1_z}$
as ambient space for $\Delta(z)$ just as one thinks of a different
copy of the Weierstrass embedding plane ${\bf P^2_{x,y,z}}$ over each
$u\in {\bf P^1_u}$ as ambient space for $E_u$ so that in both cases
one really ends up with a fibration (where the fibres are disjoint)}
${\bf H_3}$ with vertices $\infty, 0, 1, z$; again the construction
degenerates for $z$ being one of the three points $\infty, 0,1$.

So the quantum regime of the universal local structure provided by the
non-compact $M$-theory conifold $X_7$ (of quantum moduli space ${\bf P^1_C}$)
corresponds
to the variation of $\De(z)$ over $z\in {\bf P^1_C}=\pa {\bf H_3}$
(cf. $\vol \, \De(z)=\L(z)$ in (\ref{volume tetrahedron})).

We will
compare to corresponding expressions in our set-up the pair
$a, \, a_D$ and the K\"ahler potential (in the S/W set-up)
as relations of dual torus periods ($\F$ prepotential)
\beqa
\label{prepotential Kahlerpotential}
a_D=\pa \F / \pa a \;\;\; &,& \;\;\;
K=-\Im \, a \, \bar{a}_D\\
\label{SW torus periods}
a=\oint_{\al} \la \; &,& \; a_D=\oint_{\be} \la
\eeqa
The quantum coordinate is
not $a^2$ but rather the corrected (cf. remark 3 below) quantity
\beqa
\label{SW uF relation}
\frac{1}{2\pi i}u=\frac{1}{8} (\F-\frac{1}{2}a \pa_a \F )
\eeqa
and in the stringy realization the quantum coordinate $u$ becomes
purely geometrical$^{\ref{refine},}$\footnote{The prepotential
$F(X^0, X^1, \dots , X^n)$ of the periods $X^i$
is related to the prepotential $\F(t^1, \dots , t^n)$ of the K\"ahler
coordinates $t^A=X^A/X^0$ via $F=(X^0)^2 \F$, giving the relation
$\frac{1}{2}(X^0)^{-1}\pa_{X^0} F=\F - \frac{1}{2} t^a \pa_a \F$.
The period $a_D$ is related to the conifold via
$\Xi_6\sim a_D \sim x_+ \sim \tilde{u}-1$ with
$\Xi_6$ the period related to the $6$-cycle in type IIA,
respectively the vanishing ${\bf S^3}$ of the conifold in type IIB.
The type IIA perspective on the conifold, the
period $a_D$ and its relation to the dilogarithm
are discussed further in [\ref{5dim}].}
\beqa
\label{SW u geometric period}
u=\Xi^2=\int_{{\cal C}_3}\Omega
\eeqa
which is made possible by going to the mirror description in type IIB.

With the modification $R$ of $Li$ (cf. (\ref{Rogers})),
whose relevance will emerge below repeatedly,
we can compare to (\ref{SW uF relation}) as one has
(so $a, \F, u$ are related to\footnote{we call here the $u$ modulus
of (\ref{u eta variables}) $z$ to avoid mix-up with the S/W $u$}
$i\Phi = \log z, Li, R$)
\beqa
\label{fR analogy}
R=Li - \frac{1}{2}\log z \, \pa_{\log z}Li
\eeqa

So the distinctive feature of the S/W solution,
that a quantity in the bulk of the quantum moduli space has
a purely geometric expression like the mentioned periods
(typical for string dualities), resembles the way how in our $N=1$ set-up
the quantum corrected quantity $W(u)$ becomes
an integral of classical geometry on a 'dual object'
(lying over a respective point in the quantum moduli space),
i.e. $\L=\int_{\De}vol$ (respectively its complexified extensions
described below which suggest the whole point of view).\\ \\
\noindent
{\em Remarks}

1) The membrane anomaly becomes manifest
in the described global quantum model
\beqa
\label{mapping of moduli spaces local version}
\sum _{{\bf Z_3}} \int_{{\bf S^3_{D_i}}}C
=\pi  \llra  \sum _{{\bf Z_3}} \tors(\ga_i)=\pi
\eeqa
(using ${\bf S^3_Q}$ for the ${\bf S^3_D}$
(phases)$^{\ref{interpret footn}}$).
This points to
a connection\footnote{One may look at an anlogue of
${\bf P^2_H}\ra \De$, a $T^2\cong U(1)^3/U(1)_D$ fibration
${\bf P^2_C}\ra \De$ (cf. [\ref{KKL}]).}
between the direct manifestation (\ref{euclid angle sum})
of the anomaly in the dual hyperbolic model
and the proof (\ref{mem anom proof}).

2) The different
{\em three-dimensional} structures we encounter
(membrane instantons and the modulus $\Phi=\int_{Q}C+i\Y$; the
Heisenberg bundle $\underline{\H}$
as built up from $Li$, $\log$ and $1$; the (solid) tetrahedron
$\De (z)$ and its volume\footnote{or
some hyperbolic 3-manifold $M_3$
with its volume and Chern-Simons form $\C^{CS}$, cf. below})
have corresponding two-dimensional structures
in the S/W set-up (world-sheet instantons;
$H^1(E_u)$ as built up by $(a_D , a)$; $E_u$).

3) In the S/W set-up there is also the relation
with a flux superpotential which we contemplated for our case in
subsect. \ref{flux supo subsect}.
For this recall the stringy realization of the $N=2 \ra N=1$ mass breaking.
According to [\ref{TV}] the quantum corrected version $W=mu=m<tr \Phi^2>$
of the classical ($u\approx a^2/2$)
mass deformation in the field theory\footnote{giving mass
to the chiral multiplet $\Phi$ of the vector multiplet, and so
the breaking $N=2 \ra N=1$. As near $\u=\pm 1$ a monopole
and a dyon become massless one gets by including the light states
$W=mu+(a_D-a_0)\phi\tilde{\phi}$
which leads to monopole condensation
and locking on $\u=\pm 1\llra a_D=a_0$. }
is realised as a flux
induced superpotential $W=\int_W \Om \we H_3$ in the type IIB string,
essentially because $u$ occurs among the Calabi-Yau periods
(cf. below). The stringy realization proposed in [\ref{TV}]
of this scenario started from the type IIA superpotential
\beqa
\label{IIA flux superpotential}
W_{flux}\sim \int_X H_2\we t \we t \sim
(\int_{{\bf P^1_b}}H_2) \cdot \vol(K3)=n_{flux}\, \pa_s \F
\eeqa
where\footnote{\label{refine}Actually
analytic continuation shows that this
expectation has to be refined [\ref{CKLTh}]: $W=mu$ is then given by
$W=mu\sim 2 i \Xi^2_{\infty} + \Xi^4_{\infty}=2it + \pa_s \F$.}
the entries $S$, $\frac{\pa \F }{ \pa S}$ of the type IIA period vector
correspond to $\vol({\bf P^1_b})$, $\vol(K3)$.

4) Having emphasized analogies between $X_7$ and the
S/W set-up let us point also to a difference.
In the S/W set-up at the three special points
$u=\infty, +1, -1$ BPS states become massless, the $W$-boson, a monopole
and a dyon, respectively. Being BPS states, in the string theory embedding
the relation between mass and volume is saturated, the respective
cycles of homology classes $N, N^+, N^-$
shrink at the special points and fulfill
\beqa
\label{CMS}
N = N^+ + N^-
\eeqa
Note that not all of the three special
points are on the same footing but some of them ($u=\pm 1$) are more equal
(in the stringy representation in type II
these correspond to ${\bf S^3}$'s in the mirror Calabi-Yau
whereas $u=\infty$ corresponds to ${\bf S^2}\x {\bf S^1}$, indicating
hypermultiplets and a vector multiplet, respectively.
In the type II conifold transition
the two small resolutions are also more equal, i.e. the type II
reduction breaks $\Si_3$ to ${\bf Z_2}$). These two points
$u=\pm 1$ lie on the curve of marginal stability. The potential decay
of BPS states when crossing such a curve were considered in investigations
about singularities of special Lagrangian three-cycles [\ref{Joyce}] from the
perspective of transitions that occur for corresponding supersymmetric
three-cycles in the Calabi-Yau manifold. In [\ref{Joyce}] two different types
of such singularities are considered, modelled (in ${\bf C^3}$) respectively
on a $T^2$-cone and two real 3-planes. The latter case was exemplified above in
(\ref{CMS}) and considered also in [\ref{KM}], [\ref{Denef 1}]
(and [\ref{AW}] when considering the cone over ${\bf P^3}$)
whereas the former is related to the case considered
in [\ref{AVII}], [\ref{AW}] (the case of the cone over ${\bf S^3}\x{\bf S^3}$)
and the present paper. Here the corresponding
relation between the homology classes of the three respective cycles
which become nullhomologous at the three special points is
\beqa
D_1+D_2+D_3=0
\eeqa
\subsection{Global interpretation}

{\em \label{global}Compact $G_2$ holonomy manifolds}

Now consider compact $G_2$ holonomy manifolds $X_7$,
$K3$ fibered over ${\bf S^3}$
(replacing the previous local fibre $K3^{decomp}={\bf R^4}$),
with singularities
not just of codimension $7$ (and potentially $4$) but also codimension
$6$. The latter case where the discriminant in the base
${\bf S^3}$ of $X_7$ is of codimension two, the 'discriminant link'
$l=\cup_j^h \ga_j$ (a union of $h$ circles),
will be especially
relevant to make the connection to the hyperbolic 3-manifold $M_3$.

More precisely, we will be concerned on the one hand with the case of a
codimension $7$ singularity of the classical geometry locally
modelled after the cone over $Y={\bf S^3}\x{\bf S^3}$, or even a
situation with many, say $h_X$, local 'ends' modelled
that way (having one ${\bf S^3}$ as base of the
$K3$ fibration
brings a certain asymmetry into the description). On the other hand
codimension $7$ singularities arise as the cone over ${\bf P^3_C}$,
the ${\bf S^2}$-twistor space over ${\bf S^4}$.
Consider here a component in the discriminant link given by
just an unknot $\ga$ and let $\S\subset {\bf S^3}$ be a spanning Seifert
surface, not touching, say, the other link components.
The total cycle traced out\footnote{Fibre singularities
relate to the cohomology of a total space: for an elliptic $K3$ for example
one gets ${\bf S^2}$'s building up $H^2(K3)$
(besides base and fibre)
from paths $P$ connecting points $p, p^{'}$
(of codimension $2$ in the base as is our link $l\subset {\bf S^3}$)
in the base ${\bf P^1}$ (so $\pa P = p^{'}-p$)
over which an ${\bf S^1}$ in the fibre shrinks.}
by following
the cycle ${\bf S^2_{x,y,z}}$ in the $K3$ fibre, which vanishes
over $\ga=\pa \S$,
through the whole $\S=D^2_{v,w}$ leads to an
${\bf S^4}=\{ x^2+y^2+z^2+v^2+w^2=1 \}$ contributing to
$b_4(X)=b_3(X)$.
\\ \\
\noindent
{\em \label{Hyperbolic 3-manifolds}Hyperbolic 3-manifolds}

Above
we considered the 3-dimensional structure given by the ideal tetrahedron
$\De$ in ${\bf H_3}$ and studied its volume. Actually one will consider a
two-fold generalisation. One refines the volume invariant
and generalises the tetrahedron to smooth manifolds. We recall
an universal cohomological interpretation of the
dilogarithm superpotential starting from the hyperbolic simplex volume
computation related to its single-valued cousin $\L$ which
points to the consideration of the complexified Chern-Simons invariant of
hyperbolic 3-manifolds (the volume combined with the Chern-Simons
invariant).

Concerning the first issue one pairs the volume with the
Chern-Simons invariant as suggested by
the cohomological interpretation of the occurrence of $Li$ (or $R$)
in the hyperbolic volume computations
(cf. app. \ref{cohomological interpretation} and below)
together with the Chern-Simons reformulation of
three-dimensional gravity [\ref{Witten CS gravity}].
Concerning the second issue one will consider general hyperbolic
3-manifolds $M$ and the way the refined (complexified)
Chern-Simons invariant varies over the hyperbolic deformation moduli
space of $M$; this shows [\ref{global paper}] (from a triangulation
by simplices) how the dilogarithm occurs in this variation.

Concerning the generalisation to smooth manifolds
note that just as in the
case of the upper half-plane one can study now discrete torsion-free
subgroups $\Gamma$ of the full group $PSl(2,{\bf C})$ of
orientation-preserving isometries of ${\bf H_3}$
and look for the corresponding (orientable) hyperbolic
three-manifold $M$ (complete Riemannian
manifold of constant curvature $-1$ of finite volume) given
by the non-compact quotient $\Gamma \backslash {\bf H_3}$ (any such
$M$ arises this way)\footnote{Cf. that a closed
surface of genus $g>1$ admits a metric of constant curvature $-1$ and
is isometric to $\Ga \back {\bf H_2}$. Note that by
the Mostow/Prasad rigidity theorem  two hyperbolic threefolds
of finite volume with isomorphic fundamental groups are actually
isometric, the volume is a topological invariant.}.
The geodesic simplices we studied occur, just as in the well known
upper half-plane case, as (parts of) fundamental domains for suitable
group actions and the sum of their volumes gives
the volume of the quotient manifold.

{\em Cohomological interpretation}\\
For $E\ra M$ a differentiable $Gl(n,{\bf C})$ bundle with {\em flat}
connection $\th$ one finds from the Bockstein exact sequence
that $c_2(E)$ lies in the image of the Bockstein homomorphism $\be$
\beqa
H^3(M, {\bf C/Z})\buildrel \be \over \lra H^4(M,{\bf Z}) \lra
H^4(M,{\bf C})
\eeqa
The second {\em Cheeger Chern-Simons class}
(app. \ref{cohomological interpretation})
gives a canonical choice of a preimage
\beqa
\label{CCS2}
\hat{C}_2(\th)\in H^3(M, {\bf C/Z})
\eeqa
With $\om$ a ${\bf C}$-valued $Sl(2,{\bf C})$ invariant three-form on
$Sl(2,{\bf C})/SU(2)={\bf H_3}$ one finds a ${\bf C/Z}$ valued
Eilenberg-MacLane cochain ${\cal I}(\om)$ with
\beqa
\hat{C}_2={\cal I}(\om)(g_1,g_2,g_3)=\int_{\Delta(z)}\om
\eeqa
One finds then\footnote{with
$\Re \, \hat{C}_2$ evaluated on $H_3(Sl(2,{\bf R})^{\de})$
for $z\in {\bf R} $.
One can give a similar interpration
for $\Re \log $ and its relation to
$\hat{C_1}\in H^1(Gl({\bf C}), {\bf R})$
just as ${\cal L}$ represents part of
$\hat{C_2}\in H^3(Gl({\bf C}), {\bf R})$.}
(with $L(z)=R(z)-\frac{\pi^2}{6}$, cf. (\ref{Rogers}))
\beqa
\label{hatC2 evaluation}
2\Re \, \hat{C}_2=\frac{1}{2\pi^2}L(z)   \;\;\;
\;\;\; (\mod \, 1/24) \;\;\;\; , \;\;\;\;
2\Im \, \hat{C}_2=\frac{1}{2\pi^2}{\cal L}(z)
\eeqa

{\em Transition to hyperbolic 3-manifolds}

Let $M$ now be a closed 3-manifold of hyperbolic
structure given by $M\cong \Ga_{hol}\back {\bf H_3}$
or equivalently by the holonomy representation
$h: \pi_1(M)\ra (P)Sl(2, {\bf C})$, respectively by a flat
$(P)Sl(2, {\bf C})$ bundle over $M$; this is pulled
back from the universal bundle $U$ over the classifying space
$BSl(2,{\bf C})^{\de}$ by a base map
$m:M\ra BSl(2,{\bf C})^{\de}$ so one can
evaluate\footnote{for $M$
closed $CS(M)$ is essentially the $\eta$ invariant, this is
suitably extended for $M$ non-compact}
$\hat{C}_2\in H^3(BSl(2,{\bf C})^{\delta})$
(cf. app. \ref{cohomological interpretation})
on the class in
$H_3(BSl(2,{\bf C})^{\delta})$ given by
$M$
\beqa
\label{real and imag invariant}
\Re \, \hat{C}_2(M)\sim  CS (M)\;\;\;\; , \;\;\;\;
\Im \, \hat{C}_2(M)\sim  \vol (M)
\eeqa

So the proper cohomological interpretation of
$\vol \, \Delta(z) = \L(z)$
leads to the consideration of hyperbolic 3-manifolds
$M$ for which the second Cheeger
Chern-Simons class is given by
(\ref{real and imag invariant})
with universal evaluation (\ref{hatC2 evaluation}).
For a hyperbolic 3-manifold $M$ the invariant
$\int_{M_3} {\cal C}^{CS}+i \, vol$ is studied
(in this complex Chern-Simons theory (as in 3D
gravity) one has naturally the
complex pairing of volume and the $CS$ 3-form field,
cf. [\ref{global paper}]).

The hyperbolic deformation moduli space is defined via
periods of the generator loops $m_i, l_i$ (for $h$ a suitable one-form)
for the (assumed) toroidal ends
\beqa
\label{dual hyp coord}
v_i({\bf u})=\pa \G / \pa u_i \;\;\; &,& \;\;\;
K({\bf u}):=2\pi \sum_i \length (\ga_i)=-\sum_i \Im u_i \bar{v_i}\\
\label{hyp coord}
u_i=\pm \int_{m_i}2h^* \; &,& \; v_i=\pm \int_{l_i}2h^*
\eeqa
In other words there exists again a prepotential $\G$
and again the expression
\beqa
\label{f analyt}
f=\frac{1}{4}(2-{\bf u} \pa_{{\bf u}})\G
\eeqa
has a purely geometrical description (Dehn filling the ends via
solid tori will be involved)
\beqa
\label{f pure geom}
f=\vol(M)+i {\cal C}^{CS}(M)=\int_M\om
\eeqa
(\ref{dual hyp coord})-(\ref{f pure geom}) compare to
(\ref{prepotential Kahlerpotential})-(\ref{SW u geometric period})
($a, \F, u$ relate to ${\bf u}, \G, f$).
One defines invariants
\beqa
\label{vol and len}
I(M)&=&\exp \{\int_M \frac{2}{\pi}\, vol+ i {\cal C}^{CS} \}
    =\exp \{ \int_M \om \}\nonumber\\
\la(\ga)&=&\exp \{ \mbox{length}(\ga)+i\, \mbox{tors}(\ga) \}
        =\exp \{ \int_{\ga}2h^* \}
\eeqa
generalising the occurrence of $Li$ and $ \log$, or
their real cousins $\L$ and $\Re \log$
as three- resp. one-dimensional volumes (\ref{volume tetrahedron}),
(\ref{length}) in $\Delta(z)$.
So one has corresponding triples
\beqa
\label{3 vectors}
\small{\left( \begin{array}{c}Li(y) \\ \log y \\ 1 \end{array} \right)}
\;\; , \;\;
\small{\left( \begin{array}{c}\hat{C}_2 \\ \hat{C}_1 \\ 1 \end{array} \right)}
\;\; , \;\;
\small{\left( \begin{array}{c}I(M_3) \\ \la(\ga_1) \\ 1 \end{array} \right)}
\;\; \; \;\;
\Bigl( \; \mbox{and} \;\;\
\small{\left( \begin{array}{c} \L \\ \Re \log  \\ 1 \end{array} \right)}
\;\; , \;\;
\small{\left( \begin{array}{c}\vol(\De)\\ \length(\ga)\\ 1 \end{array} \right)}
\; \Bigr)
\eeqa
\\
\noindent
{\em Interpretation}

Now having generalised our local
$G_2$ holonomy manifold $X_7$ to a global manifold (compact and
$K3$ fibered over ${\bf S^3}$) and furthermore having generalised
the hyperbolic geometry of the simplex $\De(z)$
to smooth hyperbolic 3-manifolds let us indicate a potential connection.
The quantum expression comprising all the corrections
may have again a purely geometric description
(when going to the dual description provided by
the hyperbolic 3-manifold $M$, a 'thinned out' (spectral) version of the dual
7-fold just as the S/W curve is the $K3$-integrated-out version of
the mirror CY). In the dual evaluation
the membrane instanton superpotential $W=Li(z)$, generalising $\L$,
occurs as a complexified volume of a simplex in
hyperbolic 3-space (respectively of a 3-manifold).
So just as the periods of the S/W curve were periods
of the type IIB Calabi-Yau (mirror to the original theory in type IIA)
now the dual 3-manifold $M^3_{\bf u}$ (analogue of the S/W curve $E_u$)
and its 'period' $f({\bf u})=\int_{M^3_{\bf u}}\vol +i \C^{CS}$
(evaluated in the local case by $R(e^{{\bf u}})$, i.e.
essentially by $Li(e^{{\bf u}})$) reflects a
($W_G$ ?) 'period' (evaluated locally
by $W\sim Li(e^{\Phi})$).

Supersymmetric (associative) ${\bf S^3}$'s, which
sit in $X$ locally like in ${\bf S^3}\x {\bf R^4}$,
contribute to $H^3(X)$; in the dual
3-manifold $M_3$ the hyperbolic moduli space has dimension $h$, the
number of ends, i.e. the number of link components in the description
of the discriminant of a $K3$ fibration
$X\ra {\bf S^3}$ (responsible for the codimension 6 singularities);
first these numbers and then the moduli of $X_7$ and $M_3$
have to be brought into relation for the dual description. Relating
the moduli spaces$^{\ref{dual footn}}$
(cf. (\ref{CY period reduction})) might include relations
\beqa
\label{mapping of moduli spaces}
\log u=\int_{{\bf S^3_{Q_i}}}-\Upsilon + i C \vspace{2cm} &\llra& \vspace{2cm}
\log \la_i=\int_{\ga_i}length + i \, tors\nonumber\\
\frac{\pa W}{\pa \log u_k}=\log v_k
&\llra&
\frac{\pa \G}{\pa u_i}=v_i
\eeqa
(the coordinate
$u  = \exp \{ i \int_Q C+i \Upsilon \}$
for the non-contractible ${\bf S^3}$ (and $\Phi_k=\log u$)
are replaced with
$\la
= \exp \{ \length (\ga)+ i \, \tors (\ga) \}$
for the non-contractible ${\bf S^1}$ (and ${\bf u}=\log \la$)).

The base part ${\bf S^3}-l$, over which
the fibre is non-degenerate, is a 3-manifold $M_3$.
Concerning the moduli spaces we want to compare note that
for the superpotential we are interested in the number $h_X$ of
these ${\bf S^3}$
(related to codimension $7$ singularities $C({\bf S^3}\x {\bf S^3})$)
whereas in the description of the hyperbolic
3-manifold and the complexified Chern-Simons invariant
we are interested in the number $h=h_M$ of ends of $M_3$
(or components of the discriminant link describing
codimension $6$ singularities,
related to ${\bf S^4}$ or the deformed $C({\bf P^3_C})$).
The 3-manifold $M_3$ having three representations
(the quotient $\Ga \back {\bf H_3}$,
a triangulation $M_3=\cup_k^n \De(z_k)$
and the link complement ${\bf S^3}-l$ of $l=\cup_j^h \ga_j$)
one now connects its second and third representation:
the question of translating the
different dimensions of moduli spaces is then
captured by the reshuffling of the
different summation boundaries in
$K=\sum_i^h (\length + i \, \tors) (\ga_i)$ and $f=\sum_k^n R(z_k)$
(where $R$ is essentially $Li$
and for $X_7$, very naively, $W\approx \sum_j^{h_X}Li(u_j)$)
inherent in
(\ref{dual hyp coord})-(\ref{f pure geom}) [\ref{global paper}].

The comparison (considered in more detail
elsewhere [\ref{global paper}]) will describe
the actual form which the analogy between a local description of a
singularity of a $G_2$ manifold by $X_7={\bf R^4} \x {\bf S^3}$ and
the Dehn filling of an end of a hyperbolic 3-manifold with the solid
torus $\T={\bf D^2} \x {\bf S^1}$ takes, i.e. the
mapping between the moduli spaces of $X_7$ and $M_3$
(including the prepotential of hyperbolic
deformation space) and the conncection between the
membrane instanton superpotential and (possibly
$G$-flux superpotentials resp.)
the complex $CS$ theory on $M_3$.

I thank A. Klemm, C. Vafa and E. Witten for remarks.

\newpage

\appendix

{\large {\bf Appendix}}

\section{\large {\bf \label{triality appendix}
The triality symmetry group on the moduli space}}
\resetcounter

$\Sigma_3$, the permutation group of
three elements, is built up from ${\bf Z_2}$
and an invariant subgroup ${\bf Z_3}=\{ e \;, \; \be \; , \; \be^2 \}$;
one has the relation $\al \be^i \al=\be^{-i}$ (read $i$ mod $3$).
\beqa
1 \lra {\bf Z_3} \lra \Sigma_3 \lra {\bf Z_2} \lra 1
\eeqa
The non-trivial coset consists of the three order two elements
${\al \; , \; \al\be \; , \; \al\be^2}$
\beqa
\label{group elements}
\begin{array}{ccccc}
e    & & \be      & &  \be^2    \\
\al  & & \al \be  & &  \al \be^2
\end{array}
\eeqa
There are three conjugacy classes (CC) given by the elements of
order one, two and three, respectively. We denote
a conjugacy class by $c$, and
the number of its elements by $n_c$.
We index the classes by the common order of its
elements, so $n_{c_1}=1, n_{c_2}=3, n_{c_3}=2$.

Some representations are:
the trivial representation ${\bf 1}=triv$;
the sign character
\beqa
\sign:\;\; \Sigma_3\ra \Sigma_3/{\bf Z_3}={\bf Z_2}=\{ \pm 1\}
\eeqa
and the fundamental
representation ${\bf 3}=fund$ induced by permutation of $\{ 1,2,3\}$.
In general, for a representation $R$, one has the following
projection operators:
first the 'invariant projector'
$P^+(v)=\sum_{\ga \in \Sigma_3}\ga v$
($v$ in the representation space of $R$)
which gives an $\Sigma_3$ invariant element; and analogously
the 'anti-invariant projection'
\beqa
\label{proj anti inv}
P^-(v)=\sum_{\ga \in \Sigma_3}\sign(\ga)\ga v
\eeqa
which transforms with the sign character
(both may be normalized by $1/6$).

Note that the representation ${\bf 3}$ is not irreducible.
Think of it in real three-space to see the invariant (Euler) axis
$\sum_i e_i$ and the $2 \pi /3$ rotation in the orthogonal
('barycentric') plane.
So it decomposes into a sum of the trivial representation and
a two-dimensional irreducible representation,
called ${\bf 2}$ (we will also denote ${\bf -1}=\sign$ and
${\bf -k}={\bf -1} \otimes {\bf k}$)
\beqa
\label{3is2+1}
{\bf 3}={\bf 2} \oplus {\bf 1}
\eeqa
Let us denote the degree and character of a representation ${\bf d}$
by $\deg_{{\bf d}}$ and $\chi_{{\bf d}}$, respectively.
The representations ${\bf 1}, {\bf -1}, {\bf 2}$
exhaust the irreducible representations as $3=\sharp$ CC or
\beqa
|\Sigma_3|=\deg^2_{{\bf 1}}+\deg^2_{{\bf -1}}+\deg^2_{{\bf 2}}
\eeqa
At this point it might be appropriate to give the character table
\begin{center}
\begin{tabular}{c|ccc}
  $$ & $\chi(c_1)$ & $\chi(c_2)$ & $\chi(c_3)$ \\
\hline
  ${\bf 1}$  & 1 & 1  &  1\\
  ${\bf -1}$ & 1 & -1 &  1 \\
  ${\bf 2}$  & 2 & 0  & -1 \\
\end{tabular}
\end{center}
Let us furthermore point to the following facts which we will use later
\beqa
\label{2x2is4}
{\bf 2} \ox {\bf 2}&\cong &{\bf 2} \o+ {\bf 1} \o+ {\bf -1}\\
\label{skew}
\Lambda^2 {\bf 3}&\cong &{\bf 2} \o+ {\bf -1}\\
\label{sym2}
\Sym^2 {\bf 3}&\cong &{\bf 2} \o+ {\bf 2} \o+ {\bf 1} \o+ {\bf 1}
\eeqa
For (\ref{2x2is4}) note\footnote{or: ${\bf 2} \otimes {\bf 2}$ is
represented by the span of
$e_i\ox e_j$ with $i=1,2; j=1,2$ ; clearly
the diagonal provides a ${\bf 2}$; the ${\bf +1}$ and ${\bf -1}$
are spaned by $e_1\ox e_1+\frac{e_1\ox e_2+e_2\ox e_1}{2}+e_2\ox e_2$
and $e_1\ox e_2 - e_2\ox e_1$, respectively.}
that the multiplicities $m_{{\bf d}}=1$
of each of our three building blocks ${\bf 1}, {\bf -1}, {\bf 2}$
occuring as isotypic components
${\bf d}$ in ${\bf 2} \ox {\bf 2}$ follow
from the character relations
\beqa
m_{{\bf d}}=\frac{1}{|\Sigma_3|}
\sum_{\ga \in \Sigma_3}\chi_{{\bf 2} \ox {\bf 2}}(\ga) \;
\overline{\chi_{{\bf d}}}(\ga)
=\frac{1}{|\Sigma_3|}
\sum_{c\in CC}n_c \chi_{{\bf 2} \ox {\bf 2}}(c)\;
\overline{\chi_{{\bf d}}}(c)
\eeqa
(\ref{skew}) follows by inspection\footnote{for
$\Lambda^2 {\bf 3}=\o+_{i\in {\bf Z_3}} e_i\wedge e_{i+1} {\bf C}$
the split (\ref{3is2+1}) leads now to the
{\em anti}-invariant line $(\sum_{i\in {\bf Z_3}} e_i\wedge e_{i+1}) {\bf C}$}
and (\ref{sym2}) from\footnote{or: among the
$e_i \cdot e_j$ $(i\leq j)$ of $\Sym^2 {\bf 3}$
the diagonal and the $i<j$ part
span each a ${\bf 3}$}
$\Sym^2 {\bf 3}\cong ({\bf 3}\ox {\bf 3})/\Lambda^2 {\bf 3}$.

Note also that if a system $(z_i)_{i\in {\bf Z_3}}$
spans a ${\bf -3}$, i.e. $\al z_i=-z_{\al i}$, then the system
of $w_i:=z_{i+1}-z_{i-1}=\be^2 z_i - \be z_i$ spans a ${\bf 2}$
(by $\sum _i w_i=0$ and $\al w_i=w_{\al i}$
from $\al \be^2=\be\al$)
\beqa
\label{minus transform}
\oplus_i z_i {\bf C}\cong {\bf -3} \;\; \Longrightarrow
\sum_i w_i {\bf C}\cong {\bf 2}
\eeqa
The $f_i$ and $\al_i$ transform essentially (shifted by $\al\ra \al \be^2$)
under ${\bf 3}$ and $-\, {\bf 3}$; then (\ref{minus transform}) leads
to the introduction of the $(\log )\, y_i$ (similar the relation of the
$(\log )\, \eta_i$ to the $(\log ) \, y_i$).
\\ \\
\noindent
{\em Some representation theory for $\Si_3$ acting on ${\bf P^1}$}

For a $\Sigma_3$ action on ${\bf P^1_C}$ consider the induced
operation on
functions\footnote{functions are considered for now just formally,
regardless of poles or the question of single-valuedness}
$\Lambda^0 {\bf P^1}$
on ${\bf P^1}$. Consider now a ${\bf Z_3}$-orbit of an
${\bf Z_2}$-anti-invariant function $f$, i.e. of a function
with $f(\al z)=-f(z)$, like the logarithm (here $\al z =1/z$ like for
the $Sl_2$ action).
One has
$\oplus_{i\in {\bf Z_3}} f(\be_i \cdot){\bf C}
\subset \Lambda^0 {\bf P^1}({\bf -3})$, so
\beqa
\oplus_i \log \be^i z \, {\bf C}\cong {\bf -3}
\eeqa
\newpage

Now
(\ref{betaprod}) implies for some $z:=\eta_i$
\beqa
\label{const combi}
\label{P+ log}
P^+_{{\bf Z_3}}\log z = \sum_{i\in {\bf Z_3}} \; \log \be^i z &=& \pm \pi i\\
\label{P- log}
P^- \log z =\sum_{\ga \in \Sigma_3} \sign(\ga)\, \log \ga z &=& \pm 2 \pi i \\
\label{minussigns}
\sum_{i\in {\bf Z_3}} \; \log \; (-1)^{\delta_{ij}} \; \be^i \; z &=& 0
\;\;\;\;\;\;\;\;\;\;\;\;\;\ (j\in {\bf Z_3})\\
\label{regroup}
\sum_{i\in {\bf Z_3}} \; d \log  \; \be^i \; z &=& 0 \\
\label{real parts}
\sum_{i\in {\bf Z_3}} \; \Re \log  \; \be^i \; z &=& 0
\eeqa
One has the exact sequence (by (\ref{P+ log}) the
left term are the constants ($=$ ker d) in the middle term)
\beqa
\label{factor sequence}
\begin{array}{ccccccccc}
0 & \lra & (\sum_i \log \be^i z ) {\bf C} & \lra &
\oplus_i \log \be^i z \, {\bf C} & \buildrel d \over \lra &
\sum_i \, (d \log \be^i z \, {\bf C}) & \lra & 0 \\
 & & \| & & \| & & \| & & \\
0 & \lra & {\bf -1} & \lra & {\bf -3} & \lra & {\bf -2} & \lra & 0
\end{array}
\eeqa
Similarly one has an interpretation of the {\em real} vector spaces with
$\Sigma_3$ action (note (\ref{real parts}))
\beqa
\label{Imlog}
\oplus _i \, \Im \log \be^i z \, {\bf R} &\cong & {\bf -3} \\
\label{Relog}
\sum _i \, (\Re \log \be^i z \, {\bf R}) &\cong & {\bf -2}
\eeqa

\subsection{\label{explicit check}Anti-invariance of ${\L}$: first argument}
We show how
$d Li = \log \be z \, d \log z $ and
$\Im \log \be z \, \Re \log z$ behave under $e- \sign(\ga) \ga$.
The complete parallelism
shows (cf. footn. \ref{real int const}) that
$(\Im \int d Li ) -  \Im \log \be z \, \Re \log z$
vanishes for all $e- \sign(\ga) \ga$ transformations,
i.e. the anti-invariant transformation behaviour.\\
\underline{$\be \cong \scriptsize{\left(\begin{array}{ccc}
\log z & \log \be z & \log \be^2 z\\
\log \be z &  \log \be^2 z  & \log z \end{array}\right)}$:}
$\;\;\;\;\;\;\;\;$ (the last equalities from (\ref{P+ log}))
\beqa
(e-\be ) \, dLi &=&\log \be z \; d \log z
- \log \be^2 z \; d \log \be z\nonumber\\
&=&d \; (\log \be z \cdot \log z) - \log z \; d \log \be z
- \log \be^2 z \; d \log \be z \nonumber\\
&=&d \; (\log \be z \cdot \log z) + \frac{1}{2} \, d \log^2 \be z
\mp \pi i \, d \log \be z\nonumber\\
(e-\be ) \,  \Im \log \be z \, \Re \log z &=&
\Im \log \be z \, \Re \log z
- \Im \log \be^2 z \, \Re \log \be z\nonumber\\
&=& \Im (\log \be z \cdot \log z)- \Im \log z \, \Re \log \be z
  -  \Im \log \be^2 z \, \Re \log \be z\nonumber\\
&=& \Im (\log \be z \cdot \log z)
+ \frac{1}{2} \Im \log^2 \be z
\mp \pi \Im \log \be z\nonumber
\eeqa
\underline{$\be^2 \cong \scriptsize{\left(\begin{array}{ccc}
\log z & \log \be z & \log \be^2 z\\
\log \be^2 z &  \log  z  & \log \be z \end{array}\right)}$:}
$\;\;\;\;\;\;\;\;$ (the last equalities from (\ref{regroup}))
\beqa
(e-\be^2 ) \, dLi
&=& \log \be z \; d \log z - \log  z \; d \log \be^2 z\nonumber\\
&=& d \, ( \log \be z \cdot \log z) - \log  z \; d \log \be z
  -  \log  z \; d \log \be^2 z\nonumber\\
&=& d \, ( \log \be z \cdot \log z) +  \frac{1}{2} \, d \log^2  z\nonumber\\
(e-\be^2 ) \,  \Im \log \be z \, \Re \log z
&=& \Im \log \be z \, \Re \log z - \Im \log  z \, \Re \log \be^2 z\nonumber\\
&=& \Im ( \log \be z \cdot \log z ) - \Im \log  z \, \Re \log \be z
- \Im \log  z \, \Re \log \be^2 z\nonumber\\
&=& \Im ( \log \be z \cdot \log z ) + \frac{1}{2} \, \Im \log^2  z\nonumber
\eeqa
\underline{$\al \cong \scriptsize{\left(\begin{array}{ccc}
\log z & \log \be z & \log \be^2 z\\
-\log z &  -\log  \be^2 z  & -\log \be z \end{array}\right)}$:}
$\;\;\;\;\;\;\;\;$ (the last equalities from (\ref{P+ log}))
\beqa
(e+\al ) \, dLi
&=& \log \be z \; d \log z + \log \be^2 z \, d \log z\nonumber\\
&=& -\frac{1}{2} \, d \log^2  z
\pm \pi i d  \log z \nonumber\\
(e+\al ) \,  \Im \log \be z \, \Re \log z
&=& \Im \log \be z \, \Re \log z +  \Im \log \be^2 z \, \Re \log z \nonumber\\
&=& -\frac{1}{2} \, \Im \log^2  z
\pm \pi  \Re \, \log z \nonumber
\eeqa
\underline{$\al \be \cong \scriptsize{\left(\begin{array}{ccc}
\log z & \log \be z & \log \be^2 z\\
-\log \be z &  -\log  z  & -\log \be^2 z \end{array}\right)}$:}
\beqa
(e+\al\be ) \, dLi
&=& \log \be z \; d \log z + \log z \; d \log \be z\nonumber\\
&=&d \, ( \log \be z \cdot \log z )\nonumber\\
(e+\al \be ) \,  \Im \log \be z \, \Re \log z
&=& \Im \log \be z \, \Re \log z + \Im \log  z \, \Re \log \be z\nonumber\\
&=& \Im ( ( \log \be z \cdot \log z )\nonumber
\eeqa
\underline{$\al \be^2 \cong \scriptsize{\left(\begin{array}{ccc}
\log z & \log \be z & \log \be^2 z\\
-\log \be^2 z &  -\log  \be z  & -\log z \end{array}\right)}$:}
$\;\;\;\;\;\;\;\;$ (the last equalities from (\ref{regroup})
and (\ref{real parts}))
\beqa
(e+\al\be^2 ) \, dLi &=& \log \be z \; d \log z
+ \log \be z \; d \log \be^2 z\nonumber\\
&=& -\frac{1}{2} d \, ( \log^2 \be z )\nonumber\\
(e+\al \be^2 ) \,  \Im \log \be z \, \Re \log z
&=& \Im \log \be z \, \Re \log z
+ \Im \log \be z \, \Re \log \be^2 z\nonumber\\
&=&-\frac{1}{2} \Im \, ( \log^2 \be z )\nonumber
\eeqa

\newpage

\subsection{\label{anti-invariant} Anti-invariance of ${\cal L}$:
second argument}

To investigate the potential anti-invariant transformation behaviour
of ${\cal L}(z)$ let us take up now our representation-theoretic
considerations from section \ref{nonlin action}.

The proper reason for the anti-invariance of
${\cal L}=\Im Li(z) - \Im \log \be z \, \Re \log z$
is the following fact: when one operates on bilinear product
expressions like $\log \be^i z \, \log \be^j z$
with either $d$ or $\Im$ one finds as image elements in their
respective target spaces (of expressions $\log \be^i z \, d \log \be^j z$
and $\Im \log \be^i z \, \Re \log \be^j z$) just the symmetric combinations
by reason of the (pseudo-)derivative nature of these operations
\beqa
\label{pseudo derivative}
d(f\cdot g) &=& df \cdot g + f\cdot dg\nonumber\\
\Im (fg)    &=& \Im f \, \Re g + \Re f \, \Im g
\eeqa
This is for $Li = \int \log \be z \, d\log z $ an indication that the
integral can not be done elementary (the integrand is not symmetric,
thereby not naturally a derivative of the presumptive candidate
functions, cf. footn. \ref{anti-sym footn}).
Now both terms, whose imaginary parts add up to ${\cal L}(z)$, i.e.
$ \int \log \be z \, d\log z$ and $\psi=\log \be z \, \Re \log z$
(or equally well $i \, \Im \log \be z  \log z $),
constitute the one missing piece which, when linearly combined with the
elementary expressions $\log \be^i z \, \log \be^j z$,
gives after application of $d$ and $\Im$ respectively not just
the symmetric elements ($\im \, d$ and $\im \, \Im$
in (\ref{Li exact diagram}) and (\ref{companion diagram}), respectively)
of their natural target space but all elements;
furthermore both of these missing 'non-symmetric' elements are built from
the same underlying element\footnote{of the tensor product which one has to
take, instead of the symmetric product used above in the elementary
expressions, to be able to apply $d$ or $\Im$ to individual factors}
$\log \be z \otimes \log z $
\beqa
\label{psi}
d \, Li &=& \log \be z \; d\log z \nonumber\\
\Im \, \psi&=& \Im \log \be z \; \Re \log z
\eeqa
We will see in a moment that the non-vanishing elements in the
respective one-dimensional quotient space (target space modulo image)
they generate transform with the sign character
(\ref{dLi rep}), (\ref{psi rep}).
The common origin of the non-trivial terms and the complete
parallelism of the (pseudo-)derivative operations mentioned above
shows then that the classes, when lifted back to the proper elements,
acquire exactly {\em the same} correction terms which gives finally the
anti-invariance of their
difference.\footnote{\label{real int const} when going
back and forth in $\Im \circ d^{-1}$ the interrelations are kept, i.e.
the integration constants are real
(actually rational multiples of $(\pi i )^2$)
(note that $ker \, d ={\bf C}, ker \, Im = {\bf R}$
on holomorphic functions)}

Now consider the following exact diagram
\beqa
\label{Li exact diagram}
\begin{array}{ccccccccc}
 & & 0          & & & & & & \\
 & & \downarrow & & & & & & \\
 & & {\bf C}    & & & & & & \\
 & & \downarrow & & & & & & \\
0& \lra & \Sym^2 fund & \lra & \Sym^2 fund \oplus Li \, {\bf C} & \lra &
Li \, {\bf C} & \lra & 0 \\
 & & \downarrow  d & &\downarrow  d & & \downarrow  d & & \\
0& \lra & \im \, d  & \lra &
\oplus_{i,j} \log \be^i z \, d \log \be^j z \, {\bf C} & \lra &
[ d \, Li ] {\bf C} & \lra & 0 \\
 & & \downarrow & & & & & & \\
 & & 0          & & & & & &
\end{array}
\eeqa
>From consideration of the lower horizontal exact sequence one finds
\beqa
\label{dLi rep}
[ d \, Li ] {\bf C}\cong {\bf -1}
\eeqa
For, by (\ref{factor sequence}), the space of elementary bilinear expressions
$\log \be^i z \, \log \be^j z$ gives a
$\Sym^2 ({\bf -3})=\Sym^2 \, {\bf 3}$;
concerning $\im \, d$ note that
the kernel of constants in the vertical short exact sequence
is, by (\ref{const combi}),
$(\sum_i \log \be^i z )^2{\bf C}\cong \Sym^2 ({\bf -1})={\bf 1}$;
so by (\ref{sym2})
\beqa
\im \, d \cong {\bf 2} \oplus {\bf 1}\oplus {\bf 2}
\eeqa
The middle term in the lower sequence,
is given by\footnote{the tensor (instead of the symmetric) product
applies as the symmetry between the factors is broken}
${\bf -3} \otimes {\bf -2}={\bf 3} \otimes {\bf 2}
={\bf 2} \otimes {\bf 2} \oplus {\bf 2}$.
Thereby $[ d \, Li ] {\bf C}\cong
({\bf -3} \otimes {\bf -2})/\im \, d =
({\bf 2} \otimes {\bf 2} \oplus {\bf 2})/
({\bf 2} \oplus {\bf 1}\oplus {\bf 2})$;
(\ref{2x2is4}) now
implies (\ref{dLi rep}).

Now consider the following companion exact diagram
\beqa
\label{companion diagram}
\begin{array}{ccccccccc}
 & & 0          & & & & & & \\
 & & \downarrow & & & & & & \\
 & & {\bf C}    & & & & & & \\
 & & \downarrow & & & & & & \\
0& \lra & \Sym^2 fund & \lra & \Sym^2 fund \oplus \psi \, {\bf C} & \lra &
\psi \, {\bf C} & \lra & 0 \\
 & & \downarrow  \Im & &\downarrow  \Im & & \downarrow  \Im & & \\
0& \lra & \im \, \Im  & \lra &
\oplus_{i,j} \Im \log \be^i z \, \Re \log \be^j z \, {\bf R} & \lra &
[ \Im \psi ] \, {\bf R} & \lra & 0 \\
 & & \downarrow & & & & & & \\
 & & 0          & & & & & &
\end{array}
\eeqa
By completely parallel arguments, where now
the second embodiment of ${\bf -2}$ in (\ref{Relog})
replaces the first one (\ref{factor sequence}) used before,
here too one finds (as representations over ${\bf R}$)
\beqa
\label{psi rep}
[ \Im \; \psi ] \, {\bf R} \cong {\bf -1}
\eeqa

\newpage
\subsection{\label{formal invar}Formal anti-invariance
of a modified superpotential}

To understand the anti-invariance property of $\tilde{W}$ in (\ref{Wtilde})
one would like to see the corresponding symmetry becoming manifest.
This can be done on the derivative level: $dW/dz$ becomes an elementary
logarithmic function just as the correction terms
$dC/dz$, with the only difference that
$C$ (in contrast to $W$) is already itself an elementary function.
To avoid an additional transformation factor
$d(\ga z) / dz$ obscuring the transformation
properties, we actually consider the one-form
$d\tilde{W}$ which again transforms with the sign character:
i.e. we are using the $\Sigma_3$ equivariant
map\footnote{meaning
$(df)(\ga \cdot)=(\frac{df}{dz}\; dz)(\ga \cdot)=\frac{df}{dz}(\ga \cdot)\;
d(\ga \cdot)=d (f(\ga \cdot))$}
$d\, : \, \La^0 \, {\bf C_z} \, \lra \,  \La^1 \, {\bf C_z}$.

One finds (cf. (\ref{dWtilde evaluation})) for $d\tilde{W}$
the manifestly anti-invariant expression
\beqa
\label{dtildeW}
6d\tilde{W}
=\sum_{i\in {\bf Z_3}} \; \log \frac{\be^{i+1} z}{\be^{i-1} z}\;
d\log \be^i z \;\;\;
\Bigl( =\sum_{i\in {\bf Z_3}} \; d\log \frac{\be^{i-1} z}{\be^{i+1} z}\;
\log \be^i z \Bigr)
\eeqa
The first of the six terms is just the original term we started with
\beqa
\label{origterm}
dW=\log \be u \;  d\log u
\eeqa
>From (\ref{origterm}) one can read off directly that $W$ is not anti-invariant
(compare to (\ref{dtildeW}).
Note that our solution (\ref{dtildeW})
is actually indeed of the form
(\ref{tildeWeta})
(by integration), i.e.
\beqa
\label{antiinvarproj}
6d\tilde{W}(\cdot )
=\sum_{\ga \in \Sigma_3}\sign(\ga)(dW)(\ga \cdot)
=\sum_{\ga \in \Sigma_3}\sign(\ga)d(W(\ga \cdot))
\eeqa
Matching to (\ref{dtildeW}) is obvious
for $\ga \in {\bf Z_3}$ by (\ref{origterm}) which also gives
$(dW)(\al z)=\log  \be^2 z \; d\log  z$.
(A non-zero integration
constant would
violate $\tilde{W}(\ga \cdot)=\sign (\ga) \tilde{W}(\cdot)$.)

Now, to prove (\ref{dtildeW}), one finds from (\ref{C eval2})
\beqa
6dC&=&-\log z \; d\log z  \;\;\;\; -2 \log \be z \; d\log z\nonumber\\
   & &-2 \log z \; d\log \be z +3 \log \be z \; d\log \be z
      \; +2 \log \be^2 z \; d\log \be z\nonumber\\
   & &\;\;\;\;\;\;\;\;\;\;\;\;\;\;\;\;\;\;\;\;\;\;\;\;\;\;
      +2\log \be z \; d\log \be^2 z + \log \be^2 z \; d\log \be^2 z
\eeqa
Combining with (\ref{origterm})
one gets\footnote{\label{anti-sym footn}Note
the anti-symmetry of the coefficient matrix which
guarantees the non-triviality of the expression (of course, being
non-symmetric is enough; if one starts form an elementary expression
$F=\sum_{i,j}a_{i,j}\log \be^i z \log \be^j z$ one ends up with a symmetric
quantity $dF=\sum_{i,j}(a_{i,j}+a_{j,i})\log \be^i z d\log \be^j z$.
The missing symmetry did account already for the non-triviality of the
original term (\ref{origterm}); cf. (\ref{Li exact diagram})).}
(\ref{dtildeW})
after a regrouping in the 'verticals' via (\ref{regroup})
\beqa
\label{dWtilde evaluation}
6d\tilde{W}&=&
      \;\;\;\;\;\;\;\;\;\;\;\;\;\;\;\;\;\;\;\;\;\;\;\;\;\;
      + \log \be z \; d\log z \;\;
      -\log \be^2 z \;\; d\log z\nonumber\\
   & &- \log z \; d\log \be z
      \;\;\;\;\;\;\;\;\;\;\;\;\;\;\;\;\;\;\;\;\;\;\;\;\;\;
      \;\; + \log \be^2 z \; d\log \be z\nonumber\\
   & &+ \log  z \; d\log \be^2 z
      -\log \be z \; d\log \be^2 z
\eeqa

\newpage

\section{\label{monodromy appendix}The monodromy representation}
\resetcounter

The polylogarithms are (with $Li(x):=Li_2(x)\; , \;
Li_1(x)=\log \be x\; , \;
Li_0(x)=x \cdot \be x$)
\beqa
\label{Li relations}
Li_k(x)=\sum_{k\geq 1} \frac{x^n}{n^k}\;\;\;\;\;\;\;\; , \;\;\;\;\;\;\;\;
\frac{d}{dx}Li_{k+1}(e^x)=Li_k(e^x)
\eeqa
To express the multi-valuedness
of $W=Li_2$  define the matrix differential form
[\ref{Hain}]
\beqa
\Omega=  {\small \left( \begin{array}{ccc}
                  0 & d\log \be z &   0     \\
                  0 &      0      & d\log z \\
                  0 &      0      &   0
                  \end{array} \right)    }
\eeqa
The one-forms $\om_i= d\, \log \be^i z$ are
related with the loops $l_i$ by
$\frac{1}{2\pi i } \int _{l_j}\om_k=\delta_{jk}$.
Now consider for a (multi-valued) function
$F: {\bf P^1}\backslash \{ 0,1,\infty\}\ra gl(3, {\bf C})$
the matrix differential equation
\beqa
d F = F \cdot \Omega
\eeqa
A fundamental solution is provided by the principal branch
(on $|z-1/2|< 1/2$) of
\beqa
\label{fundam solut}
L(z)=  \left( \begin{array}{ccc}
                  1 &  \log \be z &   Li(z)    \\
                  0 &      1      &   \log z   \\
                  0 &      0      &   1
                  \end{array} \right)
\eeqa
Analytic continuation of the principal branch of $L(z)$ about a loop $l$
in ${\bf P^1}\backslash \{ 0,1,\infty \}$ (based at $1/2$, say) leads to
another fundamental solution $M(l)L(z)$ where
\beqa
M: \pi_1 ({\bf P^1} \backslash  \{ 0,1,\infty \}) \ra Gl(3,{\bf C})
\eeqa
defines the monodromy representation.
One finds for the images of the generator loops $l_i(t)$ ($i=0,1$)
the representing matrices\footnote{\label{pi footn}Multiplying the
rows $r^{(j)}$ by $(2\pi i ) ^{j-1}$ the factor $2\pi i $
can be put in (\ref{fundam solut}). The $r^{(j)}$ are
multi-valued but the ${\bf Q}$-linear span of the
$(2\pi i ) ^{j-1} r^{(j)}$ is well-defined (the monodromy
representation is then rational.)}
$M(l_i)
$
in (\ref{monodromy matrices}).

In a column {\em vector picture} the three columns $c_k$, $k=1,2,3$, of $L$
fulfill $d \; c_k = c_k \cdot \Omega$ and
one gets from (\ref{monodromy matrices})
the monodromies (\ref{multi valued}) for $c_3$.
Similarly in the row picture
the rows $r^{(j)}$ ($j=1,2,3$) in (\ref{fundam solut})
are {\it flat} sections$^{\ref{pi footn}}$ of
a meromorphic connection $\nabla$ (on the trivial ${\bf C^3}$
bundle over ${\bf P^1}$) given for a section
$s=(s_1,s_2,s_3): {\bf P^1}\backslash \{ 0,1,\infty \}\ra {\bf C^3}$ by
\beqa
\label{connect vect pict}
\nabla s &=& d s - s \Omega
         = (d s_1 \;\; , \;\; ds_2 - s_1 \, d\log \be z \;\; , \;\;
              ds_3 - s_2 \, d\log z)
\eeqa
The {\em Heisenberg picture}
involves the complexified Heisenberg group.
Consider first the situation over the reals with
the following central extension of the group $({\bf R^2}, +)$
by $({\bf S^1}, \cdot )$
\beqa
\label{real Heisenberg sequence}
1\lra {\bf S^1} \lra \H \lra {\bf R^2} \lra 0
\eeqa
So the normal subgroup ${\bf S^1}$ of $\H$ constitutes the centre
and one has the group law
\beqa
(X,\la)\cdot (Y,\mu)=(X+Y, e(X,Y)\la\mu)
\eeqa
with a skew-multiplicative\footnote{So one has
$e(X+X',Y)=e(X,Y)e(X',Y)$, similarly in $Y$ and
$e(Y,X)=e(X,Y)^{-1}, e(X,X)=1$.}
pairing\footnote{$X\ra e(X,\cdot)$ will then provide an isomorphism of
${\bf R^2}$ with its character group.}
$e:{\bf R^2} \x {\bf R^2} \ra {\bf S^1}$
given by $e(X,Y)=e^{2\pi i A(X,Y)}$ for
$A:{\bf R^2} \x {\bf R^2} \ra {\bf R}$
a non-degenerate, bilinear, skew-symmetric pairing.
With the parameterisation
$\la=e^{2\pi i c}, \mu =e^{2\pi i d}$ one finds as multiplication law
on ${\bf R}\x {\bf R^2}$
\beqa
\label{real Heisenberg group law}
(x_1,x_2 \, | \, c) \cdot (y_1,y_2 \, | \, d) = (x_1+y_1,x_2+y_2 \, |
\, A(X,Y)+c+d )
\eeqa
Choosing for $A$ the pairing
$A(X,Y)=x_1 y_2 - x_2 y_1$
(for $X=(x_1, x_2), Y=(y_1, y_2)$)
one sees that the group law (\ref{real Heisenberg group law})
on triples $(x_1,x_2 \, | \, c) \in {\bf R^2}\x {\bf R}:={\cal H}'$
is induced from matrix multiplication under the following association
of ${\cal H}'$ with the upper triangular matrices
\beqa
\label{group law}
(a,b \, | \, c)\cong
\left( \begin{array}{ccc}
                  1 &      a      &   \frac{c+ab}{2}   \\
                  0 &      1      &   b   \\
                  0 &      0      &   1
                  \end{array} \right)
&\Lra & (a,b\, | \, c) \cdot (u,v \, | \, w)
= (a+u, b+v \, | \, av-bu+c+w)\nonumber
\eeqa
Note that one has a slightly different induced group law by the
following association
\beqa
\label{Heisenberg group law}
(a,b \, | \, c)\cong
\left( \begin{array}{ccc}
                  1 &      a      &   c   \\
                  0 &      1      &   b   \\
                  0 &      0      &   1
                  \end{array} \right)
&\Lra& (a,b \, | \, c) \cdot (u,v \, | \, w) = (a+u, b+v \, | \, av+c+w)
\eeqa
Define the {\em complexified} Heisenberg group
${\bf {\cal H}_C}$ (with the ${\bf S^1}$ from
$e^{2\pi i (\cdot )}$ replaced by ${\bf C^*}$)\\
where  ${\bf {\cal H}_C}$ is ${\bf C^3}$ with this
composition
(and so with inverse $(a,b \, | \, c)^{-1}=(-a,-b \, | \, ab-c)$)\\
which makes ${\bf {\cal H}_C}$ a non-commutative group
with normal subgroups
$(*,0 \, | \, *')$ and $(0, * \, | \, *')$,
both isomorphic to $({\bf C^2}, +)$, whose
intersection $(0,0 \, | \, *)$ is the centre of ${\bf {\cal H}_C}$.\\
{\em The adjusted imaginary part ${\cal L}$ of $W$}

Consider the Heisenberg bundle (\ref{proper heisenberg bundle})
with section $s$ (where $e({\bf {\cal H}_Z}(a,b \, | \, c))=(e^{a}, e^{b})$)
\beqa
\label{fibre diagram}
\begin{array}{ccc}
 & & (2\pi i )^2{\bf Z}\backslash {\bf C_c} \\
 & & \downarrow \\
\underline{{\cal H}} & \lra & {\bf {\cal H}_Z}\backslash {\bf {\cal H}_C}  \\
\;\;\; s \uparrow \, \downarrow \, pr & & \downarrow \, e \\
{\bf P^1}\backslash \{ 0,1,\infty \} & \buildrel (1-z,z) \over \lra
& {\bf C^*} \x {\bf C^*}
\end{array}\\
\label{general section}
s(z)={\bf {\cal H}_Z}(- \log \be z , \log z \, | \, c)
\eeqa
The ${\bf {\cal H}_Z}$ coset expresses the
fact that $z\ra c$ is not given as a function (the monodromy increments
of $Li$). This comes as the right vertical
sequence in (\ref{fibre diagram}) does not split, i.e. there is no
map $\al$ with
$(0,0 \, | \, c)\ra {\bf {\cal H}_Z}(0,0 \, | \, c)\buildrel \al \over
\ra (0,0 \, | \, c)$.
For the imaginary part of the fibre
$(2\pi i )^2{\bf Z}\backslash {\bf C_c}=
\Bigl( (2\pi)^2{\bf Z}\backslash {\bf R}_{\Re c}\Bigr)
\oplus i {\bf R}_{\Im c}$ there is such a map.
The function $f$ on ${\bf {\cal H}_C}$
\beqa
\label{Heisenberg function projection}
f(u,v,w)=\Im \; w - \Re \; u \; \Im \; v
\eeqa
is invariant under action of ${\bf {\cal H}_R}$ from the left,
so a fortiori under ${\bf {\cal H}_Z}$ which according to the remark
after (\ref{flat section}) represents the monodromy increments.
So [\ref{Hain MacPherson}] the combination
\beqa
\label{adjustment combination}
{\cal L} : {\bf P^1}\backslash \{ 0,1, \infty \} \ni z
\buildrel \Lambda \over \ra
{\bf {\cal H}_Z}(-\frac{\log \be z}{2\pi i} , \frac{\log z}{2\pi i} ,
- \frac{Li(z)}{(2\pi i )^2}) \in
{\bf {\cal H}_Z}\backslash {\bf {\cal H}_C}
\buildrel -(2\pi i )^2 f \over \lra  {\bf R}
\eeqa
(making factors $2\pi i $ manifest)
gives a $\pi_1$-invariant, i.e. single-valued function
(\ref{adjust imag part}).\\
\noindent
{\em Some expressions related to $Li$ and
 \label{Rogers CFT section}the Rogers CFT relation}

The ${\bf Z_2}$ anti-projector $P^-_{{\bf Z_2}}f=
\frac{1}{2}\sum_{i \in {\bf Z_2}}\sign (\ga^i)\ga^i f=(f-\ga f )/2$
does not reproduce $Li$
(not anti-invariant; $\ga \in \Si_3\backslash {\bf Z_3}$),
so we introduce the Rogers function $R$ (for $\ga = \al \be$)
\beqa
\label{Rogers}
R(z)=\frac{1}{2}\Bigl( Li(z)-Li(1-z) \Bigr) +\frac{\pi^2}{12}
=Li(z)-\frac{1}{2}\log \be z  \log z
\eeqa
\beqa
\label{Rogers relations}
Li(z)&=&\int_0^z \log \be w \, d \log w
\;\;\;\;\; \;\;\;\; \;\;\;\; \;\;\;\; \;\;\;\; , \;\;\;\;
d\, Li (z)= \log \be z \, d \log z
\nonumber\\
{\cal L}(z)&=&\Im Li(z) - \Im \log \be z \Re \log z
\;\;\;\; , \;\;\;\;
d \, i{\cal L}(z)=\frac{1}{2}
\Bigl( \Re \log \be z \, d \log z -\Re \log z \, d \log \be z \Bigr)
\nonumber\\
R(z)&=&Li(z)-\frac{1}{2}\log \be z  \log z
\;\;\;\;\;\; \;\;\;\; \;\;\;\; , \;\;\;\;
d\, R(z)=\frac{1}{2}
\Bigl( \log \be z \, d \log z - \log z \, d \log \be z \Bigr)
\eeqa
For background on (\ref{Rogers CFT relation})
recall that Calabi-Yau hypersurfaces in weighted projective space
have a Gepner point in their moduli space with the underlying exactly
solvable RCFT a tensor product of $N=2$ superconformal minimal models
of central charge\footnote{For
$\sum_{i=0}^4 z_i^{a_i}=0$ in ${\bf P^4}_{(w_i)}(d)$
with $a_i=\frac{d}{w_i}$ the CFT is a suitably
interpreted tensor product of five $SU(2)$ theories of
level $a_i-2$ and chiral primary operators with integral anomalous
dimensions come from operators in the
$SU(2)_{k=a_i-2}$ factors with anomalous
dimensions $\De_j^{(k)}=\frac{j(j+2)}{4(k+2)}$ ($j=0, \dots , k$).}
$c=\frac{3k}{k+2}$ (the central charge of an integrable level $k$
representation of the affine Kac-Moody Lie algebra of $Sl(2)$).
Recall the character $\chi_n(\th)=\sin(n\pi \th) / \sin (\pi \th)$
of the $n$-dimensional representation of $SU(2)$.
The characters
$\chi_i(\tau ,z)=tr_{\H_i}q^{L_0-\frac{c}{24}}q^{2\pi i u J_0}$
transform like
$\chi_i (\frac{-1}{ \tau}, \frac{z}{\tau})
=e^{\pi i k z^2/2}\sum_j S_{ij}\chi_j(\tau ,z)$
($Q_{ij}$ generalized quantum dimensions)
\beqa
S_{ij}=\sqrt{\frac{2}{k+2}}\sin (i+1)(j+1) \frac{\pi}{k+2}
\;\;\;\;\; , \;\;\;\;\;
Q_{ij}=\frac{S_{ij}}{S_{0j}}
=\frac{\sin (i+1)(j+1) \frac{\pi}{k+2}}{\sin (j+1) \frac{\pi}{k+2}}\nonumber
\eeqa
giving ($j$ fixed)
$\sum_{i=1}^k R(\frac{1}{Q_{ij}^2})=\frac{\pi^2}{6}
\Bigl( \frac{3k}{k+2} - 24 \De_j^{(k)} + 6 j \Bigr)$.
$j=0, N=k+2$ give (\ref{Rogers CFT relation})
[\ref{Richmond Szekeres}].\footnote{Concerning $Q_{i0}$ recall that
$Z({\bf S^2}\x {\bf S^1})=1, Z({\bf S^3})=S_{0,0}$ give for the vev
$<C>= \frac{\sin N \frac{\pi}{N+k}}{\sin  \frac{\pi}{N+k}}$
of the unknot as Wilson line in ${\bf S^3}$ (for $G=SU(N)$)
that $<C>= \frac{Z({\bf S^3}, R_2)}{Z({\bf S^3})}=\frac{S_{0,1}}{S_{0,0}}
$ for $G=SU(2)$.}

\newpage

\section{\label{tetrahedron} Volume of a hyperbolic ideal tetrahedron}
\resetcounter
{\em Hyperbolic three-space}

As model
for the hyperbolic space ${\bf H_3}$ we take the half-space
model constructed in analogy to the upper half-plane
${\bf H}=\{ {\bf x}=x_1+x_2i\, |\, x_1,x_2\in{\bf R}, x_2>0 \}$;
we consider ${\bf C}$ embedded at $x_3=0$
so that ${\bf H_3}$ is
$\{ (w:=x_1+ix_2, t:=x_3)\in {\bf C}\x {\bf R^{>0}} \}$.
Now, ${\bf H}$ is also a homogeneous space $PSl(2,{\bf R})/SO(2)$
from the operation of $Sl(2,{\bf R})$ on $i$ by fractional linear
transformations. Consider here the following subspace of the
quaternions
\beqa
\label{H3}
{\bf H_3}=\{ {\bf x}=x_1+x_2i+x_3j\, |\, x_1,x_2,x_3\in{\bf R}, x_3>0 \}
\eeqa
$g=\tiny{\left(\begin{array}{cc}a&b\\c&d\end{array}\right)}\in
Sl(2,{\bf C})$ operates on ${\bf H_3}$
by
\beqa
g\cdot {\bf x}=(a{\bf x}+b)(c{\bf x}+d)^{-1}
\eeqa
With the norm $||c(w+tj)+d||^2=|cw+d|^2+|c|^2t^2$
in the quaternions this is given by
\beqa
\label{operation}
\left(\begin{array}{cc}a&b\\c&d\end{array}\right)(w,t)=
\frac{1}{||c(w+tj)+d||^2}
\Bigl( (aw+b)(\bar{c}\bar{w}+\bar{d}) +|c|^2t^2,t\Bigr)
\eeqa
So for
$z\in {\bf C^{\x}}\hra Sl(2, {\bf C})$ (via $z=a^2, b=c=0$)
one has $z\cdot (w, t)=(zw, |z|t)$.

Now just as for ${\bf H}={\bf H_2}$
one has here that the map $q: Sl(2,{\bf C})\ra {\bf H_3}$ given by
$g \buildrel q \over \ra g\cdot j$ induces an equivariant
diffeomorphism
$q:Sl(2,{\bf C})/SU(2)\buildrel \cong \over \lra {\bf H_3}$
\beqa
\label{H3 exact sequence}
1\lra SU(2) \buildrel r \over \lra Sl(2,{\bf C})
\buildrel q \over \lra {\bf H_3} \lra 1
\eeqa
This may be equally well expressed by considering the quotient
$PSl(2, {\bf C})/SO(3)$.

Prolonging the analogy to the real case note that
the boundary of the upper half plane is identified with
${\bf P^1_R}={\bf R}\cup \{ \infty \}$ with ${\bf R}$ the locus $x_2=0$
whereas here the boundary is
${\bf P^1_C}={\bf C}\cup \{ \infty \}$ with ${\bf C}$ the locus $x_3=0$.
The group of (orientation preserving)
isometries of ${\bf H}$ is isomorphic to $PSl(2,{\bf R})$ and
for ${\bf H_3}$ to\footnote{the latter
acts on the boundary ${\bf P^1_C}$ fractionally linear, so
acts three-fold transitively and maps (uniquely)
two quadruples of points onto another exactly if they
have the same cross ratio}
$PSl(2,{\bf C})$.

The standard hyperbolic metric $ds^2$ and the volume form $vol$ are given by
\beqa
\label{metric vol}
ds^2=\frac{dx_1^2+dx_2^2+dx_3^2}{x_3^2}\;\;\;\;\; , \;\;\;\;\;
vol =\frac{dx_1dx_2dx_3}{x_3^3}
\eeqa
For example, using the mentioned embedding
$z\in {\bf C^{\x}}\hra Sl(2, {\bf C})$,
one finds
for the length
of the geodesic line-segment $\ga_z$
from $q(e)$ to $q(z)$
(with $e={\bf 1_2}\in Sl(2, {\bf C})$)
\beqa
\label{length}
\length(\ga_z)=distance(j,|z|j)=\int_1^{|z|} \frac{dx_3}{x_3}=\Re \log z
\eeqa
\\
\noindent
{\em Volume of a hyperbolic ideal tetrahedron}

Now an ideal tetrahedron is determined (up to congruence) by the
dihedral angles $\ga_1, \ga_2,\ga_3$ of the edges incident to any vertex.
Then choosen {\em any} vertex\footnote{as opposite dihedral angles are
equal the $\ga_i$ are independent of the vertex chosen}
one has
\beqa
\label{euclidean triangle}
\sum_i \ga_i=\pi
\eeqa
We will choose the vertex at $\infty$ so that the angles become
angles of an euclidean triangle in\footnote{the corresponding face
of $\De$ is a hemisphere {\em over} ${\bf C}$
through $u,v,w$ bounded by semi-circles over ${\bf C}$}
${\bf C}$ given by the remaining
three vertices $u,v,w$.\footnote{The {\em link} $L$
(parametrizing the rays in $\De$ through $v$)
of a vertex $v$ of an ideal tetrahedron $\De$ is an euclidean
triangle (well-defined up to orientation preserving similarity) given
by the intersection of the boundary of $\De$
with a horizontal euclidean plane (a "horosphere");
$L$ determines $\De$ up to congruence.}

Now concerning the parametrization of an euclidean triangle
$\De(u,v,w)\subset {\bf C}$ (the vertices labeled in the mathematical
positive sense) note that if one associates to each vertex the ratio
of the adjacent sides
\beqa
\label{vertex invariants}
z(u)=\frac{w-u}{v-u}\; , \; z(v)=\frac{u-v}{w-v}=\be z(u)\; , \;
z(w)=\frac{v-w}{u-w}=\be^2 z(u)
\eeqa
then these {\em vertex invariants} depend only on the orientation preserving
similarity class of $\De(u,v,w)$ which in turn is determined by $z(u)$
($\arg z(u)$ is the angle of $\De(u,v,w)$ at $u$; $\Im z(u) >0$).
So after the usual normalization in our tetrahedron set-up we are
considering the angles of the euclidean triangle with
vertices $0 , 1, z$ in ${\bf C}$, the angle $\al_0$ at $0$ is $\arg z$
and the angles $\al_1$ at $1$ and
$\al_z$ at $z$ are given by  (cf. footn. \ref{arctan})
\beqa
\label{euclid triangle}
\al_0=\arg z\;\;\; , \;\;\; \al_1=\arg \be z \;\;\; , \;\;\;
\al_z=\arg \be^2 z
\eeqa
In other words this gives a geometric manifestation of the
membrane anomaly (\ref{betaprod})
\beqa
\label{euclid angle sum}
\sum_{i\in {\bf Z_3}}\Im \log \be^i z = \pi
\eeqa
(cf. (\ref{membrane anomaly}),
(\ref{the membrane anomaly})).
As $z, \be z, \be^2 z$ give the
same tetrahedron one must pick an edge of $\De$ (the dihedral angle of
the faces adjacent at this edge is then $\arg z$) to specify $z$ uniquely.

Furthermore one has with (\ref{Lobachevsky}) that
(where $\ga_{1,2,3}=\al_{0,1,z}$)
\beqa
\label{Lobachevsky volume}
\vol \, \De(z)=\sum_i \Pi (\ga_i)
\eeqa
For convenience let us choose a slightly different 'circle gauge' of the
points $z_i$: the one actual complex degree of freedom (which is left after
the $Sl(2,{\bf C})$ operation) will not be encoded in the complex number
$z$ with the other three points fixed
(leaving two real degrees of freedom); rather we gauge to a
situation where one of the points again becomes $\infty$ and the other
three points $a,b,c$ lie on the unit circle $|z|=1$ (these are three
real degrees of freedom) with the one further condition that $\Re \, b =
\Re \, c$ (leaving two real degrees of freedom). In this situation,
where we assume that the face opposite to $\infty$ lies in the
hemisphere $x_1^2+x_2^2+x_3^2=1$ ($x_3\geq 0$) and has vertices
$a, b, c$ (of $x_3=0$),
project $\Delta$
orthogonally down to the unit disk $D_{x_1,x_2}$ where you get the
picture of an euclidean triangle (of vertices $a, b, c$)
whose angles sum up to $\pi$.
Subdividing this triangle by drawing the heights from the origin on the sides
one gets six smaller right triangles. Then one computes with (\ref{metric vol})
for the volume $\vol_1$ of the region lying over one of these triangles
(with angle $\ga$, $A^2:=1-x_1^2$, $\cos \th=x_1$ and (\ref{Irel2}))
([\ref{Milnor}])
\beqa
\vol_1&=&
\int_0^{\cos \ga}dx_1\int_0^{x_1\tan \ga}dx_2
\int_{\sqrt{1-x_1^2-x_2^2}}^{\infty}\frac{dx_3 }{x_3^3}
=\frac{1}{2}\int_0^{\cos \ga}dx_1\int_0^{x_1\tan \ga}
\frac{dx_2}{1-x_1^2-x_2^2}\nonumber\\
&=&\frac{1}{4}\int_0^{\cos \ga}dx_1
\frac{1}{A}\log \frac{A+x_1\tan \ga}{A-x_1\tan \ga}
= -\frac{1}{4}\int_{\pi /2}^{\ga} d\th \log \frac{2\sin (\th + \ga)}
{2\sin (\th - \ga)}\nonumber\\
&=&\frac{1}{4}\Bigl( \Pi(2\ga)+\Pi(\pi/2-\ga)-\Pi(\pi/2-\ga)-\Pi(0)\Bigr)
=\frac{1}{2}\Pi(\ga)
\eeqa
By summing over the six partial triangles one gets thereby
(\ref{Lobachevsky volume}). This gives the connection of
(\ref{adjust-proj Z3}) with (\ref{volume tetrahedron}) in view of the
remark following (\ref{euclidean triangle}).\\

\noindent
{\em cross ratios}

The $\Sigma_3$ transformation properties of a cross ratio
can be understood as follows. For
four points $z_1, z_2, z_3, z_4$ of ${\bf P^1(C)}$ one defines their
cross ratio
\beqa
\label{cross ratio}
cr\{z_1, z_2, z_3, z_4 \}=\frac{z_1-z_3}{z_1-z_4}/\frac{z_2-z_3}{z_2-z_4}
\eeqa
For example $cr\{0, 1, \infty, z\} = z$.
Clearly a $\Sigma_4$ is operating. One has the equalities
\beqa
cr\{z_1, z_2, z_3, z_4\}=cr\{z_2, z_1, z_4, z_3\}
=cr\{z_3, z_4, z_1, z_2\}=cr\{z_4, z_3, z_2, z_1\}
\eeqa
but the index four subgroup $\Sigma_3$ operates effectively
which gives the following realisation
\beqa
\label{cross ratio table}
\begin{array}{ccccc}
x=cr\{z_1, z_2, z_3, z_4\} & & \frac{1}{1-x}=cr\{z_1, z_3, z_4, z_2\} & &
\frac{x-1}{x}=cr\{z_1, z_4, z_2, z_3\} \\
\frac{1}{x}=cr\{z_1, z_2, z_4, z_3\} & & 1-x=cr\{z_1, z_3, z_2, z_4\} & &
\frac{x}{x-1}=cr\{z_1, z_4, z_3, z_2\}
\end{array}
\eeqa
of the isomorphism $\Sigma_3  \cong Sl(2,{\bf Z})/\Gamma(2)$ in
\beqa
\label{cross ratio isomorphism}
1 \lra V \rightarrow \Sigma_4 \lra
\Sigma_3  \cong  Sl(2,{\bf Z})/\Gamma(2) \lra 1
\eeqa

\newpage

\section{\label{cohomological interpretation}Cohomological interpretation}
\resetcounter
(\ref{CCS2}) gives
for the classifying space $BGl(n,{\bf C})^{\delta}$ of flat bundles
an universal class
\beqa
\hat{C}_2 \in H^3(BGl(n,{\bf C})^{\delta}, {\bf C/Z})\cong
H^3_{EM}(Gl(n,{\bf C})^{\mbox{\small{disc}}}, {\bf C/Z})
\eeqa
where we also indicated the isomorphism of the topological homology with the
Eilenberg-MacLane group cohomology\footnote{homology is of
chain complex of elements of $G^n$ with boundary
$\partial (g_1, \dots , g_n)=(g_2, \dots , g_n)+\sum_{i=1}^{n-1}
(-1)^i (g_1, \dots , g_i g_{i+1}, \dots , g_n)+(-1)^n (g_1, \dots ,
g_{n-1})$; so $H^0(G)={\bf Z}, H^1(G)=G^{ab}=G/G^{comm}$}
of the underlying discrete group of $Gl(n,{\bf C})$.

One defines a geodesic simplex for three elements $g_i$ of $G$
by $\Delta(z)$ with (cf. (\ref{cross ratio}))
\beqa
\label{sigma def}
\si \bigl( (g_1,g_2,g_3) \bigr)=z
=cr\{\infty, g_1\infty, g_1g_2 \infty, g_1g_2g_3\infty\}
\eeqa
With $\om$ a ${\bf C}$-valued $Sl(2,{\bf C})$ invariant three-form on
$Sl(2,{\bf C})/SU(2)={\bf H_3}$ one finds a ${\bf C/Z}$ valued
Eilenberg-MacLane cochain ${\cal I}(\om)$ with
$\hat{C}_2={\cal I}(\om)(g_1,g_2,g_3)=\int_{\Delta(z)}\om$.
This is evaluated [\ref{Dup}] as $2\hat{C}_2=c$
via the exterior square version
(\ref{exterior square diagram}) of the Heisenberg bundle
\beqa
\begin{array}{ccccc}
 & & {\bf Q}\backslash {\bf C} & \lra & {\bf Q}\backslash {\bf C} \\
 & & \al \, \uparrow \, \downarrow \;\;\; & &
\;\;\;\;\;\;\; \downarrow 1\we id \\
 & \buildrel c \over \nearrow & \widetilde{{\bf C} \we_{{\bf Z}} {\bf C}} &
\lra & {\bf C} \we_{{\bf Z}} {\bf C}  \\
 & & \rho \, \uparrow \, \downarrow \;\;\; & & \downarrow \, e \\
H^3\bigl( Sl(2,{\bf C})\bigr)/ ({\bf Q/Z}) & \buildrel \si \over \lra
& {\bf P^1}\backslash \{ 0,1,\infty \} & \buildrel z \we (1-z) \over \lra
& {\bf C^*} \we_{{\bf Z}} {\bf C^*}
\end{array}
\eeqa
(for the proper target of $\si$ cf. (\ref{c diagram}))
The arrows in the lower row compose
to zero.\footnote{\label{la footnote}The target space of $e$
has to be ('log'-)interpreted so that
$(ab)\we c=a\we c + b\we c, \frac{1}{b}\we c = -(b\we c), 0=\pm 1 \we c$
hold; in particular $\la(z)=0$
for $z\in \mu_{{\bf C}}$ (the complex roots of unity) as
$z\we (1-z)=(1/n) (z^n\we (1-z))$.}
This is the commutative diagram with exact rows\footnote{the
upper row is well-defined as the part ${\bf Q/Z}$ modded out
comes from $H^3\bigl( \mu_{{\bf C}}, {\bf Z}\bigr)$ embedded diagonally
and this goes to zero under $\si$; note also that
$\la(z)=0$ for $z\in \mu_{{\bf C}}$ by footn. \ref{la footnote}}
[\ref{Dup}], [\ref{Dup Sah}]
\beqa
\label{c diagram}
\begin{array}{ccccc}
H^3\bigl( Sl(2,{\bf C}), {\bf Z}\bigr)/ ({\bf Q/Z}) & \buildrel \si \over \hra
& {\bf P_C} & \buildrel \la \over \lra & \La^2_{{\bf Z}}({\bf C^*}) \\
\downarrow  c & &\downarrow  \rho & & ||  \\
{\bf C/Q}  & \buildrel 1\we id \over \hra &
\La^2_{{\bf Z}}({\bf C}) & \buildrel e \over \lra&\La^2_{{\bf Z}}({\bf C^*})
\end{array}
\eeqa
with\footnote{in [\ref{Dup}], [\ref{Dup Sah}] actually a group
${\cal P}_{\bf C}={\bf P_C}/{\sim}$ for a certain $5$-term equivalence
relation $\sim$ is considered}
${\bf P_C}=F({\bf P^1_C})/\Sigma_3^-$, i.e. free generators from ${\bf P^1_C}$
modulo the equivalence relation given by the non-linear $\Sigma_3$
action with order two elements operating together with a minus sign
($\la$ is then still welldefined)\footnote{Note that the mapping
$\la:{\bf C}\ni z \ra z \wedge (1-z) \in {\bf C^*}\we_{{\bf Z}}{\bf C^*}$
transforms anti-invariantly (as
$-\la (z)=z\we \be z $ and $z\cdot \be z \cdot \be^2 z=-1$), i.e.
$z \wedge (1-z) \in {\bf -1}$ (cf. sect. \ref{anti-invariant}).}.
Therefore $\rho \circ \si$ comes from an element in
${\bf Q}\backslash {\bf C}$, i.e. one defines
$c=\al \circ \rho \circ \si$
(a natural continuous option for the splitting $\al$ is
given only for the imaginary part); so $c$ is essentially given by
$\rho$, i.e. the Rogers 'function' (in the end the dilogarithm).
One finds then that $2\hat{C_2}=c$,
more precisely\footnote{\label{real footnote}It suffices
to evaluate $2\Re \, \hat{C}_2$
on the cohomology $H_3(Sl(2,{\bf R})^{\de})$ of the real subgroup
(actually the universal cover $\widetilde{PSl}(2,{\bf R})^{\de}$ is concerned)
where one finds that in $H^3(Sl(2,{\bf R}), {\bf R/Z})$ it is congruent
to $\frac{1}{4\pi^2}R(z)$ modulo
$\frac{1}{24}$ ($=\frac{1}{4\pi^2}\cdot \frac{\pi^2}{6}$);
$\Im \, \hat{C}_2$ is a {\em continuous} cochain
and so uniquely determined (up to a factor) as
$H^3_{cont}(Sl(2,{\bf C}), {\bf R})\cong {\bf R}$ [\ref{Dup}].}
(\ref{hatC2 evaluation}).
\\ \\
{\em Representation via exterior squares}

We are interested in the diagram analogous to (\ref{fibre diagram})
(again $e(z,w)=(e^{2 \pi i z}, e^{2 \pi i w})$; the tilde indicates
a pullback of the $e$ projection map along the indicated base map)
\beqa
\label{exterior square diagram}
\begin{array}{ccc}
 & & {\bf Q}\backslash {\bf C} \\
 & & \;\;\;\;\;\;\; \downarrow 1\we id \\
\widetilde{{\bf C} \we_{{\bf Z}} {\bf C}} &
\lra & {\bf C} \we_{{\bf Z}} {\bf C}  \\
\downarrow \, p & & \downarrow \, e \\
{\bf P^1}\backslash \{ 0,1,\infty \} & \buildrel z \we (1-z) \over \lra
& {\bf C^*} \we_{{\bf Z}} {\bf C^*}
\end{array}
\eeqa
Note that one has the expression which is {\em not} welldefined
as a function (cf. (\ref{Rogers}))
\beqa
\frac{1}{2\pi^2}Li(z)=\frac{1}{2\pi i }\log z  \frac{1}{2\pi i }\log (1-z)
        +  \frac{-2}{(2\pi i)^2 }R(z)
\eeqa
and can define a map
$\rho: {\bf C}\ni z \lra \rho(z)\in \La^2_{{\bf Z}}({\bf C})$
which {\em is} indeed welldefined [\ref{Bloch}]
\beqa
\label{rho def}
\rho(z)=\frac{1}{2\pi i }\log z \we \frac{1}{2\pi i }\log (1-z)
        + 1 \we \frac{-2}{(2\pi i)^2 }R(z)
\eeqa
This is a section of $p$ in (\ref{exterior square diagram})
just as (\ref{flat section}) was a section
of $pr$ in (\ref{fibre diagram}).
Note that if one wants to go back from a value in
$\La^2_{{\bf Z}}({\bf C})$ to a complex number (to have a function
instead of a section of a non-trivial projection),
i.e. if one wants to define a splitting
$\al:{\bf C}\we_{{\bf Z}}{\bf C}\ra {\bf Q}\backslash {\bf C}$ to
${\bf {\bf Q}\backslash C}  \buildrel 1\we id \over \hra
{\bf C}\we_{{\bf Z}}{\bf C}$
one has {\em natural} option just for the imaginary part
(this replaces (\ref{Heisenberg function projection});
cf. also the alternative identification before (\ref{Heisenberg group law}))
[\ref{Dup}], [\ref{Dup Sah}]
\beqa
\label{Im alpha}
\Im \, \al (z\we w)=\Re \, z \, \Im \, w - \Re \, w \, \Im \,z
\eeqa
The relative minus sign
escapes the symmetry in (\ref{pseudo derivative}); for by
(\ref{companion diagram}) $\psi$, and so ${\cal L}$,
is not an imaginary part of ordinary (rather than wedge) products.
One has\footnote{Such a relation without taking the imaginary part would
be inappropriate ($\L\neq\Im R$ as
$\al \, \rho \neq \frac{1}{2\pi^2}R$).}
(cf. (\ref{adjustment combination}))
\beqa
\label{imag restrict}
\Im \, \al \, \rho=\frac{1}{2\pi^2}{\cal L}
\eeqa

\section*{References}
\begin{enumerate}

\item
\label{AMV}
M. Atiyah, J. Maldacena and C. Vafa,
{\em "An M-theory Flop as a Large N Duality"},
J.Math.Phys. {\bf 42} (2001) 3209,
hep-th/0011256.

\item
\label{AW}
M. Atiyah and E. Witten,
{\em "M-Theory Dynamics On A Manifold Of $G_2$ Holonomy"},
hep-th/0107177.

\item
\label{Acha}
B. S. Acharya,
{\em "On Realising N=1 Super Yang-Mills in M theory"},
hep-th/0011089.

\item
\label{BGGG}
A. Brandhuber, J. Gomis, S.S. Gubser and  S. Gukov,
{\em "Gauge Theory at Large N and New $G_2$ Holonomy Metrics"},
Nucl.Phys. {\bf B611} (2001) 179,
hep-th/0106034.

\item
\label{TV}
T.R. Taylor and C. Vafa,
{\em "RR Flux on Calabi-Yau and Partial Supersymmetry Breaking"},
Phys. Lett. {\bf B474} (2000) 130, hep-th/9912152.

\item
\label{BW}
C. Beasley and  E. Witten,
{\em "A Note on Fluxes and Superpotentials
in M-theory Compactifications on Manifolds of $G_2$ Holonomy"},
JHEP 0207 (2002) 046,
hep-th/0203061.

\item
\label{W}
E. Witten,
{\em "Phases of $N=2$ Theories In Two Dimensions"},
Nucl. Phys. {\bf B403} (1993) 159, hep-th/9301042.

\item
\label{CIV}
F. Cachazo, K. Intriligator and C. Vafa,
{\em "A Large N Duality via a Geometric Transition"},
Nucl. Phys. {\bf B 603} (2001) 3, hep-th/0103067.

\item
\label{G}
S. Gukov,
{\em "Solitons, Superpotentials and Calibrations"},
Nucl. Phys. {\bf B574} (2000) 169,
hep-th/9911011.
S. Gukov, C. Vafa and E. Witten,
{\em "CFT's From Calabi-Yau Four-folds"},
Nucl. Phys. {\bf B584} (2000) 69; Erratum-ibid. {\bf B608} (2001) 477,
hep-th/9906070.

\item
\label{AS}
B.S. Acharya and B. Spence,
{\em "Flux, Supersymmetry and M theory on 7-manifolds"}, hep-th/0007213.

\item
\label{CK1}
G. Curio and A. Krause,
{\em Four-Flux and Warped Heterotic M-Theory Compactifications},
Nucl. Phys. {\bf B602} (2001) 172,
hep-th/0012152.

\item
\label{CK2}
G. Curio and A. Krause,
{\em G-Fluxes and Non-Perturbative Stabilisation of
Heterotic M-Theory},
Nucl. Phys. {\bf B643} (2002) 131,
hep-th/0108220.

\item
\label{AVI}
M. Aganagic and C. Vafa,
{\em Mirror Symmetry, D-Branes and Counting Holomorphic Discs},
hep-th/0012041.

\item
\label{AVII}
M. Aganagic and C. Vafa,
{\em "Mirror Symmetry and a $G_2$ Flop"}, hep-th/0105225.

\item
\label{AKV}
M. Aganagic, A. Klemm and C. Vafa,
{\em "Disk Instantons, Mirror Symmetry and the Duality Web"},
Z. Naturforsch. {\bf A57} (2002) 1,
hep-th/0105045.

\item
\label{Brand}
A. Brandhuber,
{\em "$G_2$ Holonomy Spaces from Invariant Three-Forms"},
hep-th/0112113, Nucl.Phys. {\bf B629} (2002) 393.

\item
\label{Wi 5brane}
E. Witten,
{\em "Five-Brane Effective Action In M-Theory"},
hep-th/9610234, J.Geom.Phys. {\bf 22} (1997) 103.

\item
\label{Hain}
R. Hain,
{\em "Classical Polylogarithms"},
alg-geom/9202022.
D. Ramakrishnan,
{\em "A regulator for curves via the Heisenberg group"},
Bull. Amer. Math. Soc. ${\bf 5}^2$ (1981) 191.

\item
\label{Hain MacPherson}
R. Hain and R. MacPherson,
{\em "Higher logarithms"},
Ill. Journ. Math. ${\bf 34}^2$ (1990) 392.

\item
\label{Bloch}
S. Bloch,
{\em "Applications of the dilogarithm function"},
Intl. Symp. on Algebraic Geometry, Kyoto (1981) 103.

\item
\label{Milnor}
J. Milnor,
{\em "Hyperbolic Geometry: The first 150 Years"},
Bull. Amer. Math. Soc. {\bf 6} (1982) 9.
J. G. Ratcliffe,
{\em "Foundations of Hyperbolic Manifolds"},
Graduate Texts in Mathematics {\bf 149}, Springer 1994.

\item
\label{Dup}
J.L. Dupont,
{\em "The dilogarithm as a characteristic class for flat bundles"},
Journ. of Pure a. Appl. Alg. {\bf 44} (1987) 137.

\item
\label{Dup Sah}
J.L. Dupont and C.-H. Sah,
{\em "Scissors congruences II},
Journ. of Pure and Appl. Alg. {\bf 25} (1982) 159.

\item
\label{HM}
J.A. Harvey and  G. Moore, {\em "Superpotentials and Membrane Instantons"},
hep-th/9907026.

\item
\label{SW1}
N. Seiberg and E. Witten,
{\em "Monopole Condensation, And Confinement In N=2 Supersymmetric
Yang-Mills Theory"},
Nucl. Phys. {\bf B426} (1994) 19, Erratum-ibid. B430
(1994) 485, hep-th/9407087.

\item
\label{Matone}
M. Matone,
{\em "Instantons and recursion relations in N=2 Susy gauge theory"},
Phys. Lett. {\bf B357} (1995) 342,
hep-th/9506102.

\item
\label{CKLTh}
G. Curio, A. Klemm, D. L\"ust and S. Theisen,
{\em "On the Vacuum Structure of Type II String Compactifications
on Calabi-Yau Spaces with H-Fluxes"},
Nucl. Phys. {\bf B609} (2001) 3, hep-th/0012213.

\item
\label{Hor/Wit}
P. Horava and E. Witten,
{\em "Heterotic and Type I String Dynamics from Eleven Dimensions"},
Nucl.Phys. {\bf B460} (1996) 506, hep-th/9510209;
{\em " Eleven-Dimensional Supergravity on a Manifold with Boundary"},
Nucl.Phys. {\bf B475} (1996) 94,
hep-th/9603142.
E. Witten,
{\em "Strong Coupling Expansion Of Calabi-Yau Compactification"},
Nucl.Phys. {\bf B471} (1996) 135,
hep-th/9602070.

\item
\label{AchW}
B. Acharya and E. Witten,
{\em "Chiral Fermions from Manifolds of $G_2$ Holonomy"},
hep-th/0109152.

\item
\label{CKKL}
G. Curio, A. Klemm, B. K\"ors and D. L\"ust,
{\em "Fluxes in Heterotic and Type II String Compactifications"},
Nucl. Phys. {\bf B620} (2002) 237,
hep-th/0106155.

\item
\label{flop}
B.R. Greene, K. Schalm and G. Shiu,
{\em "Dynamical Topology Change in M Theory"},
J.Math.Phys. {\bf 42} (2001) 3171,
hep-th/0010207.

\item
\label{TianYau}
G. Tian and S.T. Yau,
{\em "Three dimensional algebraic manifolds with $c_1=0$ and
$\chi=-6$"},
in {\em Mathematical Aspects of String theory}, Proc. San Diego (1986) 543.

\item
\label{V flux supo}
C. Vafa,
{\em "Superstrings and Topological Strings at Large N"},
J. Math. Phys. {\bf 42} (2001) 2798,
hep-th/0008142.

\item
\label{W QCD str}
E. Witten,
{\em "Branes And The Dynamics Of QCD"},
Nucl. Phys. {\bf B507} (1997) 658,
hep-th/9706109.

\item
\label{5dim}
G. Curio,
{\em "Superpotential of the $M$-theory conifold and
type IIA string theory"},
hep-th/0212233.

\item
\label{KKL}
G. Curio, B. Kors and D. L\"ust,
{\em "Fluxes and Branes in Type II Vacua and M-theory Geometry
with G(2) and Spin(7) Holonomy"},
hep-th/0111165.

\item
\label{Joyce}
D. Joyce, {\em "On counting special Lagrangian homology 3-spheres"},
hep-th/9907013.

\item
\label{KM}
S. Kachru and J. McGreevy,
{\em "Supersymmetric Three-cycles and (Super)symmetry Breaking"},
Phys. Rev. {\bf D61} (2000) 026001,
hep-th/9908135.

\item
\label{Denef 1}
F. Denef,
{\em "Supergravity flows and D-brane stability"}
JHEP 0008 (2000) 050,
hep-th/0005049.
{\em "(Dis)assembling Special Lagrangians"},
hep-th/0107152.

\item
\label{Witten CS gravity}
E. Witten,
{\em "2+1 dimensional gravity as an exactly soluble system"},
Nucl. Phys. {\bf B 311} (1988) 46.

\item
\label{global paper}
in preparation.

\item
\label{Richmond Szekeres}
B. Richmond and G. Szekeres,
{\em "Some formulas related to dilogarithms, the zeta function and the
Andrews-Gordon identities"},
J. Austral. Math. Soc. {\bf 31} (1981) 362.
A.N. Kirillov and Yu.N. Reshetikhin,
{\em "Exact solutions of the integrable XXZ Heisenberg model with
arbitrary spin: I, II"},
J. Phys. {\bf A}: Math. Gen. {\bf 20} (1987) 1565, 1587.

\item
\label{Dupont Sah Conformal}
J.L. Dupont and C.H. Sah,
{\em "Dilogarithm Identities in Conformal Field Theory and Group
Homology"},
Commun. Math. Phys. {\bf 161} (1994) 265, hep-th/9303111.

\item
\label{CGT}
R. Caracciolo, F. Gliozzi and R.Tateo,
{\em "A topological invariant of RG flows in 2D integrable quantum
field theories"},
Int. J. Mod. Phys. {\bf B13} (1999) 2927, hep-th/9902094.

\end{enumerate}
\end{document}